\pdfoutput=1
\documentclass[journal=jacs,manuscript=article,layout=traditional,bookmarks=false]{achemso}

\usepackage[version=3]{mhchem} 
\usepackage[bookmarks=false]{hyperref}
\usepackage{orcidlink}
\usepackage{amsmath}
\usepackage{bm}
\hypersetup{
  colorlinks   = true, 
  urlcolor     = blue, 
  linkcolor    = blue, 
  citecolor   = blue 
}

\mathchardef\mhyphen="2D 

\author{Qichen Song\,\orcidlink{0000-0002-1090-4068}}
\affiliation{Department of Chemistry and Chemical Biology, Harvard University, Cambridge, Massachusetts 02138, United States}
\email{qichensong@g.harvard.edu}
\author{Sorren Warkander\,\orcidlink{0000-0001-9969-5604}}
\affiliation{Chemical Sciences Division, Lawrence Berkeley
National Laboratory, Berkeley, California 94720, United States}
\author{Samuel C. Huberman\,\orcidlink{0000-0003-0865-8096}}
\affiliation{Department of Chemical Engineering, McGill University, Montreal, Quebec H3A 0C5, Canada}
\email{samuel.huberman@mcgill.ca}
\title[carrier and phonon]
{Probing carrier and phonon transport in semiconductors all at once through frequency-domain photo-reflectance} 

\abbreviations{FDTR, CCR, CTR}

\begin{document}

\begin{tocentry}
\centering
\includegraphics[width=0.9\textwidth]{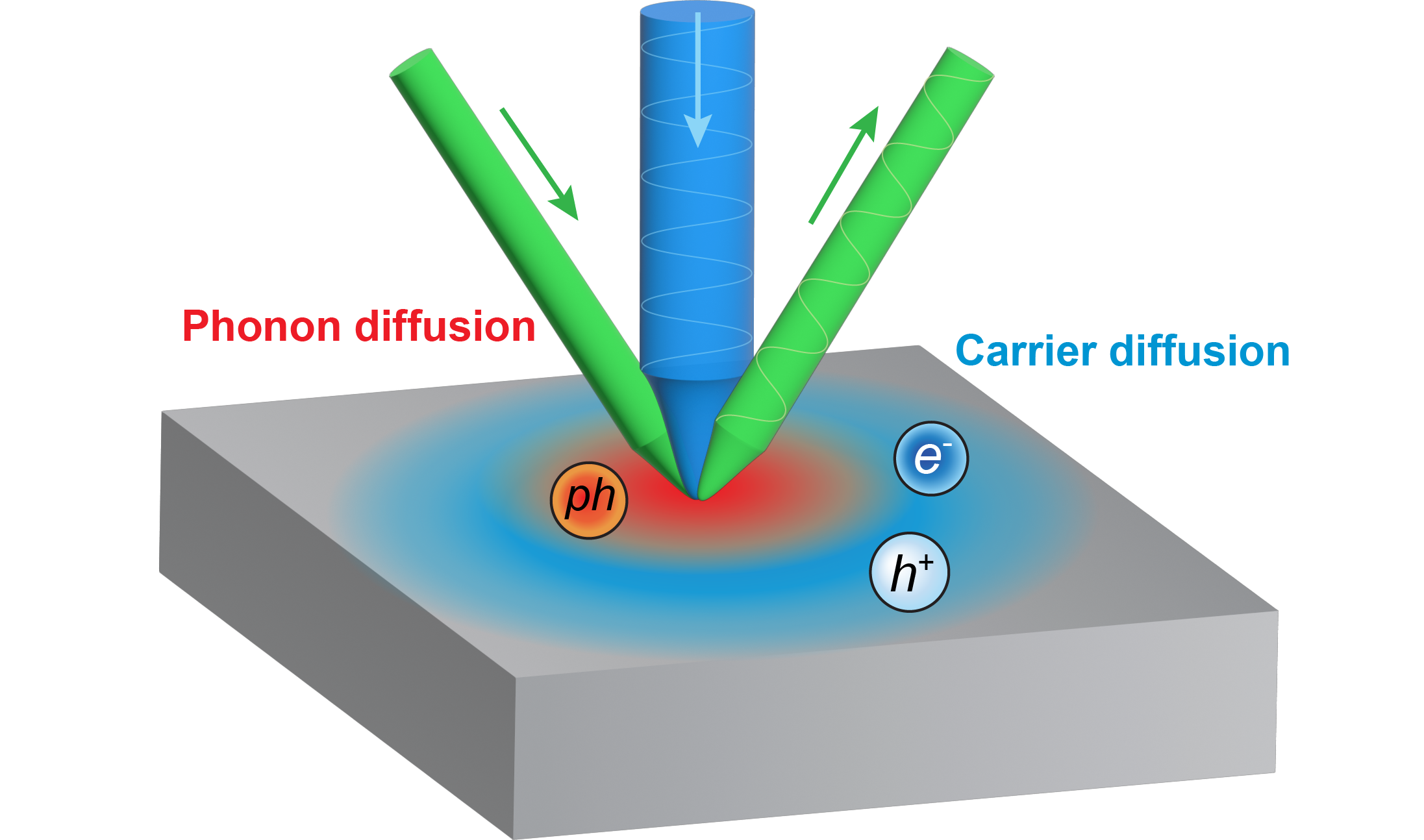}
\end{tocentry}

\begin{abstract}
Semiconductor devices favor high carrier mobility for reduced Joule heating and high thermal conductivity for rapid heat dissipation. The ability to accurately characterize the motion of charge carriers and heat carriers is necessary to improve the performance of electronic devices. However, the conventional approaches of measuring carrier mobility and thermal conductivity require separate and independent measurement techniques. These techniques often involve invasive probing, such as depositing thin metal films on the sample as Ohmic contacts for characterizing electrical transport or as optical transducers for characterizing thermal transport, which becomes more cumbersome as the geometry of the semiconductor devices becomes small and complicated. Here we demonstrate a non-contact frequency-domain pump-probe method that requires no sample pretreatment to simultaneously probe carrier and phonon transport. We find that the optical reflectance depends on both excess carriers and phonons in response to exposure to a modulated continuous-wave pump laser source. By modeling the ambipolar diffusion of photo-induced excess carriers, energy transfer between electrons and phonons, and phonon diffusion, we are able to extract temperature-dependent electrical and thermal transport coefficients in Si, Ge, SiGe and GaAs. The intrinsically weak perturbation from the continuous-wave pump laser avoids invoking the highly nonequilibrium transport regime that may arise when using a pulsed laser and allows for accurate assessment of electrical transport in actual devices. Our approach provides a convenient and accurate platform for the study of the charge transport and energy dissipation in semiconductors.
\end{abstract}

\section{Introduction}

In the age of artificial intelligence, we witness an ever-growing demand for computation. 
Modern computer chips consist of semiconductor transistors as the fundamental building block and the transistor density has grown rapidly from several hundreds to over one hundred million per mm$^2$ during the past half century\cite{burg2021moore,cao2023future}. 
However, the power density has grown concurrently, leading to the formidable challenge of effectively dissipating the increasing amount of heat\cite{pop2006heat,lundstrom2022moore}. To suppress the heat generation and enhance the heat dissipation, materials with high carrier mobility which causes less Joule heating and high phonon thermal conductivity which lowers the chip temperature rise are desirable.
The ability to accurately characterize the carrier and phonon transport in semiconductor devices is vital for the discovery of new materials with desired properties and the rational design of electronic devices with boosted efficiency. Conventionally, carrier transport and phonon transport are separately characterized with different techniques. For instance, the carrier mobility can be measured by the
time-of-flight technique\cite{PhysRev.119.1226} or Hall effect measurement\cite{PhysRev.96.28}, and
the thermal conductivity can be measured using 3$\omega$ method\cite{cahill1990thermal}, time-domain thermoreflectance (TDTR)\cite{paddock1986transient,capinski1996improved,cahill2004analysis} or frequency-domain thermoreflectance (FDTR)\cite{rosencwaig1985detection,schmidt2009frequency}. However, these transport characterization techniques all require the deposition of a metallic film on the sample as an electrode or a heater, which significantly lowers the sensitivity to the transport coefficients in thin-film samples. 

Direct optical pump-probe measurement on the bare sample is a truly non-invasive method, at the cost of creating a more complicated coupled transport scenario where the electronic and lattice degrees of freedom are both excited.
Transient reflectance over several tens of picoseconds after the pump laser pulse encodes transient carrier density and temperature rise. For example, in silicon, the carrier and temperature contribute to the transient reflectance at 760 nm with opposite signs \cite{tanaka1997subpicosecond}. 
The transient grating (TG) technique\cite{PhysRevB.25.2645,choudhry2021characterizing} has been used to probe the ambipolar mobility and thermal conductivity simultaneously in cubic boron arsenide, where a pulsed probe and a continuous-wave (CW) probe are used to probe the carrier ambipolar diffusion at short times ($\sim$100 ps) and the phonon transport in the long times ($\sim$100 ns)\cite{PhysRevLett.110.025901,shin2022high}. 
TDTR without using metal transducer\cite{wang2016thermal,warkander2022transducerless} has recently been used to measure the thermal conductivity in semiconductors and the contribution of the carrier to the signal is modeled and subtracted.
However, the pulsed pump beam leads to highly nonequilibrium electrical transport, where the large amount of excited carriers experience complicated high-order recombination processes and strong electron-phonon interactions\cite{PhysRevX.6.021003,wilson2020parametric}, making it difficult to study electrical transport properties intrinsic to the materials with enough fidelity. In contrast, conducting the pump-probe measurement in the frequency-domain using modulated CW lasers only weakly perturbs the material with an amplitude uniform in time, which is closer to the actual transport regime in devices. 
Transducerless FDTR using CW lasers has been used to measure several selected materials with minimal carrier effect such that the signal is still proportional to temperature\cite{qian2020accurate,zhang2017anomalous}. However, the temporal behavior of optical reflectance under modulated laser irradiation encodes information about the dynamics of charge carriers\cite{opsal1987temporal} yet the ability to simultaneously probe electrical and thermal transport in frequency-domain photo-reflectance measurement has, to the best of our knowledge, never been explored.

In this paper, we present a method of probing the carrier and phonon transport in semiconductors in the frequency domain, where the time-harmonic responses of carrier and phonon coexist. The electron-phonon interaction is found to be insignificant below the characteristic frequency for electron-phonon coupling determined by the material properties. The carrier diffusion is nearly in-phase, whereas the phonon diffusion is responsible for a larger phase lag that varies more rapidly with modulation frequency. This difference in dependence on modulation frequency enables the extraction of the ambipolar diffusion coefficient and phonon thermal conductivity in a single-shot frequency-sweep measurement. Our work provides a simple approach to characterize the carrier and phonon transport in semiconductors, which can be readily applied to the study of actual devices. 

\section{Results and discussion}

\begin{figure}[t]
  \includegraphics[width=0.9\textwidth]{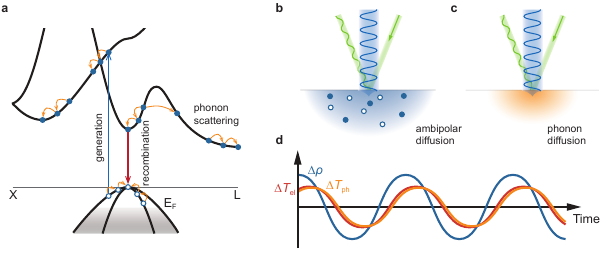}
  \caption{Schematic for the frequency-domain photo-reflectance. a. The carrier dynamics induced by pump photons in \textit{p}-type Ge. The incident photon first generates an electron-hole pair across the direct band gap. The excess free carriers (electrons/holes) interact with phonons and quickly thermalize achieving the electron temperature $T_\mathrm{el}$. The phonons are excited by those hot carriers with temperature $T_\mathrm{ph}$. b-c. The pump and probe laser geometry in experiment. The reflectance for the probe beam is modulated by the carriers and phonons. The reflected probe beam consists DC and AC parts. The AC part contains the information of transport processes including carrier diffusion (b) and phonon diffusion (c). d. The real parts of the AC signals due to excess carriers, electron temperature and phonon temperature dynamics. The power of the modulated pump is assumed to be $P(t)\propto e^{i\omega t}$ and its real part is a cosine wave $\mathrm{Re}\{ P(t)\} \propto \cos(\omega t)$. In semiconductors with strong electron-phonon interaction under modest modulation frequency, the electron and phonon temperatures are approximately the same. The carriers normally diffuse faster than phonons, thus having a smaller phase lag.}
  \label{fig1}
\end{figure}

We use a 458 nm CW pump laser with its power being modulated at the given frequency $f$. The pump-induced change in reflectance of a 532 nm probe laser is modulated by the relevant excited degrees of freedoms in the material at the same frequency such that $\Delta R(t) = \Delta R e^{i\omega t}$, where $\omega = 2\pi f$ is the angular frequency. Because of the optically-induced nature, we call $\Delta R$ the photo-reflectance. Details of the experimental setup can be found in the Methods section.
Fig.~\ref{fig1} presents the carrier dynamics in semiconductors under the irradiation of a modulated laser. The absorbed pump photons excite electrons across the band gap to generate electron-hole pairs. 
The excess electrons/holes then start to diffuse away from the center of the pump laser spot (Fig.~\ref{fig1} b).
When colliding with other carriers and phonons, the excited electrons/holes quickly thermalize (within 1 ps in silicon\cite{PhysRevLett.72.1364}) among themselves via electron-electron scattering\cite{PhysRevB.26.2147,PhysRevB.56.15252} and electron-phonon scattering\cite{PhysRevB.97.064302}, particularly with phonons that have low energy and high momentum to populate different valleys. The thermalized carriers follow a quasi-Fermi-Dirac distribution with an effective electron temperature $T_\mathrm{el}$, referred to as hot electron temperature\cite{PhysRevB.19.5928}. 
The hot electrons/holes gradually transfer their energy to the lattice via electron-phonon interaction\cite{PhysRevB.97.064302}, which heats up the phonon subsystem. The phonons thermalize with other phonons through anharmonic phonon-phonon scattering achieving a local temperature $T_\mathrm{ph}$ and diffuse towards colder areas, depicted in Fig.~\ref{fig1} c.
In parallel to these scattering events, the electrons and holes experience frequent recombination events that annihilate electron-hole pairs (Fig.~\ref{fig1} a). 

To quantify the pump-induced dynamics,
we model the transport processes of excited charge and energy carriers
using the ambipolar diffusion equation with carrier generation and recombination, the electron heat equation and the phonon heat equation with electron-phonon energy transfer,
\begin{equation}
    \frac{\partial \rho}{\partial t} = \nabla\cdot \mathbf{D}_a \cdot \nabla \rho -\frac{\rho}{\tau}+\frac{P}{E_\mathrm{photon}},
    \label{eq1}
\end{equation}
\begin{equation}
    C_\mathrm{el}\frac{\partial T_\mathrm{el}}{\partial t}=\nabla\cdot\bm{\kappa}_\mathrm{el} \cdot \nabla T_\mathrm{el}
    - g_\mathrm{e\mhyphen p}\left(T_\mathrm{el}-T_\mathrm{ph}\right)+\frac{\rho E_\mathrm{gap}}{\tau}+\frac{P\Delta E}{E_\mathrm{photon}},
\label{eq2}
\end{equation}
\begin{equation}
    C_\mathrm{ph}\frac{\partial T_\mathrm{ph}}{\partial t}=\nabla\cdot\bm{\kappa}_\mathrm{ph} \cdot\nabla T_\mathrm{ph} + g_\mathrm{e\mhyphen p}\left(T_\mathrm{el}-T_\mathrm{ph}\right),
    \label{eq3}
\end{equation}
where $\rho$, $T_\mathrm{el}$ and $T_\mathrm{ph}$ are the excess carrier density, electron temperature and phonon temperature, respectively. $\mathbf{D}_a$, $\bm{\kappa}_\mathrm{el}$ and $\bm{\kappa}_\mathrm{ph}$ are the ambipolar diffusivity, electron thermal conductivity and phonon thermal conductivity tensors. $g_\mathrm{e\mhyphen p}$ is the electron-phonon coupling factor. $\tau$ is the effective lifetime of the excess carrier, predominantly determined by the radiative recombination process caused by above-bandgap emission at modest defect concentrations\cite{schroder2015semiconductor}. $E_\mathrm{photon}$ is the energy of a pump photon. 
$P=\frac{2P_0\alpha}{\pi \sigma_x\sigma_y} e^{-2x^2/\sigma_x^2}e^{-2y^2/\sigma_y^2}e^{-\alpha z}e^{i\omega t}$ is the absorbed laser power density.
Here $\alpha$ is the inverse penetration depth for the pump beam. $P_0 = P_\mathrm{pump}(1-R)$ is absorbed pump power, with $P_\mathrm{pump}$ the peak pump power, $R=\frac{(n-1)^2+\kappa^2}{(n+1)^2+\kappa^2}$ the reflectance at the pump wavelength, and $n$ and $\kappa$ the real and imaginary parts of the refractive index. $\sigma_x$ and $\sigma_y$ are the $1/e^2$ beam radii for the pump beam in $x$- and $y$- directions.
$E_\mathrm{gap}$ is the direct bandgap energy and the second-to-last term in Eq.~(\ref{eq2}) describes the heating due to carrier recombination.
The remaining energy transferred from a photon to the hot carrier ensemble is given by $\Delta E = E_\mathrm{photon} - E_\mathrm{gap}$.
We apply the Fourier transform to Eqs.~(\ref{eq1})-(\ref{eq3}) in the in-plane directions to solve for $\tilde{\rho}(\mathbf{q}_\parallel,z,t)$, $\tilde{T}_\mathrm{el}(\mathbf{q}_\parallel,z,t)$ and $\tilde{T}_\mathrm{ph}(\mathbf{q}_\parallel,z,t)$.
Then, the inverse Fourier transform renders the spatial profiles of $\rho(\mathbf{r},t)$, $T_\mathrm{el}(\mathbf{r},t)$ and $T_\mathrm{ph}(\mathbf{r},t)$ (Sec. II and III in SI). Note that the carrier density and temperatures are deviations from their corresponding background values (AC parts) rather than the absolute values (DC+AC parts). For conciseness, we will refer to the deviation in the excess carrier density as the excess carrier density and the deviation in temperature as temperature in the following.


The photo-reflectance of the probe beam encodes the contributions from the time-varying excess carrier density, electron and phonon temperatures. In the limit of linear response,
\begin{equation}
  \frac{\Delta R}{R} 
 = \mathrm{Re}\left\{A\overline{\rho} + B \overline{T}_\mathrm{el} + C \overline{T}_\mathrm{ph}\right\},
 \label{signal}
\end{equation}
where the overline indicates the weighted average of the physical quantity $X$ over the probe beam profile, such that $\overline{X} = \frac{2\gamma}{\pi \sigma_x^\prime \sigma_y^\prime}\int X(\mathbf{r})e^{-2(x-x_\mathrm{o})^2/\sigma_x^{\prime 2}}e^{-2(y-y_\mathrm{o})^2/\sigma_y^{\prime 2}}e^{-\gamma z}d^3\mathbf{r}$, where $\sigma_x^{\prime}$ and $\sigma_y^{\prime}$ are the probe beam waists, $(x_\mathrm{o}, y_\mathrm{o})$ is the position of probe beam center relative to the pump center and $\gamma$ is the inverse penetration depth of the probe beam. $A$ is the coefficient of carrier-induced reflectance (CCR), $B$ is the coefficient of hot electron thermoreflectance and $C$ is the coefficient of phonon thermoreflectance.
The carrier-induced reflectance originates from the dielectric response of free carriers introduced by the pump beam and the order of magnitude of CCR can be estimated using the Drude model (Sec. I in SI).
The electron and phonon thermoreflectance are caused by temperature-dependent electronic band structures and electron damping\cite{PhysRev.176.950,PhysRevB.11.5077,PhysRevMaterials.5.106001,block2023observation}.

By examining the Eqs.~(\ref{eq2}) and (\ref{eq3}), we find that the electron and phonon temperature obey a simple relation, namely $\overline{T}_\mathrm{el} \approx \overline{T}_\mathrm{ph} \left(1+\sqrt{i\omega/\omega_\mathrm{e\mhyphen p}}\right)$, 
where the characteristic frequency for electron-phonon coupling $\omega_\mathrm{e\mhyphen p} =\frac{g_\mathrm{e\mhyphen p}\kappa_\mathrm{el}}{C_\mathrm{ph}\kappa_\mathrm{ph}}$ (Sec. VII.B in supplementary material). 
This expression implies that the electron temperature has a smaller phase lag than phonon temperature when the modulation frequency is much smaller than $\omega_\mathrm{e\mhyphen p}$, as depicted in Fig.~\ref{fig1} d.
In Ge, the characteristic frequency is found to be $\frac{\omega_\mathrm{e\mhyphen p}}{2\pi} = 332$ MHz, where we take $C_\mathrm{ph}$ = $1.6 \times 10^6$ $\mathrm{J/(m^3\cdot K)}$, $\kappa_\mathrm{ph} = 60$
$\mathrm{W/(m\cdot K)}$, $\kappa_\mathrm{el} = 2$ $\mathrm{W/(m\cdot K)}$ and $g_\mathrm{e\mhyphen p} =10^{17}$ $\mathrm{W/(m^3\cdot K)}$ (according to Ref. \cite{PhysRevLett.119.136602}, $g_\mathrm{e\mhyphen p}\sim 10^{17}-10^{18}$ $\mathrm{W/(m^3\cdot K)}$ in GaAs, Si, AlAs, \textit{etc.}).
In the typical modulation frequency range in our experiment (500 Hz to 50 MHz),
we make the approximation $\overline{T}_\mathrm{el}\approx \overline{T}_\mathrm{ph}$. We can therefore combine the electron and phonon temperatures' contributions in Eq.~(\ref{signal}) into a single term $C\overline{T}_\mathrm{ph}$, where $C$ can be regarded as the effective coefficient of thermoreflectance (CTR). 
In this case, the photo-reflectance signal consists of carrier and phonon contributions only.
We call such simplified model the carrier and phonon model, in contrast to the model with the explicit inclusion of a distinct electron temperature which we call the full model (Sec. IV in SI).

\begin{figure}[ht]
  \includegraphics[width=\textwidth]{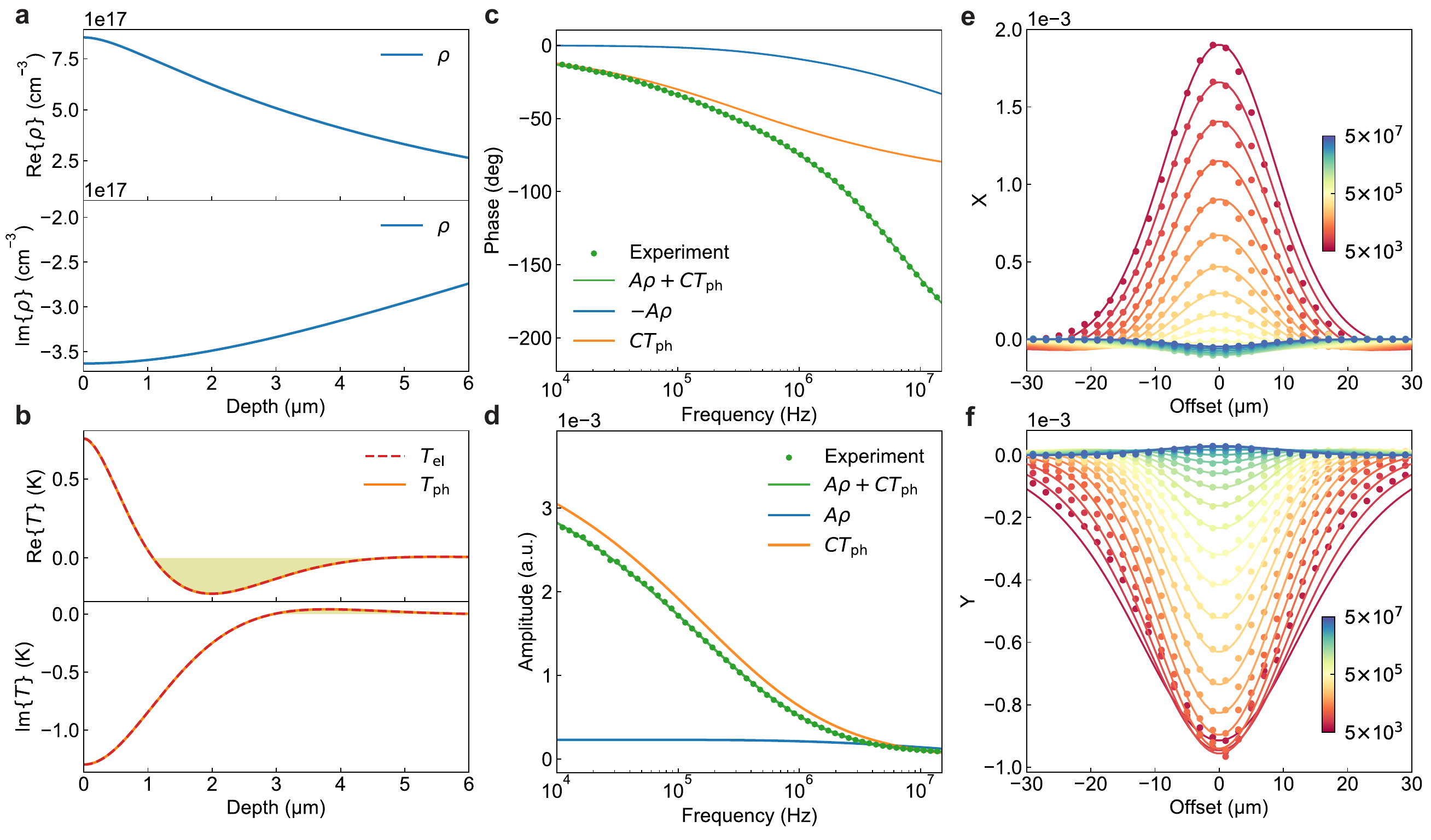}
  \caption{a. The excess carrier density profile $\overline{\rho}(z)$ and b. the electron and phonon temperature distribution $\overline{T}_\mathrm{el}(z)$ and $\overline{T}_\mathrm{ph}(z)$ in \textit{p}-type Si$_{0.98}$Ge$_{0.02}$ with modulation frequency $f=7.96$ MHz and absorbed pump power $P_0 = 78 $ mW at 273.15 K.
  The weighted average of the carrier and temperatures can be computed by $\overline{X} = \int_0^\infty \overline{X}(z)e^{-\gamma z}dz$. 
  The shaded area indicates the region where electron temperature has the opposite sign to the electron temperature at the surface. c-d. The phase and amplitude of photo-reflectance $\frac{\Delta R}{R}$ in \textit{p}-type Si$_{0.98}$Ge$_{0.02}$ at 273.15 K. Note in c we multiply the carrier part by negative one to plot the phases of different part of the signal together. e-f. The real and imaginary parts (X and Y) of the photo-reflectance as a function of the beam offset distance, where the color represents the modulation frequency.}
  \label{fig2}
\end{figure}

The average carrier density, electron and phonon temperature under modulation are all represented by complex numbers, obtained from solving the transport equations, Eqs.~(\ref{eq1}), (\ref{eq2}), and (\ref{eq3}). The Fig.~\ref{fig2} a illustrates that the real and imaginary part of the carrier density $\overline{\rho}(z)$ in Si$_\mathrm{0.98}$Ge$_\mathrm{0.02}$ decreases with depth as a consequence of ambipolar carrier diffusion. The unique feature of ambipolar diffusion under modulation is that it involves two types of diffusion lengths.
In Si$_\mathrm{0.98}$Ge$_\mathrm{0.02}$, the ambipolar diffusivity is $D_\mathrm{a} = $ 30.4 $\mathrm{cm^2/s}$ and the carrier lifetime is $\tau = $ 0.52 $\mu$s at 273.15 K, yielding the ambipolar diffusion length $\sqrt{D_\mathrm{a}\tau}=$ 39.8 $\mu$m. However, at a modulation frequency $f$ = 7.96 MHz, the decay length of the carrier density is much smaller than the ambipolar diffusion length, as it is also limited by the diffusion length over one period $\sqrt{D_\mathrm{a}/(2\pi f)} = $ 7.8 $\mu$m (Sec. II.A in SI). The excited carrier density is on the order of $10^{17}$ $\mathrm{cm}^{-3}$, much smaller than the typical excited carrier density by a pulse laser ($\sim 10^{19}$ $\mathrm{cm}^{-3}$)\cite{zhou2020direct}, which ensures the carrier transport is in the linear response regime.
The phonon temperature decays with distance to surface more rapidly than the carrier density (Fig.~\ref{fig2} b).
The phonon thermal diffusivity is given by $D_\mathrm{ph}=\frac{\kappa_\mathrm{ph}}{C_\mathrm{ph}}$ = 0.29 $\mathrm{cm^2/s}$. Hence, the phonon diffusion length over one period (also known as the thermal penetration depth) is only  $\sqrt{D_\mathrm{ph}/(2\pi f)}$ = 0.76 $\mu$m. Both the real and imaginary parts of phonon temperature decay with depth while oscillating across zero, as there is a spatially varying phase lag arising from the phonon diffusion (Sec. IX in SI).
The electron temperature has almost exactly the same distribution to the phonon temperature since $f \ll \omega_\mathrm{e\mhyphen p}/(2\pi)$, justifying our use of the simplified carrier and phonon model.

The transport coefficients used in Fig.~\ref{fig2} a and b are obtained by fitting the experimental data with the carrier and phonon model.
The phase and amplitude of the measured photo-reflectance $\frac{\Delta R}{R}$ are presented Fig.~\ref{fig2} c and d.
Using the carrier and phonon model, we fit the phase and amplitude simultaneously to utilize the full information of the signal.
We choose $A,$ $C$, $D_\mathrm{a}$ and $\kappa_\mathrm{ph}$ as fitting variables, as these variables have higher fitting sensitivities than others (see the sensitivity analysis in Sec. VI in SI). We observe that the carrier and phonon model fits experimental data well. As the modulation frequency increases, the carrier transport plays a more and more important role. At low modulation frequencies, the phase lag of the carrier density is zero and its amplitude approaches a constant.

The effective beam radius is a geometrical measure of carrier generation source, defined by $r_\mathrm{eff} = \sqrt{(\sigma_\mathrm{pump}^2+\sigma_\mathrm{probe}^2)/2}$, where $\sigma_\mathrm{pump}$ and $\sigma_\mathrm{probe}$ are the beam radii for the pump and probe beam. We find the effective beam radius is $r_\mathrm{eff}$ = 10 $\mu$m, by directly extracting it from a beam offset measurement at highest modulation frequency.
Since the effective beam radius is much smaller than the ambipolar diffusion length, the ambipolar diffusion reaches the steady state limit $\bar{\rho}(\omega = 0)=\frac{P_0}{2\sqrt{\pi}E_\mathrm{photon}D_\mathrm{a}r_\mathrm{eff}}$ at low modulation frequencies $\omega \ll \frac{4\pi^2D_\mathrm{a}}{r^2_\mathrm{eff}}$ (Sec. II.A in SI). 
In Si$_{0.98}$Ge$_{0.02}$, the phonon diffusion length is only slightly larger than the effective beam size at low modulation frequencies. With decreasing frequency, the phonon temperature has sufficient time to closely follow the modulated heating, hence the amplitude of phonon temperature increases and the phase lag decreases. As a result, at low modulation frequencies, the phase is dominantly determined by the phonon temperature. As for the overall amplitude, it is dominated by the phonon temperature, yet the carrier part cannot be neglected.
At high modulation frequencies, the phase lag of the carrier contribution starts to increase, since the diffusion length over one period $\sqrt{D_\mathrm{a}/(2\pi f)}$ begins to limit the carrier distribution. Note that the carrier and phonon terms contribute to the photo-reflectance with the opposite signs. Consequently, the amplitude of photo-reflectance is slightly lower than that of the phonon part above 50 MHz. The phase lag of the photo-reflectance becomes much larger than that of phonon temperature. Eventually the phase lag surpasses 90$^\circ$, which is the upper bound of temperature given by the Fourier heat conduction model. This is a strong evidence that the reflectance signal cannot explained by the phonon system alone. 

We also study the case where the pump and probe beams are offset from one another. In Fig.~\ref{fig2} e and f, we show the real and imaginary parts of photo-reflectance at various modulation frequencies when offsetting the pump with respect to the probe. 
At high modulation frequencies, the real and imaginary parts follow a Gaussian-like distribution with respect to the beam offset distance. This is because the carrier and phonon diffusion are both smaller than the effective beam size and the geometry of excitation source plays a more important role. At the modulation frequency decreases, both the carrier and phonon diffusion lengths increase, which is observed in the increase of the full width at half maximum of the real and imaginary components of the reflectance signals.
We note that the real and imaginary components of the reflectance signals both experience a change of sign at certain modulation frequencies, which is not explainable by the Fourier heat conduction model alone, but can be simply explained as a natural consequence of two contributions, the carrier density and phonon temperature, respectively, having opposite signs. In other words, the CCR and CTR have opposite signs.
The solid lines are solutions to the carrier and phonon model using the best fits without offset in Fig.~\ref{fig2} c and d. Although we did not directly fit the data with offset, the fact that the model and experiment match well with physically realistic parameters infers that the carrier and phonon model captures the essential physics correctly. The deviation between model and experiment at large offset distance and low modulation frequency may be the result of a slightly asymmetric tilt in beam incidence with large offset as well as the impact $1/f$ noise.

\begin{figure}[ht!]
\includegraphics[width=\textwidth]{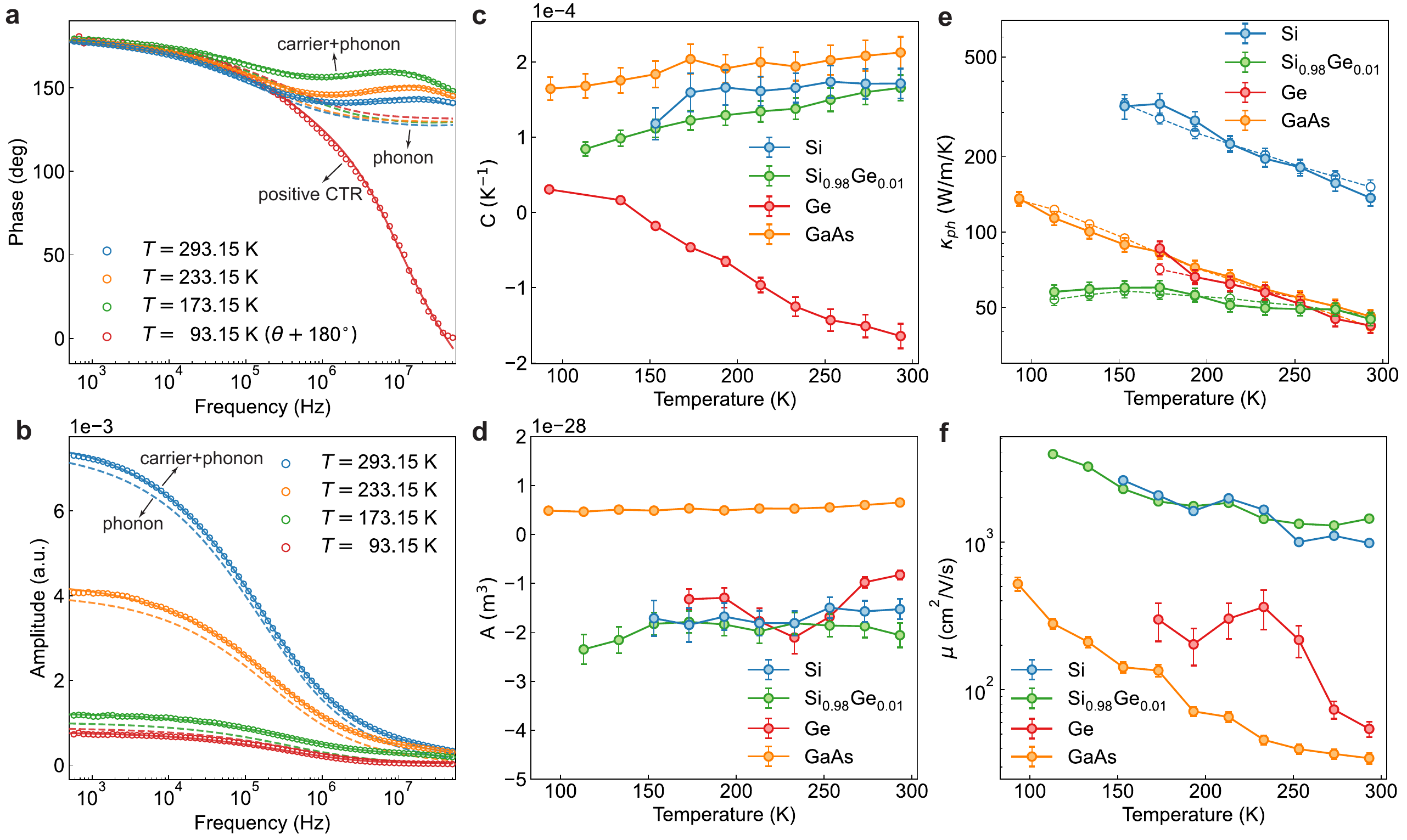}
\caption{Coefficients of reflectance and transport properties from temperature-dependent frequency-domain photo-reflectance in several semiconductors. a-b. The phase and amplitude at different temperatures in \textit{p}-type Ge. The solid line are fitted values of $A\overline{\rho}+C\overline{T}_\mathrm{ph}$ and the dashed lines are the phonon part $C\overline{T}_\mathrm{ph}$. 
Note that at 93.15 K the phases of the photo-reflectance and the phonon part are both shifted by 180 $^{\circ}$ for comparative purposes with other temperatures, as the sign of CTR at 93.15 K is positive while negative at other three temperatures.
c. The coefficients of thermoreflectance as a function of temperature. d. The coefficients of carrier-induced reflectance as a function of temperature. e. The phonon thermal conductivity from the current measurement (filled circle) and a separate measurement with a metal transducer (open circle).  f. The ambipolar mobility as a function of temperature. The lifetime is assumed to be a temperature-independent constant as it mostly depend on the band gap and impurity concentrations and the lifetimes that give the best fits across all temperatures are 1.1 $\mu$s, 0.52 $\mu$s, 5 ns, 9 ns
in Si, Si$_{0.98}$Ge$_{0.02}$, Ge, GaAs, respectively.}
\label{fig3}
\end{figure}

We further validate our model by measuring several semiconductors at various temperatures.
In Fig.~\ref{fig3} a and b, we show the phase and amplitude of photo-reflectance in Ge. 
The sign of the CCR and CTR are the same at high temperatures. The total phase lag is smaller than that of the phonon temperature, as the carrier density generally has a low phase lag. At 93.15 K, the sign of the CTR flips and CCR and CTR now have opposite sign, hence we see that the amplitude of phonon part is larger than the total amplitude. The sign change of the CTR may be due to the competition between different interband electron transition channels as the band structure changes with temperature\cite{PhysRev.176.950}. The sign flip in CTR also significantly impacts the phase of photo-reflectance. At the low temperature, the phase of photo-reflectance curves downward with increasing frequency, when compared with phonon temperature. In contrast, the phase curves upward with frequency at higher temperatures due to the sign change in CTR (Fig. S8 in SI) and eventually curves downward at even higher frequencies due to the shrinking carrier diffusion length.
In Si and Si$_{0.98}$Ge$_{0.02}$, the CCR is negative and the CTR is positive for the entire temperature range. In GaAs, both CCR and CTR are positive for the entire temperature range. According to the Drude model, the real part of refractive index $n$ decreases with increasing carrier density. Qualitatively speaking, the pump-induced free carriers can help to absorb more probe photons via free-carrier absorption\cite{PhysRevB.106.205203}, thus leading to a smaller reflectance. However, when the excited carrier are mostly localized near the surface in a length scale smaller than probe wavelength (e.g. trapped by defects), the electromagnetic field of the probe photon (Sec. I.B in SI) inside the material is only partially altered, where the region with positive $\frac{dn}{d\rho}$ is mostly involved ($n$ is the real part of refractive index). Consequently, the change in reflectance is positive, resulting in a positive CCR.
The CTR generally has a much stronger temperature dependence than the CCR, because the temperature-dependent band structure and electron-phonon scattering rate both contribute to CTR, whereas though a temperature-dependent effective mass can cause change in CCR, it is a more subtle effect than the previously mentioned effects.

The phonon thermal conductivity in Si, Si$_{0.98}$Ge$_{0.02}$, Ge and GaAs from the frequency-domain photo-reflectance agree well with the conventional FDTR measurements on gold-coated samples (the corresponding methodology can be found in Ref.\cite{schmidt2009frequency}).
The ambipolar mobility is given by $\mu =\frac{e D_a}{k_B T}$,
where $e$ is the elementary charge and the ambipolar mobility is related to the unipolar mobilities via $\mu = \frac{2\mu_n \mu_p }{\mu_n + \mu_p}$, with $\mu_n$ and $\mu_p$ the electron and hole mobilities, respectively\cite{PhysRevB.64.085307}. We find that the ambipolar mobility in Si and Si$_{0.98}$Ge$_{0.02}$ are similar while thermal conductivity in Si$_{0.98}$Ge$_{0.02}$ is much lower than that in Si, implying that the alloy scattering affects phonon transport more than carrier transport. GaAs has the lowest ambipolar mobility, as the low intrinsic hole mobility in GaAs significantly limits the ambipolar mobility despite the intrinsic high electron mobility. The mobility in Ge has a peak around 235 K. The peak can be partially due to the temperature dependent electronic band structures and partially due to the uncertainty in fitting the ambipolar diffusion coefficient. As shown in Fig.~\ref{fig3} e, there is a peak in CCR around the same temperature as well, which signifies that CCR and ambipolar diffusivity can be 
inversely correlated in fitting.




\begin{figure}[ht!]
  \centering
  \includegraphics[width=0.88\textwidth]{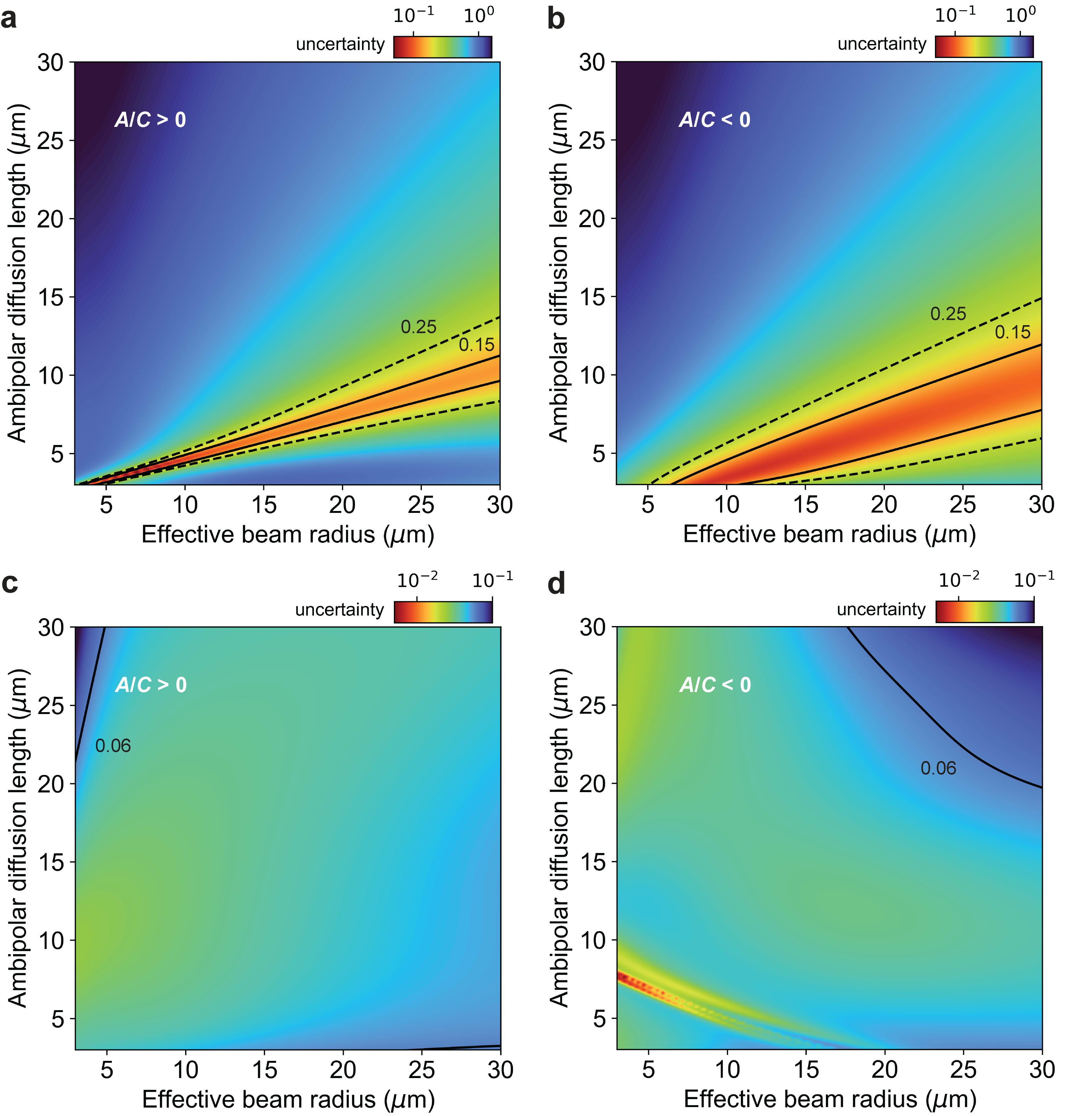}
  \caption{The uncertainty of $D_\mathrm{a}$ when fitting $D_\mathrm{a}$ and $A$, as a function of the effective radius and the ambipolar diffusion length in Ge: a. $A/C = 10^{24}$ $\mathrm{m}^{3}\cdot \mathrm{K}$ and b. $A/C = -10^{24}$ $\mathrm{m}^{3}\cdot\mathrm{K}$ and in Si$_{0.98}$Ge$_{0.02}$: c. $A/C = 10^{24}$ $\mathrm{m}^{3}\cdot\mathrm{K}$ and d. $A/C = -10^{24}$ $\mathrm{m}^{3}\cdot \mathrm{K}$. The color represents the common logarithm of the uncertainty. The uncertainty in the ambipolar diffusivity is lower in Si$_{0.98}$Ge$_{0.02}$ than Ge, since Si$_{0.98}$Ge$_{0.02}$ has a larger carrier lifetime. Note that to avoid the complexity from the choice of the scaling constant in front of the amplitude when fitting the phase and amplitude together, we consider the case of only fitting phase in the uncertainty analysis.}
  \label{fig4}
\end{figure}

To better understand the source of uncertainty in our fitting scheme,
we conduct an uncertainty analysis. We first consider the case when $A/C =  10^{-24} $ $\mathrm{m}^{3}\cdot\mathrm{K}$ and $A/C = -10^{-24}$ $\mathrm{m}^{3}\cdot\mathrm{K}$, with the remaining parameters taken from Ge's properties at 293.15 K. We then vary the pump beam radius and the ambipolar diffusion constant and calculate the corresponding uncertainty in $D_\mathrm{a}$ when fitting $D_\mathrm{a}$ and $A$ together. The carrier lifetime in Ge is fixed at 5 ns, faster than most of the modulation periods. 
The carrier density at different $\mathbf{q}_\parallel$ is expressed by $\rho(\mathbf{q}_\parallel,z)\propto \frac{P_0}{D_a\sqrt{q_\parallel^2+(D_\mathrm{a}\tau)^{-1}+i\omega D^{-1}_\mathrm{a}}}$. As a result, the carrier part of the signal has a negligible imaginary part at most modulation frequencies. When the ambipolar diffusion length is significantly larger than the beam size, the carrier density approaches the steady state limit, $\overline{\rho}\propto \frac{P_0}{D_\mathrm{a}}$. When the ambipolar diffusion length is much smaller than the beam size, the carrier density is found to satisfy $\overline{\rho}\propto \frac{P_0}{\sqrt{D_\mathrm{a}}}$ (Sec. II in SI). In both cases, the mobility and the CCR are highly anti-correlated, preventing us from fitting them together with high confidence. When the ambipolar diffusion length is just slightly smaller than the beam size ($\sqrt{D_\mathrm{a}\tau} \sim \frac{r_\mathrm{eff}}{2\pi}$), the carrier density experiences both limits when sweeping the modulation frequency. This is advantageous since the change in the way that the signal depends on diffusivity lowers the uncertainty. In Ge, the beam radius is much greater than the ambipolar diffusion length, corresponding to a region far below the zone of theoretical minimum uncertainty ($r_\mathrm{eff} = 10$ $\mu \mathrm{m}$, $\sqrt{D_\mathrm{a}\tau}= 0.08$ $\mu \mathrm{m}$ in Ge at 293.15 K) in Fig.~\ref{fig4} a. Accordingly, we observe the enlarged uncertainty in Ge in Fig.~\ref{fig3} f. Other the other hand, if we flip the sign of the CCR, as illustrated in Fig.~\ref{fig4} b, the uncertainty is found to decrease. This is because the opposite sign of CCR and CTR combine to increase the phase lag as compared with the phase lag of the phonon system alone. Consequently, in the same frequency-sweep range, the phase varies over a larger range thus leading to a higher sensitivity to and a lower uncertainty in the ambipolar diffusivity. 

In Si$_\mathrm{0.98}$Ge$_\mathrm{0.02}$, the uncertainty is found to be much lower than Ge, for either case of $A/C>0$ or $A/C<0$. The carrier lifetime in Si$_\mathrm{0.98}$Ge$_\mathrm{0.02}$ is 0.52 $\mu$s, much larger than that in Ge. When the modulation frequency is on the same order of magnitude $\omega \sim \tau^{-1}$, the imaginary part in the carrier density becomes important. As we sweep the modulation frequency from low to high values, the carrier transport can be limited by the beam size, the diffusion length over one modulation period or the ambipolar diffusion length, depending on which quantity is the smallest. The carrier part of the signal is more sensitive to the diffusivity, which leads to a lower uncertainty than Ge. When $A/C >0$, the real parts of the carrier density and the phonon temperature are both positive and the phase lag of the signal is between the phase of the phonon temperature and the carrier density. When $A/C<0$, the real parts of the carrier density and phonon temperature can cancel and the real part of the total signal can be either positive or negative. The narrow stripe with the lowest uncertainty corresponds to the near-zero real part at low modulation frequencies with a phase lag close to 90$^{\circ}$. When the ambipolar diffusion length is above or below this stripe, the phase at low frequencies either approaches the phase of the phonon temperature (0$^{\circ}$ when $\omega\to 0$) or the phase of the carrier density (180$^{\circ}$ when $\omega\to 0$). As a consequence, near the stripe, the phase is highly sensitive to the carrier diffusivity.
At large beam radius and large ambipolar diffusion length, the phonon part and carrier part of the signal may cancel (both the real and imaginary parts), leading to a near zero total signal and leading to increased uncertainty. On the other hand, when $A/C >0$, a zero component in signal case will never occur for a finite modulation frequency.

To conclude, we model the carrier- and phonon- induced frequency-domain photo-reflectance and explore the possibility of applying the model to extract transport properties in semiconductors. In common semiconductors, the CCR is almost temperature-independent and the CTR varies more with temperature. With the proper choice of modulation frequency and beam size, we can measure the ambipolar mobility, thermal conductivity, CCR and CTR at the same time with high confidence.  Our modelling can be applied to analyze the transient reflectance as well, which expand the possibility of using optical-reflectance to probe electron transport and phonon transport simultaneously.

\section{Methods}

We use a fiber laser (Coherent OBIS LX 458 nm) with digital modulation of the power as the pump and a CW laser (Coherent OBIS LS 532 nm) as the probe. Little beam asymmetry along $x$- and $y$- directions for pump and probe are found therefore we assume pump and probe beam are both isotropic. The $1/e^2$ radius of the probe is $r_\mathrm{probe} = $ 2.98 $\mu$m, determined by the knife edge method (Fig. S18). The effective beam radius can be extracted from the beam offset measurement on a sample coated with gold thin film at the highest modulation frequency, as the amplitude at such frequency is following a Gaussian distribution and the standard deviation is equal to $r_\mathrm{eff}/\sqrt{2}$. Then, the pump radius is $r_\mathrm{pump} = \sqrt{2r_\mathrm{eff}^2-r^2_\mathrm{probe}}$. The beam offset is achieved by moving the angle of the mirror to change the location of the pump beam while the probe beam is unchanged. The perturbed probe beam goes through a half-waveplate first before reaching the sample. The half-waveplate is rotated by a motor until reaching an angle that minimize the intensity difference between the perturbed (reflected off the sample) and unperturbed probe. Then, the unperturbed probe and perturbed probe are sent to a balanced photodetector. The detected signal is amplified by the lock-in amplifier (Zurich Instruments HF2LI). We take the ratio of the AC part divided by the DC part as the photo-reflectance signal. Note that we measure the pump phase before measuring the sample and make sure it is subtracted. We also make sure the perturbed and unperturbed probe beam do not have any optical path difference as it can lead to a larger phase difference at higher modulation frequencies. Other details of the setup can be found in Ref.\cite{yang2013thermal}.

The Si, Si$_\mathrm{0.98}$Ge$_\mathrm{0.02}$, Ge and GaAs are commercial available (MTI Corp.) with resistivity 1-10 $\mathrm{\Omega\cdot cm}$, $>$50 $\mathrm{\Omega\cdot cm}$, 0.0007-0.005 $\mathrm{\Omega\cdot cm}$, 0.6$\times 10^8$-2.0$\times 10^8$ $\mathrm{\Omega\cdot cm}$, respectively. For measurements with a transducer, the gold thin film is coated using an e-beam evaporator. A 3 nm Ti layer is used as the adhesion layer and another 92 nm Au layer is deposited, confirmed by the SEM image. The experimental data and best fits for phase and amplitude for all samples can be found in Sec. VIII of SI. The error bar in the transport properties for the sample with gold transducer are calculated using Monte Carlo method. The parameters in the Fourier model, except the fitting variables, are randomly sampled in 100 times according to the normal distribution with an assumed relative standard deviation of 3 \%. The fitted variables will roughly follow a normal distribution with different standard deviations.
The error bar in Fig.~\ref{fig3} and the uncertainty in Fig.~\ref{fig4} are both computed using the covariance matrix assuming a 3 \% uncertainty in all parameters except the fitting variables, detailed in Ref.~\cite{yang2016uncertainty}.

\begin{acknowledgement}
Q. S. acknowledges the support from the Harvard Quantum Initiative. The authors thank Jarad Mason for the instrument support, Aaron Schmidt for the offset measurement implementation, Rahil Ukani and Yuykung Moon for the deposition of the metal transducer and Alexei Maznev for helpful discussions. S. H. acknowledges support from the NSERC Discovery Grants Program under Grant No. RGPIN-2021-02957.
\end{acknowledgement}

\bibliography{ref}
\end{document}


\title{Supplemental Material for \textit{Probing carrier and phonon transport in semiconductors all at once through frequency-domain photo-reflectance}}

\author{Qichen Song\,\orcidlink{0000-0002-1090-4068}}
\email{qichensong@g.harvard.edu}
\affiliation{Department of Chemistry and Chemical Biology, Harvard University, Cambridge, Massachusetts 02138, USA}
\author{Sorren Warkander\,\orcidlink{0000-0001-9969-5604}}
\affiliation{Chemical Sciences Division, Lawrence Berkeley
National Laboratory, Berkeley, California 94720, United States}
\author{Samuel Huberman\,\orcidlink{0000-0003-0865-8096}}
\email{samuel.huberman@mcgill.ca}
\affiliation{Department of Chemical Engineering, McGill University, Montreal, Quebec H3A 0C5, Canada}




\maketitle
{
  \hypersetup{linkcolor=black}
  \tableofcontents
}
\clearpage

\section{The optical reflectance induced by excess carriers}
We explain the origin of the carrier-induced reflectance using Drude model, a classical model that describes the electrodynamics properties of metal. It can be modified to describe the dynamics of electrons and holes in semiconductors to external electromagnetic field.
\subsection{Drude model for carrier-induced dielectric change}

The Drude model\cite{ashcroft2022solid} states that,
under the external electric field from incident photons,
\begin{equation}
    \mathbf{E}(t) = \mathrm{Re}\{\mathbf{E}(\omega)e^{-i\omega t}\},
\end{equation}
the induced current density and electric field satisfies a linear response relation,
\begin{equation}
    \mathbf{j}(\omega) = \sigma(\omega)\mathbf{E}(\omega),
\end{equation}
where the optical conductivity is,
\begin{equation}
    \sigma(\omega) =\frac{\sigma_0}{1-i\omega\tau}.
\end{equation}
Here $\sigma_0 = \frac{ne^2\tau}{m}$ is the DC conductivity, $n$ is the free electron density, $e$ is the elementary charge, and $m$ is the electron's effective mass and $\tau$ is the relaxation time (mean time between collisions events).

The dielectric constant (in SI metric system) is given by,
\begin{equation}
   \varepsilon(\omega) = 1+\frac{i\sigma}{\epsilon_0\omega} = 1-\frac{ne^2\tau^2}{m\epsilon_0(1+\omega^2\tau^2)}+i\frac{ne^2\tau}{m\epsilon_0\omega(1+\omega^2\tau^2)},
   \label{dielectric}
\end{equation}
where $\epsilon_0$ is the vacuum permittivity.
For the probe beam (532 nm) in this work, we have $\omega\tau\gg 1$, which simplifies Eq.~(\ref{dielectric}):
\begin{equation}
    \varepsilon(\omega)\approx 1-\frac{ne^2}{m\epsilon_0\omega^2}+i\frac{ne^2}{m\epsilon_0\omega^3\tau}.
\end{equation}
We denote the real and imaginary parts $\varepsilon_r = 1-\frac{ne^2}{m\epsilon_0\omega^2}$ and $\varepsilon_i=\frac{ne^2}{m\epsilon_0\omega^3\tau}$. Under pump laser's irradiation (in our case, a blue CW laser of 458 nm), valence electrons in the semiconductor are excited across the bandgap and the stimulated electron and hole density are the same $\Delta n_\mathrm{electron} = \Delta n_\mathrm{hole}$. Since the amount of photoexcited carriers far exceeds that of the intrinsic carriers, 
we do not further distinguish the electron density $n_\mathrm{elec.}$ or hole density $n_\mathrm{hole}$ in the following discussion by denoting $\rho$ the density of either type of carrier. 
Under the probe beam, the excess electrons and holes generated by the pump laser
s irradiation both respond to the electromagnetic field of the probe photons. The change in the dielectric constant attributed to the dielectric response from the excess carriers can be expressed by\cite{lompre1984optical},
\begin{equation}
   \Delta\varepsilon_r = -\frac{\rho e^2}{\epsilon_0\omega^2}\left(\frac{1}{m_e}+\frac{1}{m_h}\right),
   \label{epreal}
\end{equation}
\begin{equation}
    \Delta\varepsilon_i= \frac{\rho e^2}{\epsilon_0\omega^3}\left(\frac{1}{m_e\tau_e}+\frac{1}{m_h\tau_h}\right),
    \label{epimag}
\end{equation}
where $\tau_e$ and $\tau_h$ are the relaxation time, $m_e$ and $m_h$ are the effective mass of electrons and holes, respectively. It should be noted that the relaxation time and effective mass
correspond to those carriers with quasi Fermi levels $E_{\mathrm{F},e/h}\approx \hbar\omega_\mathrm{probe}$, instead of those carriers near the intrinsic Fermi levels without laser excitations. The change in dielectric is related to the high-field carrier transport, rather than the transport near the Fermi level determined by the doping concentration. In addition, the above expression of the change in dielectric constant can be further modified by considering multiple bands with different effective masses.

\subsection{The coefficient of carrier-induced reflectance}
With the expressions of dielectric constant acquired, we proceed to compute the optical reflectance at the probe wavelength, which can be conveniently computed using the refractive index.
The dielectric constant and refractive index are linked by,
\begin{equation}
    \varepsilon = \tilde{n}^2,
    \label{dielectric_refractive}
\end{equation}
which suggests the differential $d\varepsilon = 2\tilde{n}d\tilde{n}$.
The refractive index has real and imaginary parts,
\begin{equation}
    \tilde{n} = n+i\kappa.
\end{equation}

With the photo-induced excess carriers,
\begin{equation}
    \tilde{n} = \tilde{n}_0 + \Delta \tilde{n} = \tilde{n}_0 + \Delta n +i\Delta \kappa,
    \label{ncomplex}
\end{equation}
where $\tilde{n}_0$ is the refractive index of the material without any excess carrier, which we can take from references or obtain using ellipsometry.
Based on Eq.~(\ref{epreal}), Eq.~(\ref{epimag}), Eq.~(\ref{dielectric_refractive}), Eq.~(\ref{ncomplex}), we derive the perturbation from excess carriers given by,
\begin{equation}
    \begin{split}
    \Delta n + i\Delta\kappa &= \left(\frac{dn}{d\rho} + i\frac{d\kappa}{d\rho}\right)\rho_0\\
    &=\frac{1}{2\tilde{n}_0}\left(\frac{d\varepsilon_r}{d\rho}+i\frac{d\varepsilon_i}{d\rho}\right)\rho_0\\
    &=\frac{1}{2\tilde{n}_0}\left[-\frac{e^2}{\epsilon_0\omega^2}\left(\frac{1}{m_e}+\frac{1}{m_h}\right)+i\frac{e^2}{\epsilon_0\omega^3}\left(\frac{1}{m_e\tau_e}+\frac{1}{m_h\tau_h}\right)\right]\rho_0,
    \end{split}
    \label{dn}
\end{equation}
where $\rho_0$ is the photoexcited carrier density due to laser irradiation. Note that for 532 nm probe, we estimate that,
\begin{equation}
    \left|\frac{dn}{d\rho}\right| \bigg{/} \left|\frac{d\kappa}{d\rho} \right|\approx\omega\tau\gg 1,
\end{equation}
thus we can neglect the contribution of carriers to the imaginary part of refractive index.

The reflectance is given by\cite{opsal1987temporal},
\begin{equation}
    R =\frac{1-\tilde{n}}{1+\tilde{n}} \frac{1-\tilde{n}^*}{1+\tilde{n}^*}=\frac{(n-1)^2+\kappa^2}{(n+1)^2+\kappa^2}
    \label{or}
\end{equation}
where $\tilde{n}^*$ is the complex conjugate of the refractive index $\tilde{n}$.
Plug Eq.~(\ref{dn}) into Eq.~(\ref{or}) and the normalized differential reflectance reads,
\begin{equation}
    \frac{\Delta R}{R} =\mathrm{Re}\left\{ \frac{4}{(\tilde{n}_0+1)(\tilde{n}_0-1)}\left(\frac{dn}{d\rho}+i \frac{d\kappa}{d\rho}\right)\rho_0\right\},
\end{equation}
where $\tilde{n}_0$ is the intrinsic (unperturbed) refractive index of the material.
Note that the carrier density can have a spatial dependence along the direction of light propagation. The probed reflectance that considers the carrier density distribution is\cite{ASPNES1969155},
\begin{equation}
    \frac{\Delta R}{R} =\mathrm{Re}\left\{ \frac{4}{(\tilde{n}_0+1)(\tilde{n}_0-1)}\left(\frac{dn}{d\rho}+i \frac{d\kappa}{d\rho}\right)\bar{\rho}_0\right\},
\end{equation}
where $\tilde{\rho}_0$ is the average excess carrier density along the direction of light propagation,
\begin{equation}
\bar{\rho}_0 \cong -2iK\int_0^\infty \rho_0(z) e^{2iKz} dz,
\end{equation}
and $K = \frac{2\pi}{\lambda }\tilde{n}$ is the complex electromagnetic wavevector. Note the factor 2 in the exponent comes from the perturbation theory analysis. 
Here we consider two extreme cases.
If $\rho_0(z)$ has a very small spatial variation over the distance of $\frac{1}{2|K|}$, it yields that the normalized differential reflectance is approximated by,
\begin{equation}
    \begin{split}
    \frac{\Delta R}{R}& \approx\mathrm{Re}\left\{ \frac{4}{(\tilde{n}_0+1)(\tilde{n}_0-1)}\left(\frac{dn}{d\rho}+i \frac{d\kappa}{d\rho}\right)\rho_0(z=0)\right\}\\
    & \approx \frac{4\frac{dn}{d\rho}}{\left[(n_0+1)^2+\kappa_0^2\right]\left[(n_0-1)^2+\kappa_0^2\right]}\left[(n_0^2-\kappa_0^2-1)\mathrm{Re}\{\rho_0(z=0)\}+2n_0\kappa_0\mathrm{Im}\{\rho_0(z=0)\}\right].
    \end{split}
    \label{slowdecay}
\end{equation}
If $\rho_0(z)$ decays rapidly with a characteristic length scale $\delta\ll\frac{1}{2|K|}$, we have,
\begin{equation}
    \begin{split}
    \frac{\Delta R}{R}& \approx\mathrm{Re}\left\{ \frac{4}{(\tilde{n}_0+1)(\tilde{n}_0-1)}\left(\frac{dn}{d\rho}+i \frac{d\kappa}{d\rho}\right)(-2iK\delta\bar{\rho}_1)\right\}\\
    &\approx\frac{8K\delta\frac{dn}{d\rho}}{\left[(n_0+1)^2+\kappa_0^2\right]\left[(n_0-1)^2+\kappa_0^2\right]}\left[(n_0^2-\kappa_0^2-1)\mathrm{Im}\{\bar{\rho}_1\}-2n_0\kappa_0\mathrm{Re}\{\bar{\rho}_1\}\right],
    \end{split}
    \label{fastdecay}
\end{equation}
where $\bar{\rho}_1 = \frac{1}{\delta}\int_0^\delta \rho_0(z)e^{2iKz}dz$ is the weighted average of the carrier density over the short distance $\delta$.

For most semiconductors of interest in this work, whose refractive indices are listed in Table~\ref{t1}, we have $n\gg\kappa$. Additionally, the imaginary part of the carrier density is zero, as we use a CW laser instead of an ultrafast pulsed laser and there is no induced carrier dynamics over the time period of $2\pi/\omega$ (the period of the electromagnetic field of the probe beam).
As a result, the Eq.~(\ref{slowdecay}) and Eq.~(\ref{fastdecay}) can be simplified as follows,
\begin{equation}
   \frac{\Delta{R}}{R} \approx
   -\frac{2e^2}{n_0(n_0^2-1)\epsilon_0m^*\omega^2}\rho_0(z=0),
   \label{case1}
\end{equation}
\begin{equation}
    \frac{\Delta{R}}{R}\approx
    \frac{8K\delta\kappa_0e^2}{(n_0^2-1)^2\epsilon_0m^*\omega^2}\bar{\rho}_1,
    \label{case2}
\end{equation}
where $m^* = \left(m_e^{-1}+m_h^{-1}\right)^{-1}$\cite{factor} is the average effective mass. The prefactors in Eq.~(\ref{case1}) and Eq.~(\ref{case2}) are the coefficients of carrier-induced reflectance (CCR) for the two cases, respectively.
Notably, Eq.~(\ref{case1}) and Eq.~(\ref{case2}) manifest that the sign of the CCR can be either negative or positive depending on if the carrier distribution is surface dominated.


\begin{table}[h!] 
    \centering
     \caption{The optical properties of selected materials at 532 nm}
     \begin{tabular}{c c c c} 
     \hline
     Material & $n$ & $\kappa$ & $\left(2|K|\right)^{-1}$ (nm)  \\ 
     \hline 
     Si &4.15 & 0.05 & 10.2 \\ 
     Ge & 4.92 & 2.36 & 7.75 \\
     Si$_{0.98}$Ge$_{0.02}$ &4.16& 0.03 & 10.2  \\
     GaAs &3.95&0.24&10.7\\
     \hline 
     \end{tabular}
    \label{t1}
\end{table}

\section{The ambipolar diffusion of photoexcited carriers}

\subsection{Bulk material}
The photoexcited carrier dynamics in an isotropic semiconductor is dictated by the following equation\cite{schroder2015semiconductor},
\begin{equation}
    \frac{d\rho}{dt}=D_\mathrm{a} \nabla^2 \rho-\frac{\rho}{\tau}-B_\mathrm{r}\rho^2-\gamma_\mathrm{A} \rho^3,
    \label{allterm}
\end{equation}
where $\rho = n-n_0 = p-p_0$, $n_0$ and $p_0$ are the electron and hole concentrations without photon excitations.  $\tau$ is monomolecular carrier lifetime.
In Ge, for example, the bimolecular recombination coefficient $B_\mathrm{r}$  is 6.41 $\times 10^{-14}\, \mathrm{cm^3\cdot s^{-1}}$ and the Auger coefficient is $\gamma_\mathrm{A} = 10^{-30}\, \mathrm{cm^6\cdot s^{-1}}$. To assess the importance of high-order recombination terms,
we estimate the lifetime due to bimolecular recombination and Auger process via,
\begin{equation}
    \tau_\mathrm{r} = \frac{1}{B_\mathrm{r} \rho_0},
\end{equation}
and
\begin{equation}
    \tau_\mathrm{A} = \frac{1}{\gamma_\mathrm{A} \rho_0^2}.
\end{equation}
Considering a typical excess carrier density $\rho_0 = 10^{18} \, \mathrm{cm}^{-3}$ in an frequency-domain photo-reflectance experiment, we estimate that $\tau_\mathrm{r} = 16$ $\mu$s and $\tau_\mathrm{A} = 1 $ $\mu$s, which are much longer than the carrier lifetime we observed experimentally, hence we only keep the first-order recombination term in Eq.~(\ref{allterm}).


Now, the equation that describes the excess carrier dynamics under laser irradiation is written,
\begin{equation}
    \frac{\partial \rho}{\partial t}=D_\mathrm{a}\nabla^2 \rho  - \frac{\rho}{\tau} + P,
\end{equation}
where $P$ is the number of photons absorbed by the material per second. 
In frequency domain and the reciprocal space, we have,
\begin{equation}
    \frac{\partial^2}{\partial z^2} \tilde{\rho}-\Lambda \tilde{\rho}+\frac{\tilde{P}}{D_\mathrm{a}}=0,
    \label{carrier-fourier}
\end{equation}
where $\Lambda = q_\parallel^2+\frac{1}{D_\mathrm{a}\tau}+i\frac{\omega}{D_\mathrm{a}}$, $\tilde{P} = \frac{A_\mathrm{photon}\alpha}{h\nu} e^{-\frac{\sigma_x^2q_x^2}{8}}e^{-\frac{\sigma_y^2q_y^2}{8}}$, $A_\mathrm{photon} = \frac{P_0}{E_p}$, $P_0$ is the absorbed pump power, and $h\nu$ is the pump photon energy. The boundary condition at the surface is,
\begin{equation}
    -D_\mathrm{a}\frac{\partial \tilde{\rho}}{\partial z}+\tilde{\rho} S=0,
\end{equation}
where $S$ is the surface recombination velocity.
The solution $\tilde{\rho}$ has the form:
\begin{equation}
    \tilde{\rho} = B e^{-\sqrt{\Lambda} z}+Fe^{-\alpha z},
\end{equation}
where  $F=\frac{A_\mathrm{photon}\alpha}{D_{a}\left(\Lambda-\alpha^2\right)} e^{-\frac{\sigma_x^2q^2_x+\sigma_y^2q^2_y}{8}}$ and $B=-\frac{D_\mathrm{a}\alpha+S}{D_\mathrm{a}\sqrt{\Lambda}+S}F$.

The signal detected by the probe beam is,
\begin{widetext}
\begin{equation}
\begin{split}
    I & =\frac{A}{(2\pi)^2} \int\int  \gamma e^{-\gamma z}\tilde{\rho}(\mathbf{q}_\parallel,z)e^{-\frac{\sigma^{\prime 2}_xq_x^2}{8}}e^{-\frac{\sigma^{\prime 2}_yq_y^2}{8}} e^{-iq_x x_\mathrm{o}}e^{-iq_y y_\mathrm{o}}dz d\mathbf{q}_\parallel \\
    &=\frac{A}{(2\pi)^2} \int \frac{A_\mathrm{photon}\alpha\gamma}{D_\mathrm{a}(\Lambda-\alpha^2)}\left(\frac{1}{\alpha+\gamma}-\frac{D_\mathrm{a}\alpha+S}{D_\mathrm{a}\sqrt{\Lambda}+S}\frac{1}{\sqrt{\Lambda}+\gamma}\right) e^{-\frac{R^2_xq_x^2}{4}}e^{-\frac{R^2_yq_y^2}{4}}e^{-iq_x x_\mathrm{o}}e^{-iq_y y_\mathrm{o}}d\mathbf{q}_\parallel.
\end{split}
\label{carrier-bulk}
\end{equation}
where 
\begin{equation}
 \tilde{\rho}(\mathbf{q}_\parallel,z) = \rho_0 \left(e^{-\alpha z}-\eta e^{-\sqrt{\Lambda}z}\right)e^{-\frac{\sigma^2_xq_x^2}{8}}e^{-\frac{\sigma^2_yq_y^2}{8}},   
\end{equation}
\end{widetext}
$\eta = \frac{D_\mathrm{a}\alpha+S}{D_\mathrm{a}\sqrt{\Lambda}+S}$, $\rho_0 = \frac{A_\mathrm{photon}\alpha}{D_\mathrm{a}(\Lambda-\alpha^2)}$, $R_i = \sqrt{(\sigma_i^2+\sigma_i^{\prime 2})/2},\,i = x, y$ is the effective beam radius, ($x_\mathrm{o},y_\mathrm{o}$) is the relative position of the probe with respect to the center of the pump beam and $A$ is the CCR.

When the penetration depth is extremely thin ($\alpha, \lambda\gg |\sqrt{\Lambda}|$) and the surface combination velocity is small ($S\to0$),
the signal is simplified to be,
\begin{equation}
    I = \frac{A}{(2\pi)^2} \int \frac{A_\mathrm{photon}}{D_\mathrm{a}\sqrt{\Lambda}}e^{-\frac{R^2_xq_x^2}{4}}e^{-\frac{R^2_yq_y^2}{4}}e^{-iq_x x_\mathrm{o}}e^{-iq_y y_\mathrm{o}}d\mathbf{q}_\parallel.
\end{equation}

The ambipolar diffusion length $l$ dedicates the carrier dynamics and it is limited by the carrier lifetime,
\begin{equation}
    l = \sqrt{D_\mathrm{a}\tau}.
\end{equation}
When the $\sqrt{D_\mathrm{a}/\omega}$ is smaller than the beam size $R_i$ but larger than diffusion length $l$, it yields,
\begin{equation}
    \Lambda\approx\frac{1+i\omega\tau}{D_\mathrm{a}\tau}.
\end{equation}
The signal now becomes,
\begin{equation}
    I(x_\mathrm{o},y_\mathrm{o}) = \frac{AA_\mathrm{photon}\sqrt{\tau}}{\pi R_xR_y\sqrt{D_\mathrm{a}}}\frac{1}{\sqrt{1+i\omega\tau}}e^{-\frac{x_\mathrm{o}^2}{R_x^2}}e^{-\frac{y_\mathrm{o}^2}{R_y^2}},
\end{equation}
whose amplitude varies with the offset distance according to a Gaussian function.

When the diffusion length $l$ is much larger than the beam size $R_i$, we consider the scenarios of low and high modulation frequency. In the low modulation frequency limit, we have
$\Lambda \approx q_\parallel^2$, which yields the simplified signal in the case of isotropic beam,
\begin{equation}
    \begin{split}
    I(x_\mathrm{o},y_\mathrm{o}) &= \frac{AA_\mathrm{photon}}{(2\pi)^2D_\mathrm{a}}\mathcal{F}\left\{\frac{1}{\sqrt{q_x^2+q_y^2}}e^{-\frac{R^2_xq_x^2}{4}}e^{-\frac{R^2_yq_y^2}{4}} \right\}(x_\mathrm{o},y_\mathrm{o})\\
    &=\frac{AA_\mathrm{photon}}{(2\pi)^2D_\mathrm{a}}2\pi \mathcal{H}\left\{\frac{e^{-r_\mathrm{eff}^2q^2/4}}{q}\right\}(r_\mathrm{o})\\
    &=\frac{AA_\mathrm{photon}}{2\pi D_\mathrm{a}}\int_0^\infty e^{-\frac{r^2_\mathrm{eff}q^2}{4}}J_0(qr_\mathrm{o}) dq,
    \end{split}
    \label{lowmod}
\end{equation}
where $\mathcal{F}$ and $\mathcal{H}$ are the Fourier and Hankel transform operator, respectively, $J_0$ is the zeroth-order Bessel function of the first kind and $r_\mathrm{o}=\sqrt{x_\mathrm{o}^2+y_\mathrm{o}^2}$ is the beam offset distance. In this limit, according to Eq.~(\ref{lowmod}) the signal is frequency independent with zero phase lag.
With no beam offset, Eq.~(\ref{lowmod}) becomes,
\begin{equation}
    I(0,0) = \frac{AA_\mathrm{photon}}{2\sqrt{\pi}r_\mathrm{eff}D_\mathrm{a}},
\end{equation}
which corresponds to the steady state limit. 
In the high modulation frequency case, Eq.~(\ref{carrier-bulk}) reduces to,
\begin{equation}
    I(x_\mathrm{o},y_\mathrm{o}) = \frac{AA_\mathrm{photon}}{\pi R_xR_y\sqrt{iD_\mathrm{a}\omega}}e^{-\frac{x_\mathrm{o}^2}{R_x^2}}e^{-\frac{y_\mathrm{o}^2}{R_y^2}}.
\end{equation}
We notice that the amplitude of the signal from carrier can be proportional to $A/D_\mathrm{a}$ or $A/\sqrt{D_\mathrm{a}}$ or somewhere in between depending on the modulation frequency used, such that $A$ and $D_\mathrm{a}$ are always inversely correlated, which can make fitting both $A$ and $D_\mathrm{a}$ challenging.

\subsection{Thin film on insulating substrate}
We can easily extend our model to the case of thin film on insulating substrate.
Assuming the first layer is optically thick (both for pump and probe beams), the laser power absorbed by the thin film is,
\begin{equation}
    P(\mathbf{r},t) = \frac{2\alpha}{\pi\sigma_x\sigma_y} e^{-2x^2/\sigma_x^2}e^{-2y^2/\sigma_y^2}e^{-\alpha z}e^{i\omega t}.
\end{equation}
Within the thin film, the solution to Eq.~(\ref{carrier-fourier}) is,
\begin{equation}
    \tilde{\rho} = De^{-\sqrt{\Lambda}z}+Ee^{\sqrt{\Lambda}z}+Fe^{-\alpha z},
\end{equation}
where $F=\frac{A_\mathrm{photon}\alpha}{D_\mathrm{a}(\Lambda-\alpha^2)}e^{-\frac{\sigma_x^2q_x^2+\sigma_y^2q_y^2}{8}}$. The coefficients $D$ and $E$ are to be determined.
The boundary conditions for the top and the bottom surfaces
of the film are,
\begin{equation}
    -D_\mathrm{a} \frac{\partial \tilde{\rho}}{\partial z}
    +\tilde{\rho}S_1 = 0,
\end{equation}
\begin{equation}
    -D_\mathrm{a} \frac{\partial \tilde{\rho}}{\partial z}
    +\tilde{\rho}S_2 = 0,
\end{equation}
where $S_1$ and $S_2$ are the surface combination velocity and interface combination velocity at the top and the bottom sides of the thin film.
These boundary conditions then yield,
\begin{equation}
    D = \frac{-(D_\mathrm{a}\alpha+S_1)(-D_\mathrm{a}\sqrt{\Lambda}+S_2)+(-D_\mathrm{a}\sqrt{\Lambda}+S_1)(D_\mathrm{a}\alpha+S_2)e^{-(\sqrt{\Lambda}+\alpha)L}}{(D_\mathrm{a}\sqrt{\Lambda}+S_1)(-D_\mathrm{a}\sqrt{\Lambda}+S_2)-(-D_\mathrm{a}\sqrt{\Lambda}+S_1)(D_\mathrm{a}\sqrt{\Lambda}+S_2)e^{-2\sqrt{\Lambda}L}}F,
\end{equation}
\begin{equation}
    E = \frac{-(D_\mathrm{a}\alpha+S_1)(D_\mathrm{a}\sqrt{\Lambda}+S_2)+(D_\mathrm{a}\sqrt{\Lambda}+S_1)(D_\mathrm{a}\alpha+S_2)e^{(\sqrt{\Lambda}-\alpha)L}}{(-D_\mathrm{a}\sqrt{\Lambda}+S_1)(D_\mathrm{a}\sqrt{\Lambda}+S_2)-(D_\mathrm{a}\sqrt{\Lambda}+S_1)(-D_\mathrm{a}\sqrt{\Lambda}+S_2)e^{2\sqrt{\Lambda}L}}F.
\end{equation}
The signal from the excess carriers in the thin film is,
\begin{equation}
    I = \frac{A}{(2\pi)^2}
    \int
    \frac{\alpha\gamma}{D_\mathrm{a}(\Lambda-\alpha^2)}
    \left(
    \frac{1}{\alpha+\gamma}+\frac{D}{\sqrt{\Lambda}+\gamma}+\frac{E}{-\sqrt{\Lambda}+\gamma} 
    \right)
    e^{-\frac{R^2_xq_x^2}{4}}e^{-\frac{R^2_yq_y^2}{4}}e^{-iq_x x_\mathrm{o}}e^{-iq_y y_\mathrm{o}}d^2\mathbf{q}_\parallel.
\end{equation}
When the surface and interface recombination velocities are small ($S_1 \approx 0$, $S_2\approx 0$), we have,
\begin{equation}
   D = \frac{\alpha}{\sqrt{\Lambda}}\frac{1+e^{-(\sqrt{\Lambda}+\alpha)L}}{-1+e^{-2\sqrt{\Lambda}L}}, 
\end{equation}
\begin{equation}
    E = \frac{\alpha}{\sqrt{\Lambda}}\frac{-1+e^{(\sqrt{\Lambda}-\alpha)L}}{-1+e^{2\sqrt{\Lambda}L}}.
\end{equation}
If we further consider the case when the penetration depth for pump and probe are much smaller than the heat diffusion length, we obtain the simplified expression for the signal,
\begin{equation}
    I = \frac{A}{(2\pi)^2}
    \int
    \frac{1}{D_\mathrm{a}\sqrt{\Lambda}}
    \frac{\sinh{2\sqrt{\Lambda}L}}{\cosh{2\sqrt{\Lambda}L}-1} 
    e^{-\frac{R^2_xq_x^2}{4}}e^{-\frac{R^2_yq_y^2}{4}}e^{-iq_x x_\mathrm{o}}e^{-iq_y y_\mathrm{o}}d^2\mathbf{q}_\parallel 
\end{equation}

\section{Electron and phonon temperature dynamics from the two-temperature model}
The electron and phonon temperatures evolves according to,
\begin{equation}
    C_\mathrm{el}\frac{\partial T_\mathrm{el}}{\partial t}=\nabla\cdot\bm{\kappa}_\mathrm{el} \cdot \nabla T_\mathrm{el} - g_\mathrm{e\mhyphen p}(T_\mathrm{el}-T_\mathrm{ph}),
    \label{eeq}
\end{equation}
\begin{equation}
    C_\mathrm{ph}\frac{\partial T_\mathrm{ph}}{\partial t}=\nabla\cdot\bm{\kappa}_\mathrm{ph} \cdot\nabla T_\mathrm{ph} + g_\mathrm{e\mhyphen p}(T_\mathrm{el}-T_\mathrm{ph}).
    \label{peq}
\end{equation}
In the frequency domain and $q_x\mhyphen q_y$ plane of the reciprocal space, they obey,
\begin{equation}
    C_\mathrm{el} i\omega \tilde{T}_\mathrm{el} =  \left(-\kappa_{\mathrm{el},\parallel}q_\parallel^2+\kappa_{\mathrm{el},z}\frac{\partial^2}{\partial z^2}\right)\tilde{T}_\mathrm{el} - g_\mathrm{e\mhyphen p}(\tilde{T}_\mathrm{el}-\tilde{T}_\mathrm{ph}),
    \label{eeqreci}
\end{equation}
\begin{equation}
    C_\mathrm{ph} i\omega \tilde{T}_\mathrm{ph} = \left(-\kappa_{\mathrm{ph},\parallel} q^2_\parallel +\kappa_{\mathrm{ph},z}\frac{\partial^2}{\partial z^2}\right)\tilde{T}_\mathrm{ph} + g_\mathrm{e\mhyphen p}(\tilde{T}_\mathrm{el}-\tilde{T}_\mathrm{ph}).
    \label{peqreci}
\end{equation}
Denote $\Lambda_\mathrm{el} = \frac{iC_\mathrm{el}\omega+\kappa_{\mathrm{el},\parallel}q^2_\parallel+g_\mathrm{e\mhyphen p}}{\kappa_{\mathrm{el},z}}$ and $\Lambda_\mathrm{ph} = \frac{iC_\mathrm{ph}\omega+\kappa_{\mathrm{ph},\parallel}q^2_\parallel+g_\mathrm{e\mhyphen p}}{\kappa_{\mathrm{ph},z}}$. The Eqs.~(\ref{eeqreci}) and (\ref{peqreci}) become,
    \begin{equation}
        \frac{\partial^2\tilde{T}_\mathrm{el}}{\partial z^2}- \Lambda_\mathrm{el}\tilde{T}_\mathrm{el}  +\frac{g_\mathrm{e\mhyphen p}}{\kappa_{\mathrm{el},z}}\tilde{T}_\mathrm{ph} = 0,
    \end{equation}
    \begin{equation}
        \frac{\partial^2\tilde{T}_\mathrm{ph}}{\partial z^2}- \Lambda_\mathrm{ph}\tilde{T}_\mathrm{ph}  +\frac{g_\mathrm{e\mhyphen p}}{\kappa_{\mathrm{ph},z}}\tilde{T}_\mathrm{el} =0.
    \end{equation}
We further define the matrix
\begin{equation}
    \mathbf{M} = 
    \begin{pmatrix}
        \Lambda_\mathrm{el}&-\frac{g_\mathrm{e\mhyphen p}}{\kappa_{\mathrm{el},z}}\\
        -\frac{g_\mathrm{e\mhyphen p}}{\kappa_{\mathrm{ph},z}}&\Lambda_\mathrm{ph}
    \end{pmatrix}.
\end{equation}
It is easy to see that the solution for temperatures satisfies,
\begin{equation}
    \begin{pmatrix}
        \tilde{T}_\mathrm{el}\\
        \tilde{T}_\mathrm{ph}
    \end{pmatrix}
    = \mathbf{u}
    \begin{pmatrix}
    B_1 e^{-\sqrt{\lambda_1}z}+C_1 e^{\sqrt{\lambda_1}z}\\
    B_2 e^{-\sqrt{\lambda_2}z}+C_2 e^{\sqrt{\lambda_2}z}
    \end{pmatrix},
\end{equation}
where $\lambda_{i}$ are the eigenvalues and $\mathbf{u}$ is the eigenvector matrix of $\mathbf{M}$,
\begin{equation}
    \mathbf{u} = 
    \begin{pmatrix}
        u_{11}&u_{12}\\
        u_{21}&u_{22}
    \end{pmatrix}.
\end{equation}
Considering the semi-infinite geometry in z direction, we have $C_i = 0$. The general solutions for temperatures now write,
\begin{equation}
    \begin{pmatrix}
        \tilde{T}_\mathrm{el}\\
        \tilde{T}_\mathrm{ph}
    \end{pmatrix}
    = \mathbf{u}
    \begin{pmatrix}
    B_1 e^{-\sqrt{\lambda_1}z}\\
    B_2 e^{-\sqrt{\lambda_2}z}
    \end{pmatrix}.
\end{equation}

Now, we consider the two temperature model with laser heating. The electron temperature is dictated by,
\begin{equation} 
    C_\mathrm{el}\frac{\partial T_\mathrm{el}}{\partial t}=\nabla\cdot\bm{\kappa}_\mathrm{el} \cdot \nabla T_\mathrm{el} - g_\mathrm{e\mhyphen p}(T_\mathrm{el}-T_\mathrm{ph})+P,
    \label{eeqp}
\end{equation}
whereas the phonon temperature $T_\mathrm{ph}$ can still be described by Eq.~(\ref{peq}).
After we apply the Fourier transform in time domain and in $x\mhyphen y$ plane, Eq.~(\ref{eeqp}) becomes,
\begin{equation}
   \frac{\partial^2\tilde{T}_\mathrm{el}}{\partial z^2}- \Lambda_\mathrm{el}\tilde{T}_\mathrm{el}  +\frac{g_\mathrm{e\mhyphen p}}{\kappa_{\mathrm{el},z}}\tilde{T}_\mathrm{ph} +\frac{\tilde{P}}{\kappa_{\mathrm{el},z}}=0,
\end{equation}
where $\tilde{P}=\left[\tilde{\rho}(0,z)\frac{E_\mathrm{gap}}{\tau} +P_0\alpha\frac{\Delta E}{E_\mathrm{photon}} e^{-\alpha z}\right]e^{-\frac{\sigma_x^2q^2_x+\sigma_y^2q^2_y}{8}}$ and $\Delta E = E_\mathrm{photon}-E_\mathrm{gap}$.
The solution for temperatures can be expressed by,
\begin{equation}
\begin{split}
    \begin{pmatrix}
        \tilde{T}_\mathrm{el}\\
        \tilde{T}_\mathrm{ph}
    \end{pmatrix}
    &
    =\mathbf{u}
    \begin{pmatrix}
    B_1 e^{-\sqrt{\lambda_1}z}\\
    B_2 e^{-\sqrt{\lambda_2}z}
    \end{pmatrix}
e^{-\frac{\sigma_x^2q^2_x+\sigma_y^2q^2_y}{8}}
    +  \mathbf{u}\frac{1 }{\kappa_{\mathrm{el},z}}\left(\frac{\rho_0E_\mathrm{gap}}{\tau}+A_\mathrm{photon}\alpha\Delta E\right)e^{-\frac{\sigma_x^2q^2_x+\sigma_y^2q^2_y}{8}}e^{-\alpha z}
    \begin{pmatrix}
        \frac{v_{11}}{\lambda_1-\alpha^2}\\
        \frac{v_{21}}{\lambda_2-\alpha^2}
    \end{pmatrix}\\
    &+   \mathbf{u} \frac{ \rho_0\eta E_\mathrm{gap}}{\kappa_{\mathrm{el},z}\tau} e^{-\frac{\sigma_x^2q^2_x+\sigma_y^2q^2_y}{8}}e^{-\sqrt{\Lambda}z}
    \begin{pmatrix}
        \frac{v_{11}}{\Lambda-\lambda_1}\\
        \frac{v_{21}}{\Lambda-\lambda_2}
    \end{pmatrix}\\
    &=\left[\mathbf{u}
    \begin{pmatrix}
    B_1 e^{-\sqrt{\lambda_1}z}\\
    B_2 e^{-\sqrt{\lambda_2}z}
    \end{pmatrix}+  \mathbf{u}A_1e^{-\alpha z}
    \begin{pmatrix}
        \frac{v_{11}}{\lambda_1-\alpha^2}\\
        \frac{v_{21}}{\lambda_2-\alpha^2}
    \end{pmatrix}+   \mathbf{u} A_2 e^{-\sqrt{\Lambda}z}
    \begin{pmatrix}
        \frac{v_{11}}{\Lambda-\lambda_1}\\
        \frac{v_{21}}{\Lambda-\lambda_2}
    \end{pmatrix}\right]\times e^{-\frac{\sigma_x^2q^2_x+\sigma_y^2q^2_y}{8}},
\end{split}
\label{gensol}
\end{equation}
where the matrix $\mathbf{v}=\mathbf{u}^{-1}$ and the coefficients,
\begin{equation}
    A_1 = \frac{1}{\kappa_{\mathrm{el},z}}\left(\frac{\rho_0E_\mathrm{gap}}{\tau}+A_\mathrm{photon}\alpha\Delta E\right),
\end{equation}
\begin{equation}
    A_2=  \frac{ \rho_0\eta E_\mathrm{gap}}{\kappa_{\mathrm{el},z}\tau}.
\end{equation}

Considering the photon emission due to surface recombination, we obtain have the boundary condition for electron temperature, at $z = 0$,
\begin{equation}
    -\kappa_{\mathrm{el},z}\frac{\partial \tilde{T}_\mathrm{el}}{\partial z}-\tilde{\rho} S E_\mathrm{gap} = 0.
\end{equation}
The phonon temperature satisfies adiabatic surface boundary condition,
$\tilde{q}_\mathrm{ph}=-\kappa_{\mathrm{ph},z}\frac{\partial \tilde{T}_\mathrm{ph}}{\partial z}=0$.
With the boundary conditions, the unknown coefficients in Eq.~(\ref{gensol}) can be determined as,
\begin{equation}
    \begin{split}
    \begin{pmatrix}
        B_1\\
        B_2 
    \end{pmatrix}=
    \frac{1}{U}
   \begin{pmatrix}
       \frac{u_{22}\tilde{\rho}(\mathbf{q}_\parallel=0,z=0) S E_\mathrm{gap}/\kappa_{\mathrm{el},z}-\alpha(u_{22}D_1-u_{12}C_1)-\sqrt{\Lambda}(u_{22}D_2-u_{12}C_2)}{\sqrt{\lambda_1}}\\
       -\frac{u_{21}\tilde{\rho}(\mathbf{q}_\parallel=0,z=0) S E_\mathrm{gap}/\kappa_{\mathrm{el},z}-\alpha(u_{21}D_1-u_{11}C_1)-\sqrt{\Lambda}(u_{21}D_2-u_{11}C_2)}{\sqrt{\lambda_2}}
   \end{pmatrix},
\end{split}
\end{equation}
where $U = u_{11}u_{22}-u_{12}u_{21}$ and
\begin{equation}
    C_1 = A_1\frac{-(u_{21}v_{11}+u_{22}v_{21})\alpha^2+u_{21}v_{11}\lambda_2+u_{22}v_{21}\lambda_1}{(\lambda_1-\alpha^2)(\lambda_2-\alpha^2)},
\end{equation}
\begin{equation}
    C_2 = A_2\frac{(u_{21}v_{11}+u_{22}v_{21})\Lambda-u_{21}v_{11}\lambda_2-u_{22}v_{21}\lambda_1}{(\Lambda-\lambda_1)(\Lambda-\lambda_2)},
\end{equation}
\begin{equation}
    D_1 = A_1\frac{-(u_{11}v_{11}+u_{12}v_{21})\alpha^2+u_{11}v_{11}\lambda_2+u_{12}v_{21}\lambda_1}{(\lambda_1-\alpha^2)(\lambda_2-\alpha^2)},
\end{equation}
\begin{equation}
    D_2 = A_2\frac{(u_{11}v_{11}+u_{12}v_{21})\Lambda-u_{11}v_{11}\lambda_2-u_{12}v_{21}\lambda_1}{(\Lambda-\lambda_1)(\Lambda-\lambda_2)}.
\end{equation}

The probed signal due to electron temperature rise can be computed by,
\begin{equation}
    \begin{split}
    I_\mathrm{el}(x_\mathrm{o},y_\mathrm{o}) &= \frac{2B\gamma}{\pi\sigma_x^\prime \sigma_y^\prime}\int d \mathbf{r} \tilde{T}_\mathrm{el}(\omega,\mathbf{r})e^{-\frac{2(x-x_\mathrm{o})^2}{\sigma_x^{\prime 2}}}e^{-\frac{2(y-y_\mathrm{o})^2}{\sigma_y^{\prime 2}}}e^{-\gamma z}\\
    &= \frac{2B\gamma}{\pi\sigma_x^\prime \sigma_y^\prime}\int d \mathbf{r} \int \frac{d\mathbf{q}_\parallel}{(2\pi)^2}\tilde{T}_\mathrm{el}(\omega,\mathbf{q}_\parallel,z)e^{i\mathbf{q}_\parallel\cdot\mathbf{r}_\parallel}e^{-\frac{2(x-x_\mathrm{o})^2}{\sigma_x^{\prime 2}}}e^{-\frac{2(y-y_\mathrm{o})^2}{\sigma_y^{\prime 2}}}e^{-\gamma z}\\
    \\
    &= B\int \frac{d\mathbf{q}_\parallel}{(2\pi)^2}\int dz \gamma e^{-\gamma z}\left[u_{11}B_1 e^{-\sqrt{\lambda_1}z}+u_{12}B_2 e^{-\sqrt{\lambda_2}z}+D_1 e^{-\alpha z}+D_2 e^{-\sqrt{\Lambda} z}\right]e^{-\frac{R_x^{ 2}q_x^2}{4}}e^{-\frac{R_y^{ 2}q_y^2}{4}}e^{-iq_x x_\mathrm{o}}e^{-iq_y y_\mathrm{o}}\\
    &=B\int\frac{d\mathbf{q}_\parallel}{(2\pi)^2}\gamma\left[\frac{u_{11}B_1}{\sqrt{\lambda_1}+\gamma}+\frac{u_{12}B_2}{\sqrt{\lambda_2}+\gamma}+\frac{D_1}{\alpha+\gamma}+\frac{D_2}{\sqrt{\Lambda}+\gamma}\right]e^{-\frac{R_x^{2}q_x^2}{4}}e^{-\frac{R_y^{2}q_y^2}{4}}e^{-iq_x x_\mathrm{o}}e^{-iq_y y_\mathrm{o}},
    \end{split}
\end{equation}
where $B$ is the coefficient of changed reflectance due to electron temperature.
The signal due to phonon temperature rise reads,
\begin{equation}
I_\mathrm{ph}(x_\mathrm{o},y_\mathrm{o})=C\int\frac{d\mathbf{q}_\parallel}{(2\pi)^2}\gamma\left[\frac{u_{21}B_1}{\sqrt{\lambda_1}+\gamma}+\frac{u_{22}B_2}{\sqrt{\lambda_2}+\gamma}+\frac{C_1}{\alpha+\gamma}+\frac{C_2}{\sqrt{\Lambda}+\gamma}\right]e^{-\frac{R_x^{2}q_x^2}{4}}e^{-\frac{R_y^{2}q_y^2}{4}}e^{-iq_x x_\mathrm{o}}e^{-iq_y y_\mathrm{o}},   
\end{equation}
where $C$ is the coefficient of thermoreflectance.

\section{Carrier and phonon model}
We find that electron-phonon coupling is usually strong in the typical range of modulation frequency in our experiment (500 Hz - 50 MHz), such that the thermal resistance for the phonon subsystem arising from the electron-phonon interactions can be neglected. The phonon temperature is well described by the following equation,
\begin{equation}
   \frac{\partial^2\tilde{T}_\mathrm{ph}}{\partial z^2}- \lambda_p\tilde{T}_\mathrm{ph}  +\frac{\tilde{P}}{\kappa_{\mathrm{ph},z}}=0,
\end{equation}
where the coefficients
$\tilde{P}=\tilde{\rho}\frac{E_\mathrm{gap}}{\tau} +P_0\alpha\frac{\Delta E}{E_\mathrm{ph}} e^{-\frac{\sigma_x^2q^2_x+\sigma_y^2q^2_y}{8}}e^{-\alpha z}$
and 
$\lambda_\mathrm{ph} = \frac{\kappa_{\mathrm{ph},\parallel}q^2_\parallel+iC_\mathrm{ph}\omega}{\kappa_{\mathrm{ph},z}}$ (rigorously proved in Sec.~\ref{eigproblem}).
The solution is given by the following,
\begin{equation}
\tilde{T}_\mathrm{ph} = \left(B_\mathrm{ph} e^{-\sqrt{\lambda_\mathrm{ph}}z}+  A_1
\frac{e^{-\alpha z}}{\lambda_\mathrm{ph}-\alpha^2}+ A_2
\frac{e^{-\sqrt{\Lambda}z}}{\Lambda-\lambda_\mathrm{ph}}\right)e^{-\frac{\sigma_x^2q^2_x+\sigma_y^2q^2_y}{8}},
\end{equation}
where
\begin{equation}
    A_1 = \frac{1}{\kappa_{\mathrm{ph},z}}\left(\frac{\rho_0E_\mathrm{gap}}{\tau}+A_\mathrm{photon}\alpha\Delta E\right),
\end{equation}
\begin{equation}
    A_2=  \frac{ \rho_0\eta E_\mathrm{gap}}{\kappa_{\mathrm{ph},z}\tau}.
\end{equation}
The zero heat flux boundary condition at the surface leads to,
\begin{equation}
    B_\mathrm{ph} = -\frac{1}{\sqrt{\lambda_\mathrm{ph}}}\left(\frac{A_1\alpha}{\lambda_\mathrm{ph}-\alpha^2}+\frac{A_2\sqrt{\Lambda}}{\Lambda-\lambda_\mathrm{ph}}\right).
\end{equation}
Finally, the signal due to phonon temperature is,
\begin{equation}
    I = \frac{C}{(2\pi)^2}
    \int\gamma\left(\frac{B_\mathrm{ph}}{\sqrt{\lambda_\mathrm{ph}}+\gamma}+\frac{A_1}{(\lambda_\mathrm{ph}-\alpha^2)(\alpha+\gamma)}+\frac{A_2}{(\Lambda-\lambda_\mathrm{ph})(\sqrt{\Lambda}+\gamma)}\right) 
    e^{-\frac{R^2_xq_x^2}{4}}e^{-\frac{R^2_yq_y^2}{4}}e^{-iq_x x_\mathrm{o}}e^{-iq_y y_\mathrm{o}}d^2\mathbf{q}_\parallel.
\end{equation}
\clearpage

\section{The experimental measurements and best fits for bare samples at various temperatures}

\begin{figure}[H]
    \centering
\includegraphics[width=\linewidth]{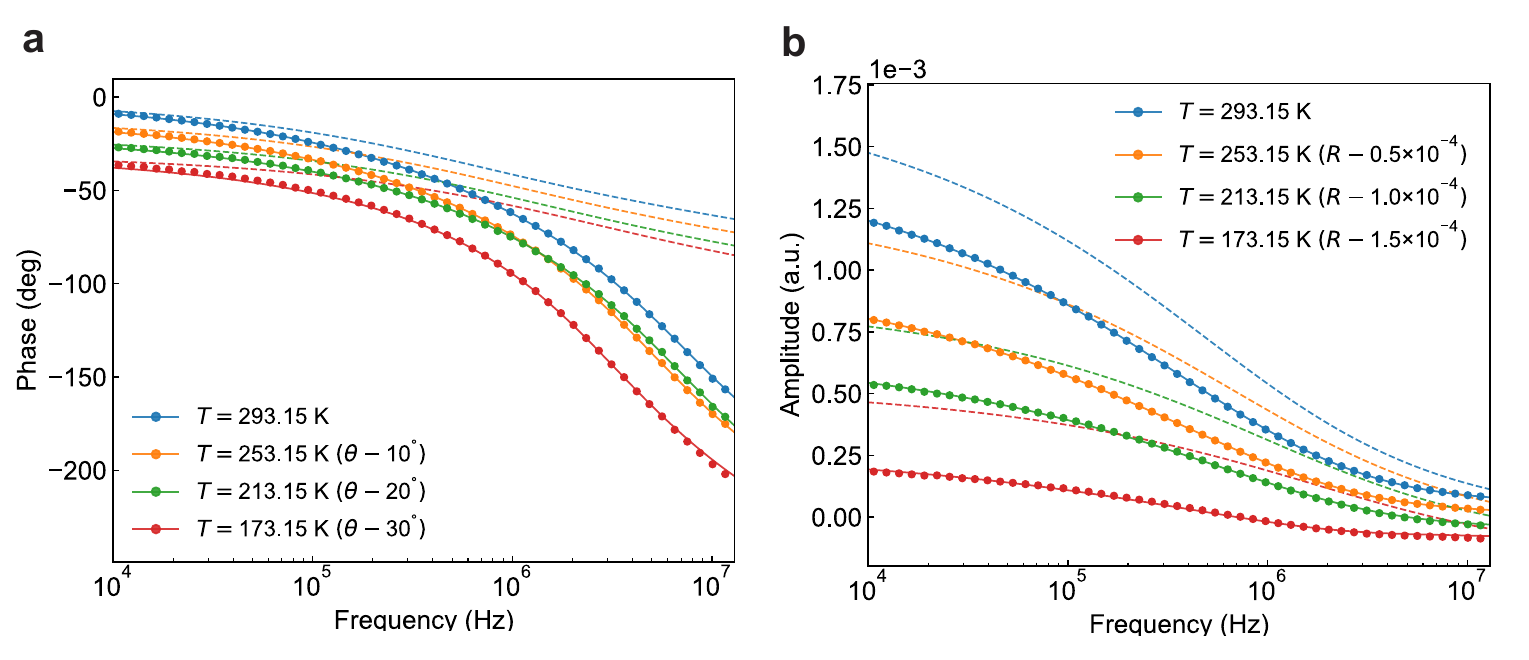}
\caption[Si temperature]{The frequency-domain photo-reflectance in Si at 173.15 K, 213.15 K, 253.15 K, 293.15 K from experiment (circle) and best fits (solid line). The phonon part of the signal is shown in the dashed line.}
\label{si temperature}
\end{figure}
\begin{figure}[H]
    \centering
\includegraphics[width=\linewidth]{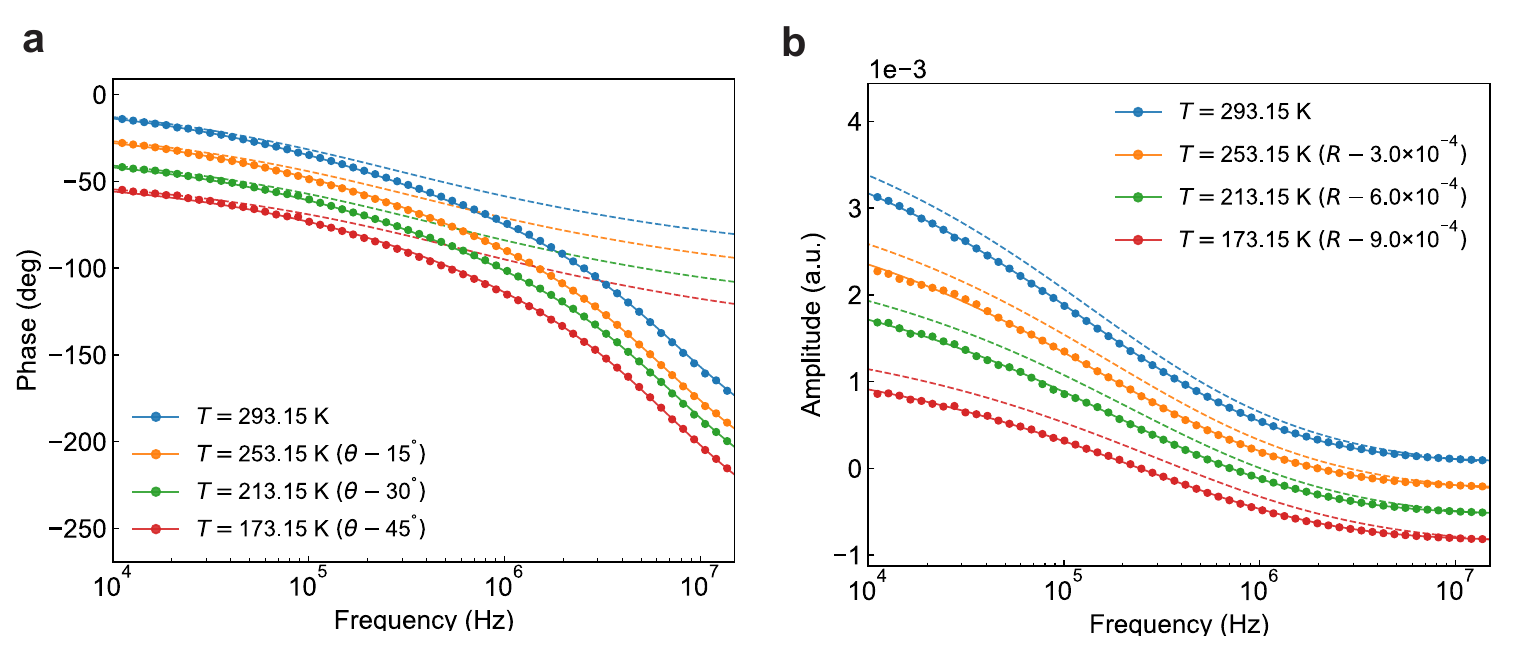}
\caption[SiGe temperature]{The frequency-domain photo-reflectance in Si$_{0.98}$Ge$_{0.02}$ at 173.15 K, 213.15 K, 253.15 K, 293.15 K from experiment (circle) and best fits (solid line). The phonon part of the signal is shown in the dashed line.}
\label{siGe temperature}
\end{figure}
\begin{figure}[H]
    \centering
\includegraphics[width=\linewidth]{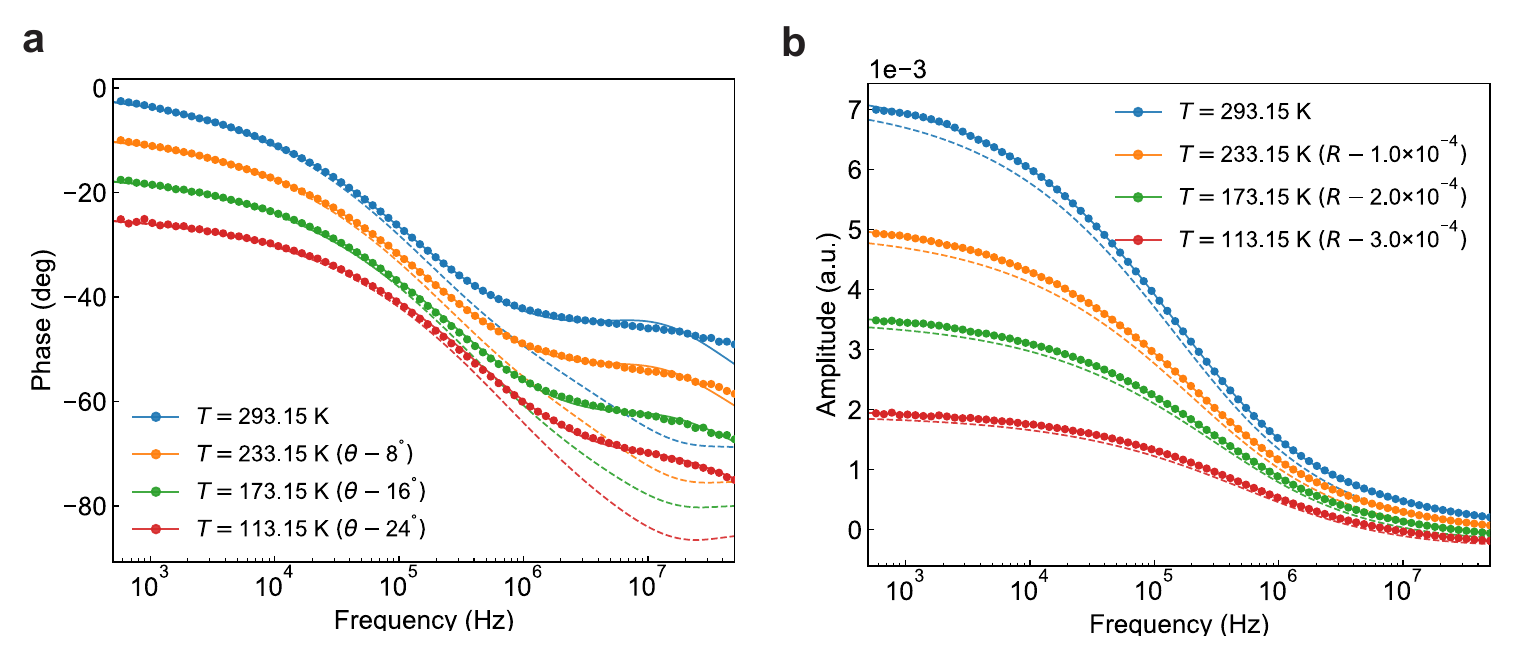}
\caption[GaAs temperature]{The frequency-domain photo-reflectance in GaAs at 113.15 K, 173.15 K, 233.15 K, 293.15 K from experiment (circle) and best fits (solid line). The phonon part of the signal is shown in the dashed line.}
\label{GaAs temperature}
\end{figure}

\clearpage

\section{Sensitivity analysis in the carrier and phonon model}
\begin{figure}[H]
    \vspace{3 cm}
    \centering
\includegraphics[width=\linewidth]{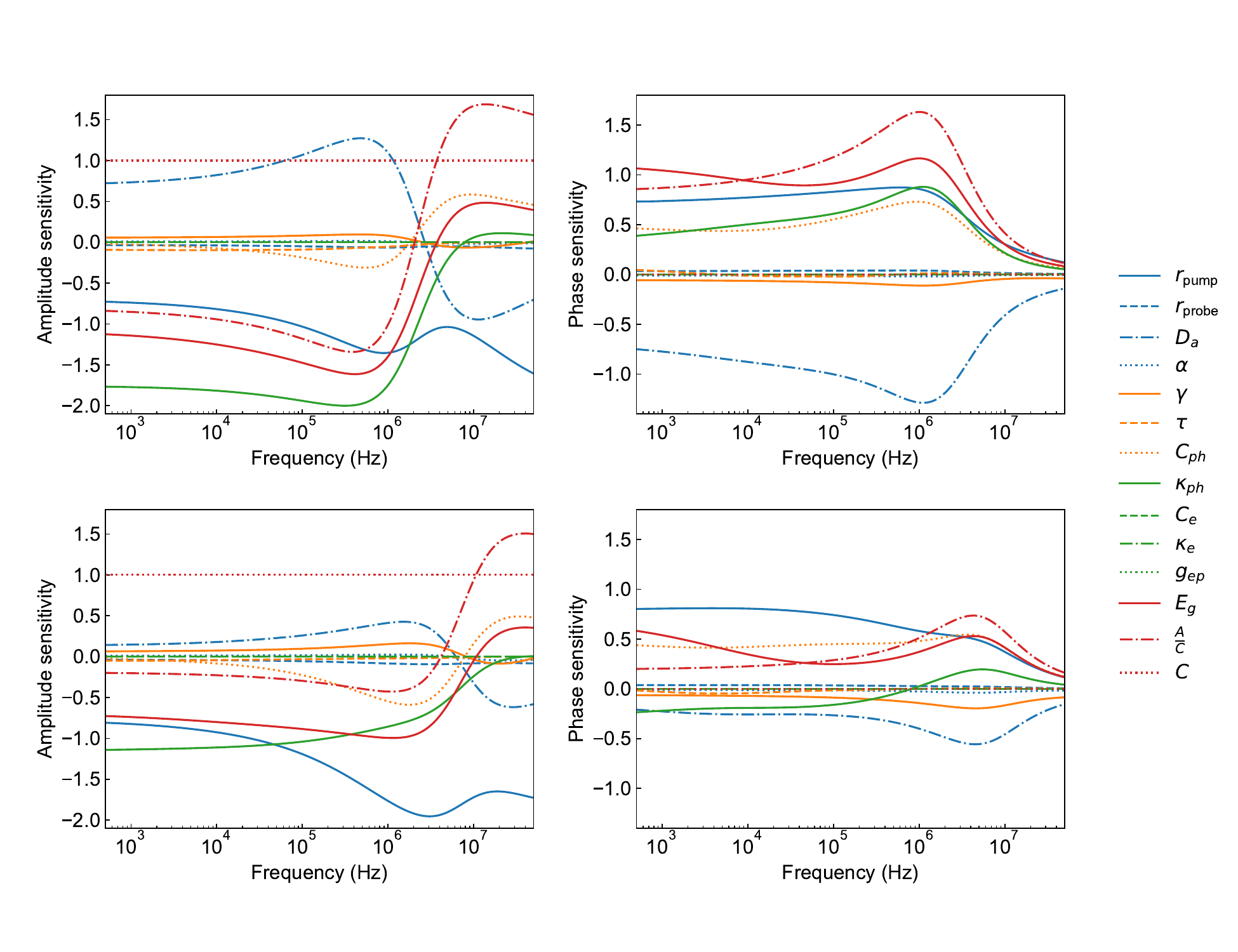}
\caption[Si_sen]{The sensitivity analysis for Si at 153 K (top row) and 293 K (bottom row).}
\end{figure}

\clearpage
\begin{figure}[H]
    \vspace{3 cm}
    \centering
\includegraphics[width=\linewidth]{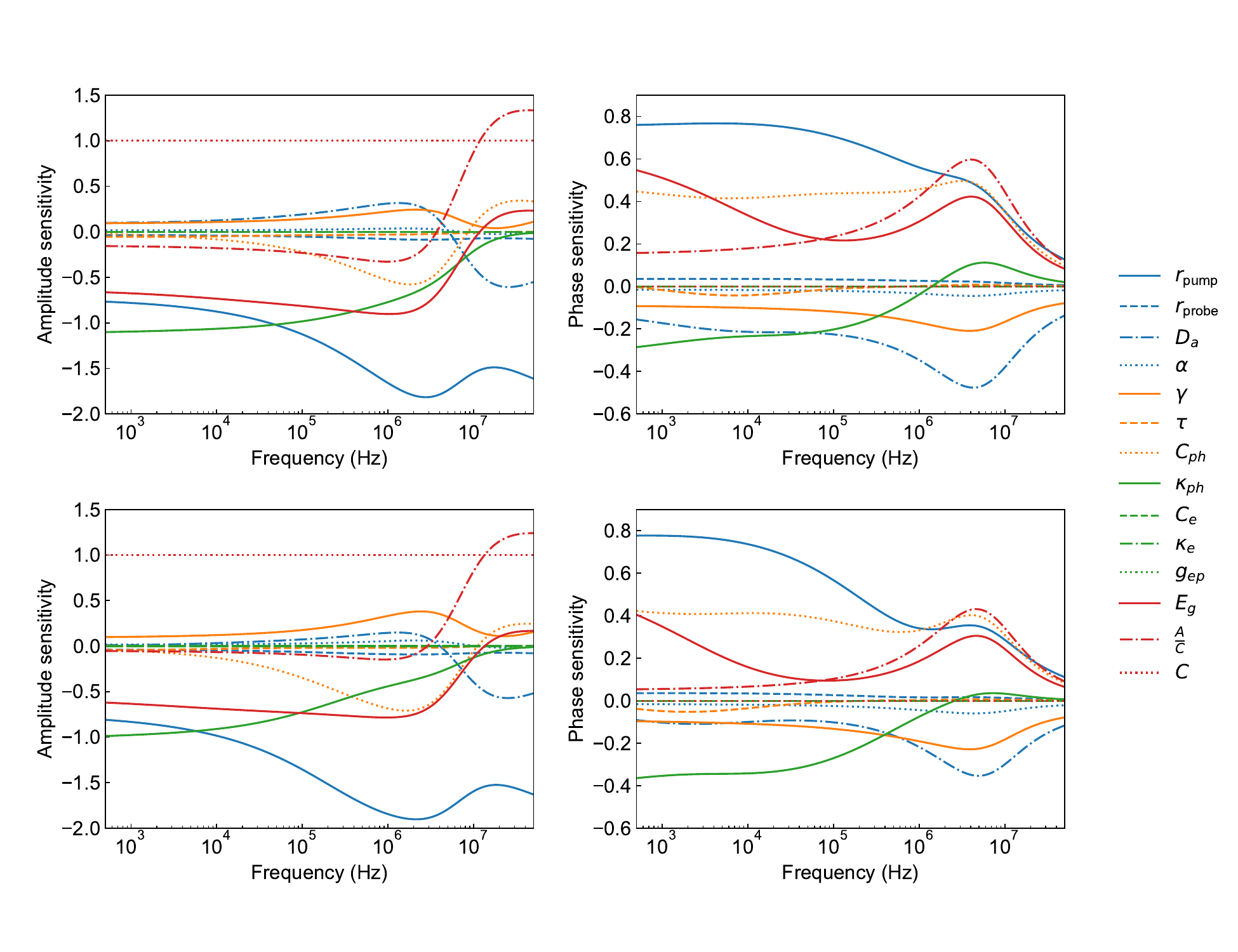}
\caption[SiGe_sen]{The sensitivity analysis for Si$_{0.98}$Ge$_{0.02}$ at 113 K (top row) and 293 K (bottom row).}
\end{figure}

\begin{figure}[H]
    \vspace{3 cm}
    \centering
\includegraphics[width=\linewidth]{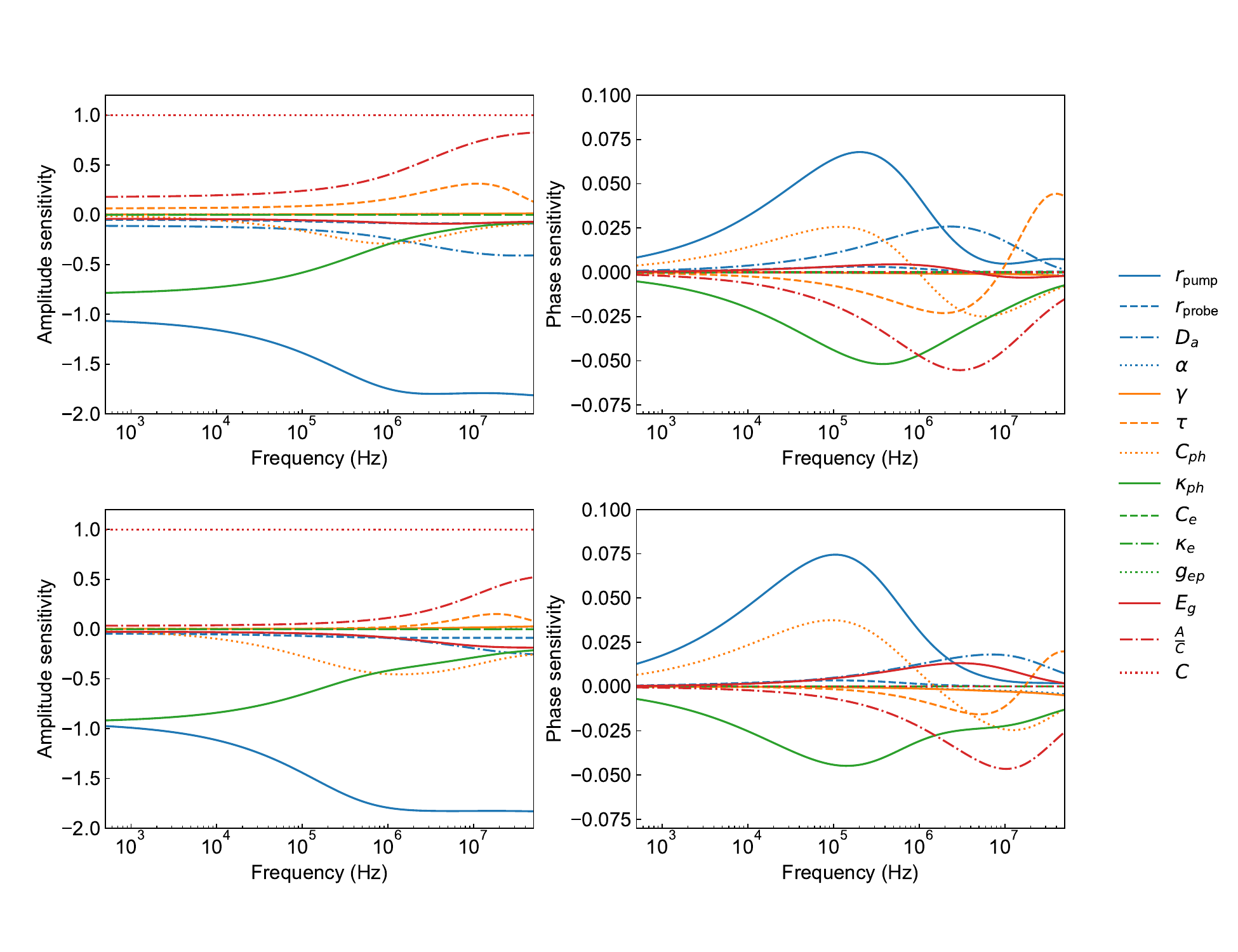}
\caption[Ge_sen]{The sensitivity analysis for Ge at 173 K (top row) and 293 K (bottom row).}
\end{figure}

\begin{figure}[H]
    \vspace{3 cm}
    \centering
\includegraphics[width=\linewidth]{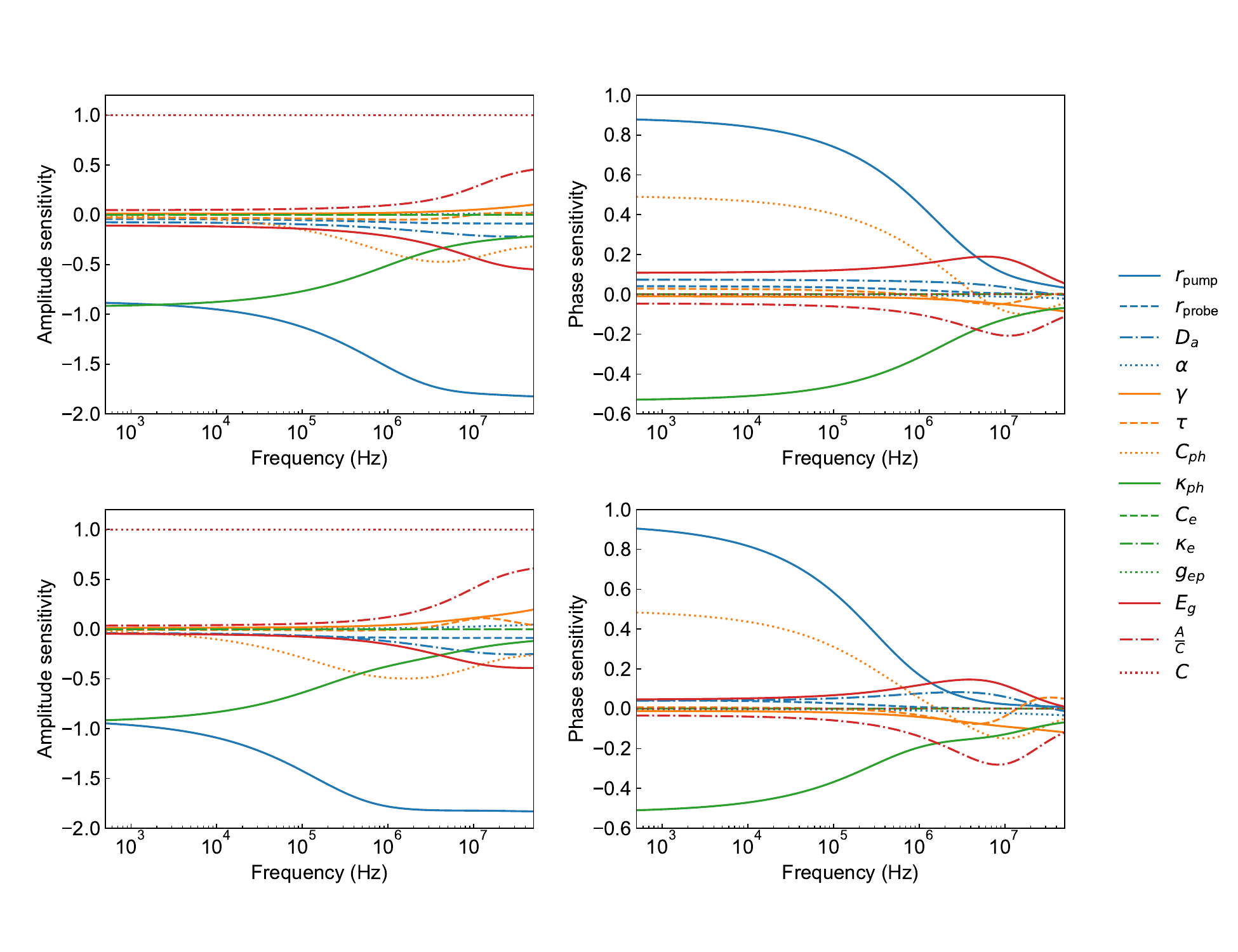}
\caption[GaAs_sen]{The sensitivity analysis for GaAs at 93 K (top row) and 293 K (bottom row).}
\end{figure}
\clearpage

\section{The curve-up and curve-down features in the phase vs frequency plots}

If we assume the photo-reflectance is solely determined by the phonon temperature rise, according to the Fourier law, the phase lag in an isotropic material should always be monotonically increasing as the modulation frequency increases. However, in our experiment, we observe curve-up features in the phase vs frequency plots starting from certain modulation frequency (for instance in Ge at 293 K, the phase lag starts to curve up around 1 MHz shown in Fig. 3a in the main text), which cannot be explained by Fourier model alone. 
These features were also experimentally reported by Beardo \textit{et. al.}\cite{beardo2021observation} and they attribute the curve-up feature to phonon hydrodynamics transport by fitting the phase from experiment to a hyperbolic heat equation. 
In the following, we show that the curve-up feature can also be explained by the two-temperature model and the carrier and phonon model.
However, some of our experimental results have curve-down features (for instance in Si$_{0.98}$Ge$_{0.02}$), which can be well explained by the carrier and phonon model, yet
entirely unexplainable using either the hyperbolic heat equation model or the two-temperature model.

\subsection{Hyperbolic heat equation}
The Cattaneo's equation\cite{cattaneo1948sulla,RevModPhys.61.41} also known as the hyperbolic heat equation, where a relaxation time $\tau_\mathrm{HHE}$ is introduced to the Fourier's law to address the issue of infinite velocity of information propagation, writes,
\begin{equation}
    \tau_\mathrm{HHE} \frac{\partial \mathbf{q}}{\partial t}+\mathbf{q}=-\bm{\kappa}\cdot \nabla T,
    \label{hhflux}
\end{equation}
where $\bm{\kappa}$ is the thermal conductivity tensor. For non-chiral thermal transport, the thermal conductivity tensor is symmetric. In particular, we consider the case where the three principle axes in the material are aligned with x, y and z axes, the in-plane thermal transport is isotropic, and the off-diagonal tensor elements are zero,
\begin{equation}
    \bm{\kappa}=
    \begin{pmatrix}
        \kappa_\parallel&0&0\\
        0&\kappa_\parallel&0\\
        0&0&\kappa_{zz}\\  
    \end{pmatrix}.
\end{equation} 
Consider the energy conservation, $C_p\frac{\partial T}{\partial t}=-\nabla\cdot \mathbf{q}$, where $C_p$ is the volumetric specific heat. We obtain the following relationship,
\begin{equation}
    \frac{\partial^2 T}{\partial t^2}+\frac{1}{\tau_\mathrm{HHE}}\frac{\partial T}{\partial t}=(\nabla \cdot \bm{c}^2\cdot \nabla) T,
\end{equation}
where $\bm{c}^2=\frac{\bm{\kappa}}{C_p \tau_\mathrm{HHE}}$ is a tensor.

In the presence of laser heating,
the hyperbolic heat equation writes,
\begin{equation}
    \frac{\partial^2 T}{\partial t^2}+\frac{1}{\tau_\mathrm{HHE}}\frac{\partial T}{\partial t}=(\nabla \cdot \bm{c}^2\cdot \nabla) T+\frac{1}{C_p \tau_\mathrm{HHE}}\left(1+\tau_\mathrm{HHE}\frac{\partial}{\partial t}\right)P.
    \label{hh}
\end{equation}
The laser heating from the modulated CW laser is, 
\begin{equation}
    P(t,\mathbf{r}) = \frac{2P_0\alpha}{\pi \sigma_x \sigma_y}e^{-\frac{2x^2}{\sigma_x^2}}e^{-\frac{2y^2}{\sigma_y^2}}e^{-\alpha z}e^{i\omega t},
\end{equation}
where $\sigma_x$ and $\sigma_y$ are the $1/e^2$ waists of the pump beam along x and y directions, respectively, $\alpha$ is the inverse penetration depth for the pump beam and $P_0$ is the amplitude of the heating power.
As a consequence, the temperature is modulated at the same frequency $\omega$ expressed as,
\begin{equation}
    T(t,\mathbf{r}) = \tilde{T}(\omega,\mathbf{r})  e^{i\omega t}.
\end{equation}

Due to the anisotropic nature of the laser heating, we apply the Fourier transform only in the x-y plane,
\begin{equation}
    \tilde{T}(\omega,\mathbf{q}_\parallel,z)=\int d^2\mathbf{r}_\parallel \tilde{T}(\omega,\mathbf{r}) e^{-i\mathbf{q}_\parallel \cdot\mathbf{r}_\parallel},
\end{equation}
where $\mathbf{q}_\parallel=(q_x,q_y)$ and $\mathbf{r}_\parallel=(x,y)$.
The Fourier transform of the temperature satisfies,
\begin{equation}
    -\left(\omega^2-\frac{i\omega}{\tau_\mathrm{HHE}}\right)\tilde{T}=\left(-c^2_\parallel q_\parallel^2+c_{zz}^2\frac{\partial^2}{\partial z^2}\right)\tilde{T}+\frac{(1+i\omega\tau_\mathrm{HHE})\tilde{P}}{C_p\tau_\mathrm{HHE}},
\end{equation}
where the Fourier transform of laser power reads $\tilde{P}(\omega,\mathbf{q}_\parallel,z)=A\alpha e^{-\frac{\sigma_x^2q^2_x}{8}}e^{-\frac{\sigma_y^2q^2_y}{8}}e^{-\alpha z}$. Denote $\lambda = \frac{1}{c_{zz}^2}\left[\left(c^2_\parallel q_\parallel^2-\omega^2\right)+\frac{i\omega}{\tau_\mathrm{HHE}}\right]$. We arrive at the following differential equation,
\begin{equation}
    \frac{\partial^2\tilde{T}}{\partial z^2}-\lambda \tilde{T}+\frac{(1+i\omega\tau_\mathrm{HHE})\tilde{P}}{\kappa_{zz}}=0.
\label{de}
\end{equation}
The solution to Eq.~(\ref{de}) follows the expression,
\begin{equation}
    \tilde{T}=Be^{-\beta z}+Fe^{-\alpha z},
\end{equation}
where $\beta=\sqrt{\lambda}$, $F=\frac{A\alpha(1+i\omega\tau_\mathrm{HHE})}{\kappa_{zz}\left(\lambda-\alpha^2\right)} e^{-\frac{\sigma_x^2q^2_x+\sigma_y^2q^2_y}{8}}$ and $B$ is a constant to be determined.
The normal heat flux at the surface (z = 0), according to Eq.~(\ref{hhflux}), satisfies,
\begin{equation}
    (i\omega\tau+1)\tilde{q}=-\kappa_{zz}\frac{\partial \tilde{T}}{\partial z}.
\end{equation}
At the surface of the material, the convection and radiation heat loss are negligible, hence $\tilde{q}\approx0$. We obtain
that, 
\begin{equation}
    B=-\frac{\alpha}{\beta} F
\end{equation}

The temperature in the real space can be attained via calculating the inverse Fourier transform,
\begin{equation}
    \tilde{T}(\omega,\mathbf{r}) = \frac{1}{(2\pi)^2}\int d^2\mathbf{q}_\parallel \tilde{T}(\omega,\mathbf{q}_\parallel,z)^{i\mathbf{q}_\parallel \cdot\mathbf{r}_\parallel}.
\end{equation}

The probed signal can be expressed by\cite{yang2016modeling},
\begin{equation}
    I(\omega) = \frac{2C\gamma}{\pi\sigma_x^\prime \sigma_y^\prime}\int d^3\mathbf{r} \,\tilde{T}(\omega,\mathbf{r})e^{-\frac{2(x-x_\mathrm{o})^2}{\sigma_x^{\prime 2}}}e^{-\frac{2(y-y_\mathrm{o})^2}{\sigma_y^{\prime 2}}}e^{-\gamma z},
\end{equation}
where $\sigma_x^\prime$ and  $\sigma_y^\prime$  are the radii for the probe beam, $\gamma$ is the probe's inverse penetration depth, and $x_\mathrm{o}$ and $y_\mathrm{o}$ are the beam offset between pump and probe beam centers along x and y directions, respectively. $C$ is the coefficient of thermoreflectance. After integration, we derive the signal as,
\begin{widetext}
\begin{equation}
    I(\omega) =\frac{CP_0}{(2\pi)^2} \int \frac{\alpha\gamma(1+i\omega\tau_\mathrm{HHE})}{\kappa_{zz}(\lambda-\alpha^2)}\left(\frac{1}{\alpha+\gamma}-\frac{\alpha}{\sqrt{\lambda}}\frac{1}{\sqrt{\lambda}+\gamma}\right) e^{-\frac{R^2_xq_x^2}{4}}e^{-\frac{R^2_yq_y^2}{4}}e^{-iq_x x_\mathrm{o}}e^{-iq_y y_\mathrm{o}}d^2\mathbf{q}_\parallel,
    \label{thhe}
\end{equation}
\end{widetext}
where $R_x = \sqrt{(\sigma_x^2+\sigma_x^{\prime 2})/2}$ and $R_y = \sqrt{(\sigma_y^2+\sigma_y^{\prime 2})/2}$ are the effective radii along x and y directions. One can further simplify the above relation by converting the Fourier transform to Hankel transform if the two effective radii are the same. At the limit of $\tau_\mathrm{HHE}\to0$, the Eq.~(\ref{thhe}) reduces to the case of Fourier's heat conduction.

We consider the case when the penetration depths for the pump and probe beams are very small ($\alpha, \gamma\gg |\sqrt{\lambda}|$) and the signal is simplified as,
\begin{equation}
    I(\omega)=\frac{CP_0}{\left(2\pi\right)^2}\int\frac{1}{\kappa_{zz}\sqrt{\lambda}}\left(1+i\omega\tau_\mathrm{HHE}\right)e^{-\frac{R^2_xq_x^2}{4}}e^{-\frac{R^2_yq_y^2}{4}}e^{-iq_x x_\mathrm{o}}e^{-iq_y y_\mathrm{o}}d^2\mathbf{q}_\parallel,
\end{equation}
When the thermal conductivity is isotropic ($\kappa_{ii}=\kappa$, $i = x, y, z$), the modulation frequency is high ($\omega\gg \frac{\kappa}{C_pR_i^2}$, $i = x, y$), and the beam offset distance is zero, the signal becomes,
\begin{equation}
    \begin{split}
    I(\omega) &= \frac{CP_0}{\left(2\pi\right)^2}\int\frac{1}{\sqrt{i\kappa C_p\omega}} 
    \sqrt{1+i\omega\tau_\mathrm{HHE}}e^{-\frac{R^2_xq_x^2}{4}}e^{-\frac{R^2_yq_y^2}{4}} d^2\mathbf{q}_\parallel\\
    &=\frac{CP_0}{\pi R_x R_y\sqrt{i\kappa C_p\omega}}\sqrt{1+i\omega\tau_\mathrm{HHE}}
    \end{split}
\end{equation}
When the $\omega\tau_\mathrm{HHE}\ll 1$,
assuming isotropic radius $R_x = R_y = r_\mathrm{eff}$, we obtain the following relation,
\begin{equation}
    \begin{split}
        I(\omega)
        &\approx \frac{CP_0}{\pi r_\mathrm{eff}^2\sqrt{i\kappa C_p \omega}}\left(1+\frac{i\omega\tau_\mathrm{HHE}}{2}\right)\\
        &\approx \frac{CP_0}{\pi r_\mathrm{eff}^2\sqrt{i\kappa C_p \omega}}e^{i\Delta \psi}
    \end{split}
\end{equation}
where $\Delta \psi = \arctan\left(\frac{\omega\tau_\mathrm{HHE}}{2}\right)$ is the phase shift to the temperature rise according to the Fourier's law.

\subsection{Two-temperature model}
\label{eigproblem}

In the phenomenological two-temperature model, the electron and phonon temperatures evolve according to,
\begin{equation}
    C_\mathrm{el}\frac{\partial T_\mathrm{el}}{\partial t}=\nabla\cdot\bm{\kappa}_\mathrm{el} \cdot \nabla T_\mathrm{el} - g_\mathrm{e\mhyphen p}(T_\mathrm{el}-T_\mathrm{ph})+P,
\end{equation}
\begin{equation}
    C_\mathrm{ph}\frac{\partial T_\mathrm{ph}}{\partial t}=\nabla\cdot\bm{\kappa}_\mathrm{ph} \cdot\nabla T_\mathrm{ph} + g_\mathrm{e\mhyphen p}(T_\mathrm{el}-T_\mathrm{ph}),
\end{equation}
where $C_\mathrm{el}$ and $C_\mathrm{ph}$ are the volumetric specific heat for electrons and phonons.

In the frequency domain and $q_xq_y$-plane in the reciprocal space, the above two equations read,
\begin{equation}
    C_\mathrm{el} i\omega \tilde{T}_\mathrm{el} =  \left(-\kappa_\mathrm{el,\parallel}q_\parallel^2+\kappa_{\mathrm{el},z}\frac{\partial^2}{\partial z^2}\right)\tilde{T}_\mathrm{el} - g_\mathrm{e\mhyphen p}(\tilde{T}_\mathrm{el}-\tilde{T}_\mathrm{ph})+\tilde{P},
\end{equation}
\begin{equation}
    C_\mathrm{ph} i\omega \tilde{T}_\mathrm{ph} = \left(-\kappa_{\mathrm{ph},\parallel} q^2_\parallel +\kappa_{\mathrm{ph},z}\frac{\partial^2}{\partial z^2}\right)\tilde{T}_\mathrm{ph} + g_\mathrm{e\mhyphen p}(\tilde{T}_\mathrm{el}-\tilde{T}_\mathrm{ph}),
\end{equation}
where $\tilde{P}(\omega,\mathbf{q}_\parallel,z)=P_0\alpha e^{-\frac{\sigma_x^2q^2_x}{8}}e^{-\frac{\sigma_y^2q^2_y}{8}}e^{-\alpha z}$ describes laser heating, with $\alpha$ being the inverse penetration depth and $\sigma_\alpha$ being the $1/e^2$ pump beam radii in x- and y-directions. Here, we have assumed that the excited electrons thermalizes among themselves efficiently via electron-electron scattering or intervalley electron-phonon scattering, which is an extremely fast process regarded as instantaneous in the timescale of our interest $\sim \omega^{-1}$. 

Denote $\Lambda_\mathrm{el} = \frac{iC_\mathrm{el}\omega+\kappa_{\mathrm{el},\parallel}q^2_\parallel+g_\mathrm{e\mhyphen p}}{\kappa_{\mathrm{el},z}}$ and $\Lambda_\mathrm{ph} = \frac{iC_\mathrm{ph}\omega+\kappa_{\mathrm{ph},\parallel}q^2_\parallel+g_\mathrm{e\mhyphen p}}{\kappa_{\mathrm{ph},z}}$. The above equations become,
    \begin{equation}
        \frac{\partial^2\tilde{T}_\mathrm{el}}{\partial z^2}- \Lambda_\mathrm{el}\tilde{T}_\mathrm{el}  +\frac{g_\mathrm{e\mhyphen p}}{\kappa_{\mathrm{el},z}}\tilde{T}_\mathrm{ph}+ \frac{\tilde{P}}{\kappa_{\mathrm{el},z}} =0,
        \label{ttel}
    \end{equation}
    \begin{equation}
        \frac{\partial^2\tilde{T}_\mathrm{ph}}{\partial z^2}- \Lambda_\mathrm{ph}\tilde{T}_\mathrm{ph}  +\frac{g_\mathrm{e\mhyphen p}}{\kappa_{\mathrm{ph},z}}\tilde{T}_\mathrm{el} =0.
        \label{ttph}
    \end{equation}

We define an auxiliary matrix,
\begin{equation}
    \mathbf{M} = 
    \begin{pmatrix}
        \Lambda_\mathrm{el}&-\frac{g_\mathrm{e\mhyphen p}}{\kappa_{\mathrm{el},z}}\\
        -\frac{g_\mathrm{e\mhyphen p}}{\kappa_{\mathrm{ph},z}}&\Lambda_\mathrm{ph}
    \end{pmatrix}.
\end{equation}
and denote
$\lambda_{i}$ ($i=1,2$) and $\mathbf{u}$ the eigenvalues and the eigenvector matrix of $\mathbf{M}$,
where explicitly,
\begin{equation}
    \mathbf{u} = 
    \begin{pmatrix}
        u_{11}&u_{12}\\
        u_{21}&u_{22}
    \end{pmatrix}.
\end{equation}
We also define the inverse of the eigenvector matrix $\mathbf{v}=\mathbf{u}^{-1}$.

Eq.~(\ref{ttel}) and Eq.~(\ref{ttph}) can now be organized as,
\begin{equation}
    \frac{\partial^2 }{\partial z^2}
    \begin{pmatrix}
        \tilde{T}_\mathrm{el}\\
        \tilde{T}_\mathrm{ph} 
    \end{pmatrix}
    - \mathbf{M}
    \begin{pmatrix}
        \tilde{T}_\mathrm{el}\\
        \tilde{T}_\mathrm{ph}
    \end{pmatrix}
    +\begin{pmatrix}
        \frac{\tilde{P}}{\kappa_{\mathrm{el},z}}\\
        0
    \end{pmatrix}=0.
    \label{eqs}
\end{equation}
Multiplying $\mathbf{v}$ on both sides, we have,
\begin{equation}
    \frac{\partial^2 }{\partial z^2}
     \mathbf{v} \begin{pmatrix}
        \tilde{T}_\mathrm{el}\\
        \tilde{T}_\mathrm{ph}
    \end{pmatrix}
    -
    \begin{pmatrix}
        \lambda_1&0\\
        0&\lambda_2
    \end{pmatrix}
    \mathbf{v}
    \begin{pmatrix}
        \tilde{T}_\mathrm{el}\\
        \tilde{T}_\mathrm{ph}
    \end{pmatrix}
    +\begin{pmatrix}
        \frac{v_{11}\tilde{P}}{\kappa_{\mathrm{el},z}}\\
        \frac{v_{21}\tilde{P}}{\kappa_{\mathrm{el},z}}
    \end{pmatrix}=0.
    \label{eqsv}
\end{equation}
Let 
\begin{equation}
    \begin{pmatrix}
        \tilde{T}^\prime_\mathrm{el}\\
        \tilde{T}^\prime_\mathrm{ph} 
    \end{pmatrix}
    =
    \mathbf{v} \begin{pmatrix}
        \tilde{T}_\mathrm{el}\\
        \tilde{T}_\mathrm{ph} 
    \end{pmatrix}.
\end{equation}
We then obtain two decoupled equations,
\begin{equation}
    \frac{\partial^2}{\partial z^2}
    \tilde{T}_\mathrm{el}^\prime-\lambda_1 \tilde{T}_\mathrm{el}^\prime +\frac{v_{11}\tilde{P}}{k_{\mathrm{el},z}} = 0,
\end{equation}
\begin{equation}
    \frac{\partial^2}{\partial z^2}
    \tilde{T}_\mathrm{ph}^\prime-\lambda_2 \tilde{T}_\mathrm{ph}^\prime +\frac{v_{21}\tilde{P}}{k_{\mathrm{el},z}} = 0.
\end{equation}

Considering the case of energy transport in isotropic materials, we can express the matrix by,
\begin{equation}
    \mathbf{M}=\frac{g_\mathrm{e\mhyphen p}}{\kappa_\mathrm{el}}
    \mathbf{M}_0
    +
    \left(q_\parallel^2+\frac{iC_\mathrm{ph}\omega}{\kappa_\mathrm{ph}}\right)
    \begin{pmatrix}
    1 & 0\\
    0 & 1
    \end{pmatrix},
\end{equation}
where
\begin{equation}
    \mathbf{M}_0
    =
    \begin{pmatrix}
        1+\delta &-1\\
        -K & K
    \end{pmatrix},
\end{equation}
$K = \frac{\kappa_\mathrm{el}}{\kappa_\mathrm{ph}}$, and $\delta = i\omega\frac{\kappa_\mathrm{el}}{g_\mathrm{e\mhyphen p}}\left(\frac{C_\mathrm{el}}{\kappa_\mathrm{el}}-\frac{C_\mathrm{ph}}{\kappa_\mathrm{ph}}\right)$. According to Glassbrenner and Slack\cite{PhysRev.134.A1058}, $\kappa_\mathrm{el}$ is zero in silicon at room temperature. However, under laser irradiation, the photo-excited hot carriers are at nonequilibrium and the diffusion of those hot carriers is also an energy transport process. Thus, it is expected the thermal conductivity of electrons is higher than that from steady-state measurement and becomes a finite value. Nevertheless, in semiconductors, the amount of free carriers per unit volume is much lower than that of phonons, thus we have $K\ll1$.
We estimate $\kappa_\mathrm{el}\sim 1$ W/(m$\cdot$K), $\kappa_\mathrm{ph}\sim 50$ W/(m$\cdot$K), $C_\mathrm{el}\sim 10^4$ J/(m$^3\cdot$ K), $C_\mathrm{ph}\sim 10^6$ J/(m$^3\cdot$ K) and $g_\mathrm{e\mhyphen p} \sim 10^{15} - 10^{16}$ W/m$^3$.
In the typical modulation frequency range for our frequency-domain photo-reflectance $\frac{\omega}{2\pi} < $ 100 MHz, $|\delta| \ll 1$.

The eigenvector matrix $\mathbf{u}$ for $\mathbf{M}$ can be obtained by diagonalizing the matrix $\mathbf{M}_0$, since shifting the diagonals with a constant does not change the eigenvectors. 
The approximated eigenvector matrix for $\mathbf{M}_0$ is found to be,
\begin{equation}
    \mathbf{u}=
    \begin{pmatrix}
        1&1\\
        -K&1
    \end{pmatrix},
\end{equation}
and the corresponding inverse matrix is,
\begin{equation}
    \mathbf{v}=\frac{1}{1+K}
    \begin{pmatrix}
        1&-1\\
        K&1
    \end{pmatrix}.
\end{equation}
The approximated eigenvalues are,
\begin{equation}
    \lambda_1 =
 q_\parallel^2+\frac{iC_\mathrm{el}\omega}{\kappa_\mathrm{el}}+\frac{g_\mathrm{e\mhyphen p}}{\kappa_\mathrm{el}}\left(1+K\right),
\end{equation}
 \begin{equation}
    \lambda_2 = 
q_\parallel^2+\frac{iC_\mathrm{ph}\omega}{\kappa_\mathrm{ph}}+\frac{g_\mathrm{e\mhyphen p}}{\kappa_\mathrm{ph}}\frac{\delta}{1+K+\delta}\approx q^2_\parallel + \frac{iC_\mathrm{ph}\omega}{\kappa_\mathrm{ph}},
\end{equation}
as for most semiconductors, we have $\frac{C_\mathrm{el}}{C_\mathrm{ph}}\ll 1$.

Plug the eigenvalues into Eq.~(\ref{eqsv}) and we arrive at the following relations,
\begin{equation}
    \frac{\partial^2\tilde{\theta}}{\partial z^2}
    -\lambda_1\tilde{\theta}
    +\frac{\tilde{P}}{\kappa_\mathrm{el}}
    =0,
    \label{Te}
\end{equation}
\begin{equation}
    \frac{\partial^2\tilde{\chi}}{\partial z^2}-\lambda_2\tilde{\chi}+\frac{\tilde{P}}{\kappa_\mathrm{ph}}=0,
    \label{Tp}
\end{equation}
where $\tilde{\theta}=\tilde{T}_\mathrm{el}-\tilde{T}_\mathrm{ph}$
is the difference between electron and phonon temperatures
and $\tilde{\chi}=\tilde{T}_\mathrm{ph}+K\tilde{T}_\mathrm{el}$.

The solution to Eq.~(\ref{Te}) follows,
\begin{equation}
    \tilde{\theta} = \frac{\tilde{P}(\omega,\mathbf{q}_\parallel,0)}{\kappa_\mathrm{el}\left(\lambda_1-\alpha^2\right)}\left(-\frac{\alpha}{\sqrt{\lambda_1}}e^{-\sqrt{\lambda_1}z}+e^{-\alpha z}\right),
    \label{deltaT}
\end{equation}
where we have applied adiabatic boundary condition at surface $\partial_z \tilde{\theta}|_{z=0} = 0$. Similarly, the solution to Eq.~(\ref{Tp}) writes,
\begin{equation}
    \tilde{\chi}=
    \frac{\tilde{P}(\omega,\mathbf{q}_\parallel,0)}{\kappa_\mathrm{ph}\left(\lambda_2-\alpha^2\right)}\left(-\frac{\alpha}{\sqrt{\lambda_2}}e^{-\sqrt{\lambda_2}z}+e^{-\alpha z}\right).
\end{equation}
The electron temperature $\tilde{T}_\mathrm{el}$ can now be expressed by,
\begin{equation}
    \tilde{T}_\mathrm{el}=\frac{1}{1+K}\left(\tilde{\theta}+\tilde{\chi}\right).
\end{equation}
In the limit of surface heating from the pump laser ($\alpha\gg |\sqrt{\lambda_i}|$, $i = 1, 2$), we have,
\begin{equation} 
\tilde{\theta}(\omega,\mathbf{q}_\parallel,z=0)=\frac{\tilde{P}(\omega,\mathbf{q}_\parallel,0)}{\alpha\kappa_\mathrm{el}\sqrt{\lambda_1}}.
\end{equation}
As a result, the electron temperature at the surface can be written as,
\begin{equation}
    \tilde{T}_\mathrm{el}\left(\omega,\mathbf{q}_\parallel,z=0\right) = \frac{\tilde{P}(\omega,\mathbf{q}_\parallel,0)}{\alpha\left(1+K\right)}\left(\frac{1}{\kappa_\mathrm{el}\sqrt{\lambda_1}}
    +\frac{1}{\kappa_\mathrm{ph}\sqrt{\lambda_2}}
    \right).
    \label{surfaceTe}
\end{equation}

When the modulation frequency is high such than the phonon diffusion length is smaller than the beam size $\left(\omega\gg \frac{\kappa_\mathrm{ph}}{C_\mathrm{ph}r_\mathrm{eff}^2}\right)$, we have $\lambda_2\approx \frac{iC_\mathrm{ph}\omega}{\kappa_\mathrm{ph}}$. For most semiconductors, the electron-phonon coupling constant $g_\mathrm{e\mhyphen p}$ is on the order of $10^{15}-10^{16}\;\mathrm{W/m^3/K}$, which leads to $\lambda_1\approx \frac{g_\mathrm{e\mhyphen p}}{\kappa_\mathrm{el}}\left(1+K\right)$.
Consequently, we can express the surface electron temperature in Eq.~(\ref{surfaceTe}) by,
\begin{equation}
    \begin{split}
    \tilde{T}_\mathrm{el}\left(\omega,\mathbf{q}_\parallel,z=0\right)&=\frac{\tilde{P}(\omega,\mathbf{q}_\parallel,0)}{\alpha\left(1+K\right)}
    \frac{1}{\kappa_\mathrm{ph}\sqrt{\lambda_2}}
    \left(1+\frac{1}{K}\sqrt{\frac{\lambda_2}{\lambda_1}}\right)\\
     &\approx\underbrace{\frac{\tilde{P}(\omega,\mathbf{q}_\parallel,0)}{\alpha\kappa_\mathrm{ph}\sqrt{\lambda_2}}}_{\tilde{T}_\mathrm{ph}(\omega,\mathbf{q}_\parallel,0)}\left(1+\sqrt{i\omega\tau_\mathrm{e\mhyphen p}}\right)
    \end{split}
    \label{Teep}
\end{equation}
where the characteristic lifetime is $\tau_\mathrm{e\mhyphen p} = \frac{C_\mathrm{ph}}{g_\mathrm{e\mhyphen p}K}$. The electron-phonon characteristic frequency can define by, $\omega_\mathrm{e\mhyphen p}=\tau_\mathrm{e\mhyphen p}^{-1}$.
The first term in the second line of Eq.~(\ref{Teep}) is identified as the phonon temperature rise in a phonon-only model and $T_\mathrm{ph}$ solves the conventional heat equation,
\begin{equation}
    C_\mathrm{ph}\frac{\partial T_\mathrm{ph}}{\partial t} = \kappa_\mathrm{ph}\nabla^2 T_\mathrm{ph} + P.
\end{equation}
This is due to the fact that the electron's heat capacity is so small that electron has minimal impact on phonon's temperature rise.

With the electron temperature solved in the reciprocal space, the electron temperature rise in the real space can be computed using the inverse Fourier transform,
\begin{equation}
    T_\mathrm{el}(\omega,\mathbf{r}_\parallel,z=0)=
    \frac{1}{\left(2\pi\right)^2}
    \int 
    \tilde{T}_\mathrm{el}(\omega,\mathbf{q}_\parallel,0) e^{i\mathbf{q}_\parallel\cdot\mathbf{r}_\parallel}
    d^2\mathbf{q}_\parallel
\end{equation}
The probed signal is given by,
\begin{equation}
    \begin{split}
    I(\omega) &=B\overline{T}_\mathrm{el}(\omega)\\
    &= B \int T_\mathrm{el}(\mathbf{r}_\parallel,z=0)\frac{2e^{-\frac{2x^2}{\sigma_x^{\prime2}}}e^{-\frac{2y^2}{\sigma_y^{\prime2}}}}{\pi\sigma_x^\prime\sigma_y^\prime}
    dxdy\\
     &=\frac{1}{\left(2\pi\right)^2}\frac{BP_0}{\kappa_\mathrm{ph}\sqrt{\lambda_2}}\left(1+\sqrt{i\omega\tau_\mathrm{e\mhyphen p}}\right)\int e^{-\frac{q_x^2R_x^2}{4}} e^{-\frac{q_y^2R_y^2}{4}} d^2\mathbf{q}_\parallel\\
     &=\frac{BP_0}{\pi R_xR_y\kappa_\mathrm{ph}\sqrt{\lambda_2}}\left(1+\sqrt{i\omega\tau_\mathrm{e\mhyphen p}}\right)
    \end{split}
\end{equation}
where $\sigma_x^\prime$ and $\sigma_y^\prime$ are the probe radius, and $R_x=\sqrt{\left(\sigma_x^2+\sigma_x^{\prime2}\right)/2}$ and $R_y=\sqrt{\left(\sigma_y^2+\sigma_y^{\prime2}\right)/2}$ are the effective radii.
Assuming an isotropic effective beam radius $R_x = R_y = r_\mathrm{eff}$, we have,
\begin{equation}
    \begin{split}
     I(\omega)&=B\underbrace{\frac{P_0}{\pi r_\mathrm{eff}^2\sqrt{i\kappa_\mathrm{ph} C_\mathrm{ph}\omega}}}_{\overline{T}_\mathrm{ph}(\omega)}\left(1+\sqrt{i\omega\tau_\mathrm{e\mhyphen p}}\right)\\
     &\approx B\overline{T}_\mathrm{ph}(\omega) e^{i\Delta \theta(\omega)}
    \end{split}
\end{equation}
where 
\begin{equation}
 \Delta\theta(\omega) = \arctan(\frac{\sqrt{\omega\tau_\mathrm{e\mhyphen p}}}{\sqrt{2}+\sqrt{\omega\tau_\mathrm{e\mhyphen p}}})
\end{equation}
is the phase shift of electron temperature $\overline{T}_\mathrm{el}(\omega)$ with respect to phonon temperature $\overline{T}_\mathrm{ph}(\omega)$ due to electron-phonon interaction. Note that sign of $\Delta \theta(\omega)$ is positive, indicating that $\overline{T}_\mathrm{el}(\omega)$ always has a smaller phase lag than $\overline{T}_\mathrm{ph}(\omega)$. Also, we find the ratio between the amplitudes of electron temperature and phonon temperature is,
\begin{equation}
    \frac{|\overline{T}_\mathrm{el}(\omega)|}{|\overline{T}_\mathrm{ph}(\omega)|}\approx \sqrt{1+\omega\tau_\mathrm{e\mhyphen p}+\sqrt{2\omega\tau_\mathrm{e\mhyphen p}}}.
\end{equation}

From Eq.~(\ref{deltaT}), we find that in the limit of zero modulation frequency, the difference between electron and phonon temperature is, 
\begin{equation}
    \begin{split}
\overline{T}_\mathrm{el}-\overline{T}_\mathrm{ph} &\approx \frac{P_0}{2\pi\sqrt{\kappa_\mathrm{el}g_\mathrm{e\mhyphen p}(1+K)}}\int_0^\infty qe^{-r_\mathrm{eff}^2q^2/4}dq\\
&= \frac{P_0}{\pi r_\mathrm{eff}^2\sqrt{k_\mathrm{el}g_\mathrm{e\mhyphen p}(1+K)}}.
    \end{split}
    \label{ttmzero}
\end{equation}
This is countering the common argument that the electron and phonon temperatures are eventually becoming the same at long enough time in a pump-probe measurement using a pulsed pump laser. The reason behind the temperature difference in zero modulation frequency is that in frequency-domain photo-reflectance, the nonequilibrium between electron and phonon always exist as they are constantly being modulated.

\subsection{Carrier and phonon model}

Oftentimes, the signal due to the carrier in the carrier and phonon model is almost perfectly in phase at modest modulation frequencies such that we can simply use a real number to represent the carrier part of the signal. Also, we find that at low modulation frequency, the phase of the signal is dominated by the phonon temperature. Yet at higher modulation frequency, the carrier plays a more tangible role, since the phonon temperature's amplitude decreases rapidly with the modulation frequency. When $\omega \gg \frac{\kappa_\mathrm{ph}}{C_\mathrm{ph}r_\mathrm{eff}^2}$, the signal from modulated carriers and phonons is,
\begin{equation}
    \begin{split}
     I(\omega)&=A\rho_0+C\frac{P_0}{\pi r_\mathrm{eff}^2\sqrt{i\kappa_\mathrm{ph} C_\mathrm{ph}\omega}}\\
     &=\frac{CP_0}{\pi r_\mathrm{eff}^2\sqrt{\kappa_\mathrm{ph}C_\mathrm{ph}\omega}}\left(\frac{\sqrt{2}}{2}-\frac{\sqrt{2}}{2}i+D\right)\\
     &\approx\frac{CP_0}{\pi r_\mathrm{eff}^2\sqrt{\kappa_\mathrm{ph}C_\mathrm{ph}\omega}} e^{i\phi}
    \end{split}
\end{equation}
where
\begin{equation}
    D = \frac{A\rho_0\pi r_\mathrm{eff}^2\sqrt{\kappa_\mathrm{ph}C_\mathrm{ph}\omega}}{CP_0}
\end{equation}
\begin{equation}
    \phi = -\arctan\left(\frac{1}{1+\sqrt{2}D}\right)\approx -\frac{\pi}{4}+\frac{\sqrt{2}}{2}D
\end{equation}
Therefore,
\begin{equation}
     I(\omega)\approx  \frac{CP_0}{\pi r_\mathrm{eff}^2\sqrt{i\kappa_\mathrm{ph} C_\mathrm{ph}\omega}}e^{i\Delta\phi(\omega)}
\end{equation}
where the phase shift from the carriers to the phase of phonon temperature is $\Delta\phi(\omega)=\frac{\sqrt{2}}{2}\frac{A\rho_0\pi r_\mathrm{eff}^2\sqrt{\kappa_\mathrm{ph}C_\mathrm{ph}\omega}}{CP_0}$. This indicates that at higher frequency, the phase correction is more significant.

We believe the above detailed discussion validates our carrier and phonon model.

\clearpage

\begin{figure}[H]

\vspace{4 cm}

    \centering
    \includegraphics[width=\linewidth]{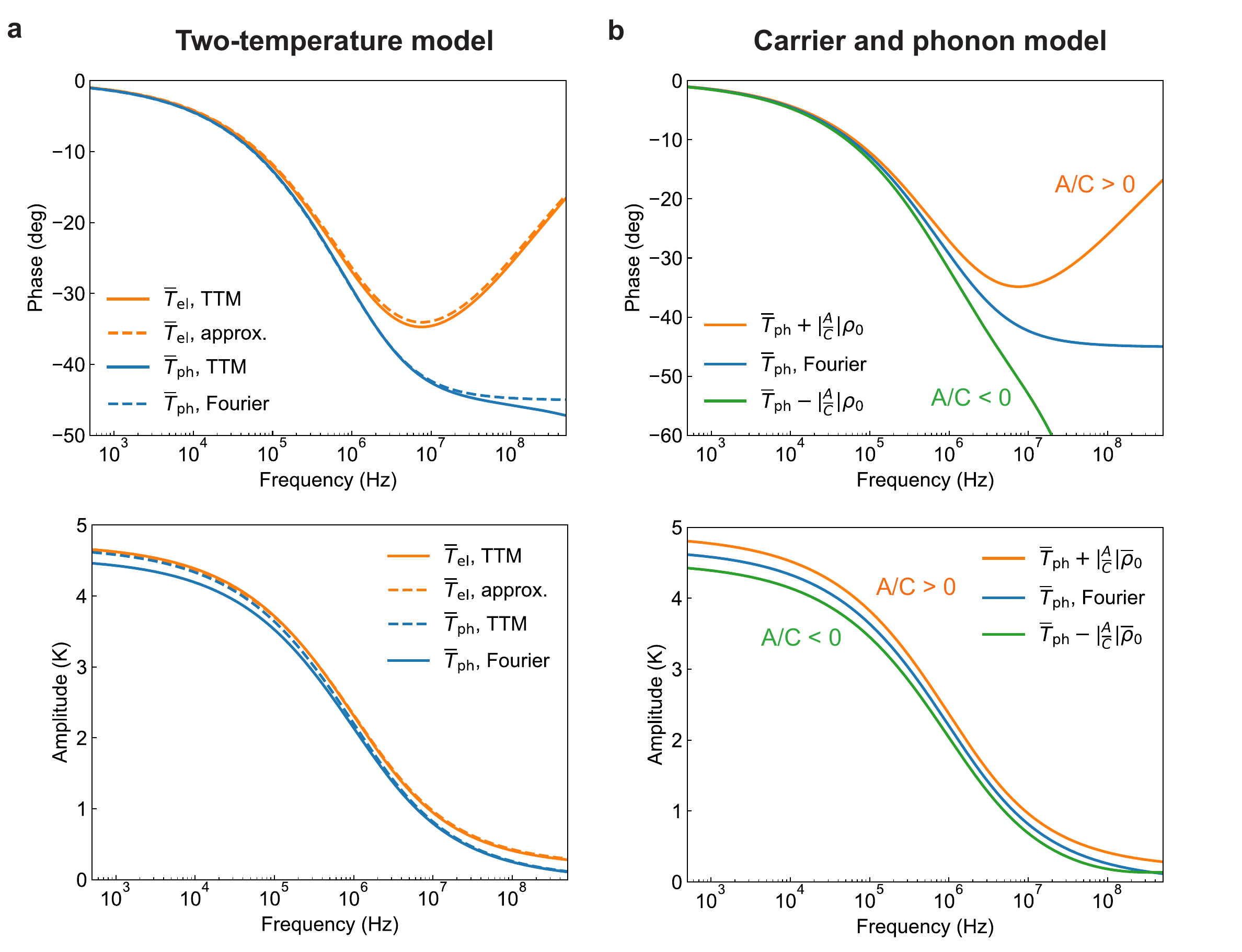}
    \caption[Curve-up features]{The curve up features in two different models. a. The phase of electron temperature curves up compared with phonon temperature. The difference between amplitudes of electron and phonon temperature is 0.2 K, well described by Eq.~(\ref{ttmzero}). b. Depending on the sign of $A/C$, the phase of the signal can curve up/down compared with the phonon temperature. $g_\mathrm{e\mhyphen p} = 5\times 10^{16}$ $\mathrm{W/K/m^3}$ is used in the calculation.}
    \label{example_curveup}
\end{figure}

\clearpage

\section{Transport properties from the FDTR measurements using gold transducer}

\begin{figure}[H]

\vspace{3 cm}

\centering
\includegraphics[width=0.85\linewidth]{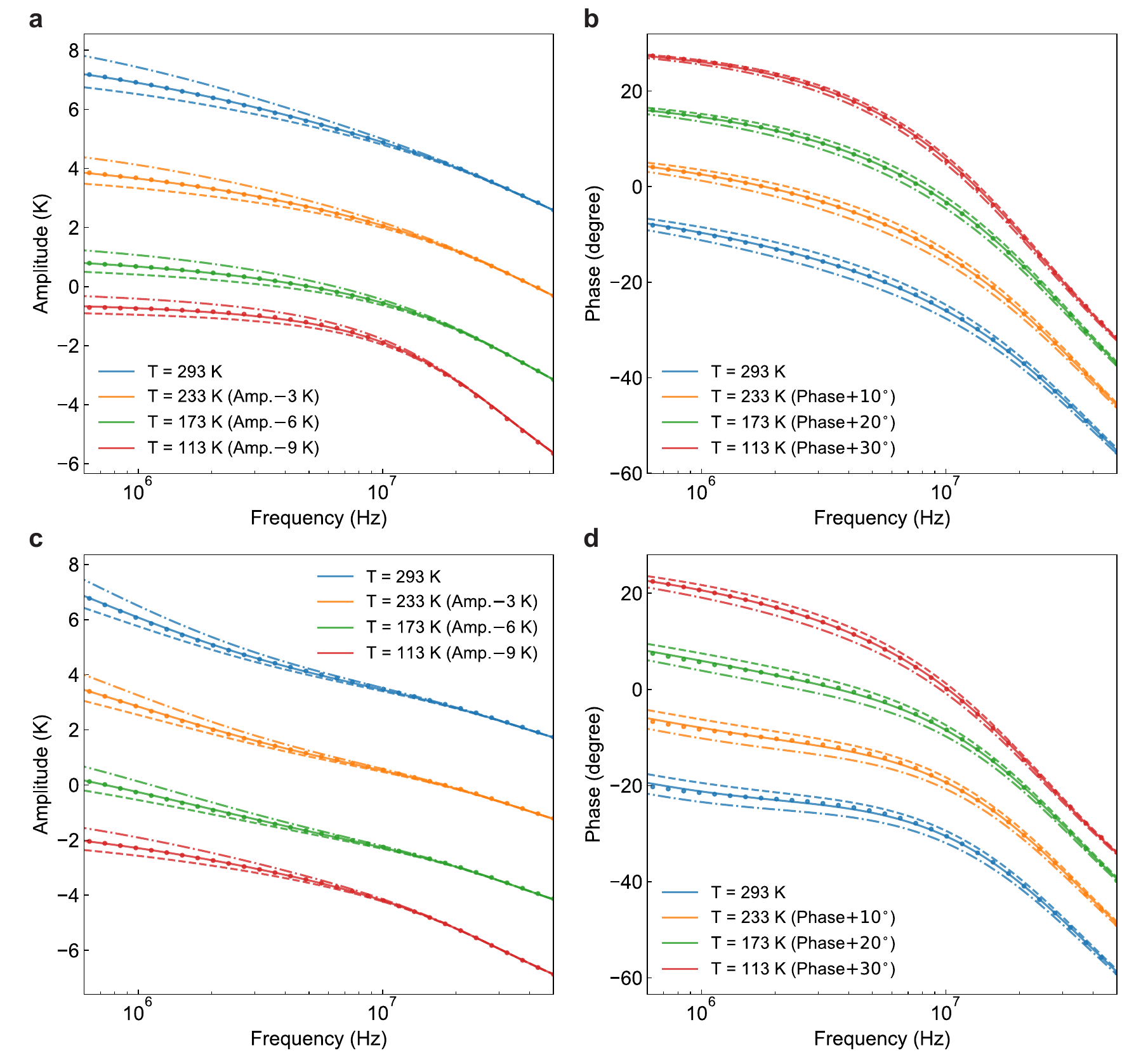}
\caption[Si fits]{The experimental data and best fits for gold-coated Si with $r_\mathrm{eff} =$ 2.93 $\mu$m (top row) and $r_\mathrm{eff} =$ 10.07 $\mu$m (bottom row). The dashed line and the dash-dotted line correspond to the fitted thermal conductivity deviated +20 \% and -20 \% from the best fit in solid line, correspondingly.}
\end{figure}

\clearpage

\begin{figure}[H]

\vspace{4 cm}
\centering
\includegraphics[width=0.85\linewidth]{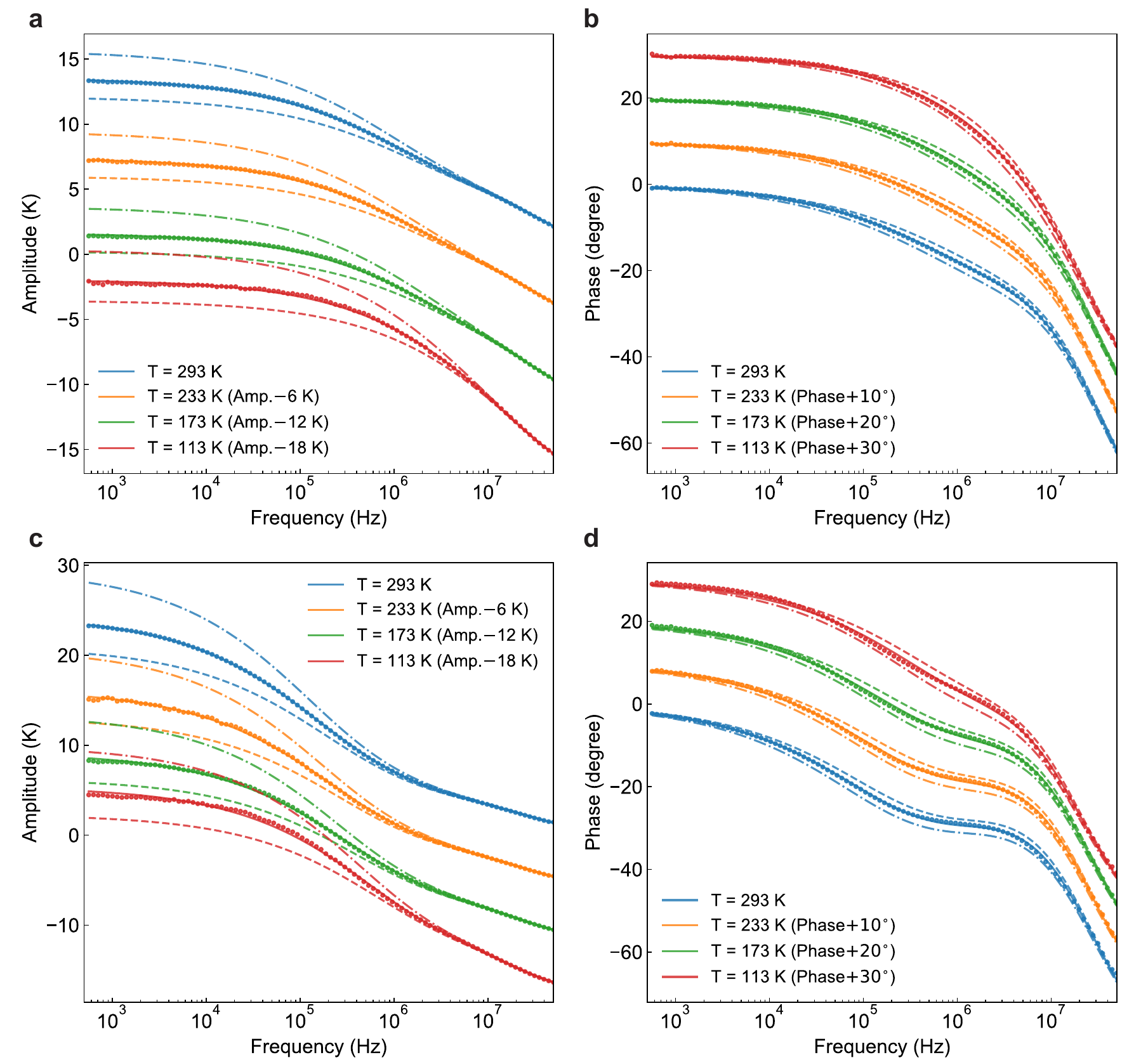}
\caption[SiGe fits]{The experimental data and best fits for gold-coated Si$_\mathrm{0.98}$Ge$_\mathrm{0.02}$ with $r_\mathrm{eff} = $ 2.91 $\mu$m (top row) and $r_\mathrm{eff} =$ 10.20 $\mu$m (bottom row). The dashed line and the dash-dotted line correspond to the fitted thermal conductivity deviated +20 \% and -20 \% from the best fit in solid line, correspondingly.}
\end{figure}

\clearpage

\begin{figure}[H]

\vspace{4 cm}
\centering
\includegraphics[width=0.85\linewidth]{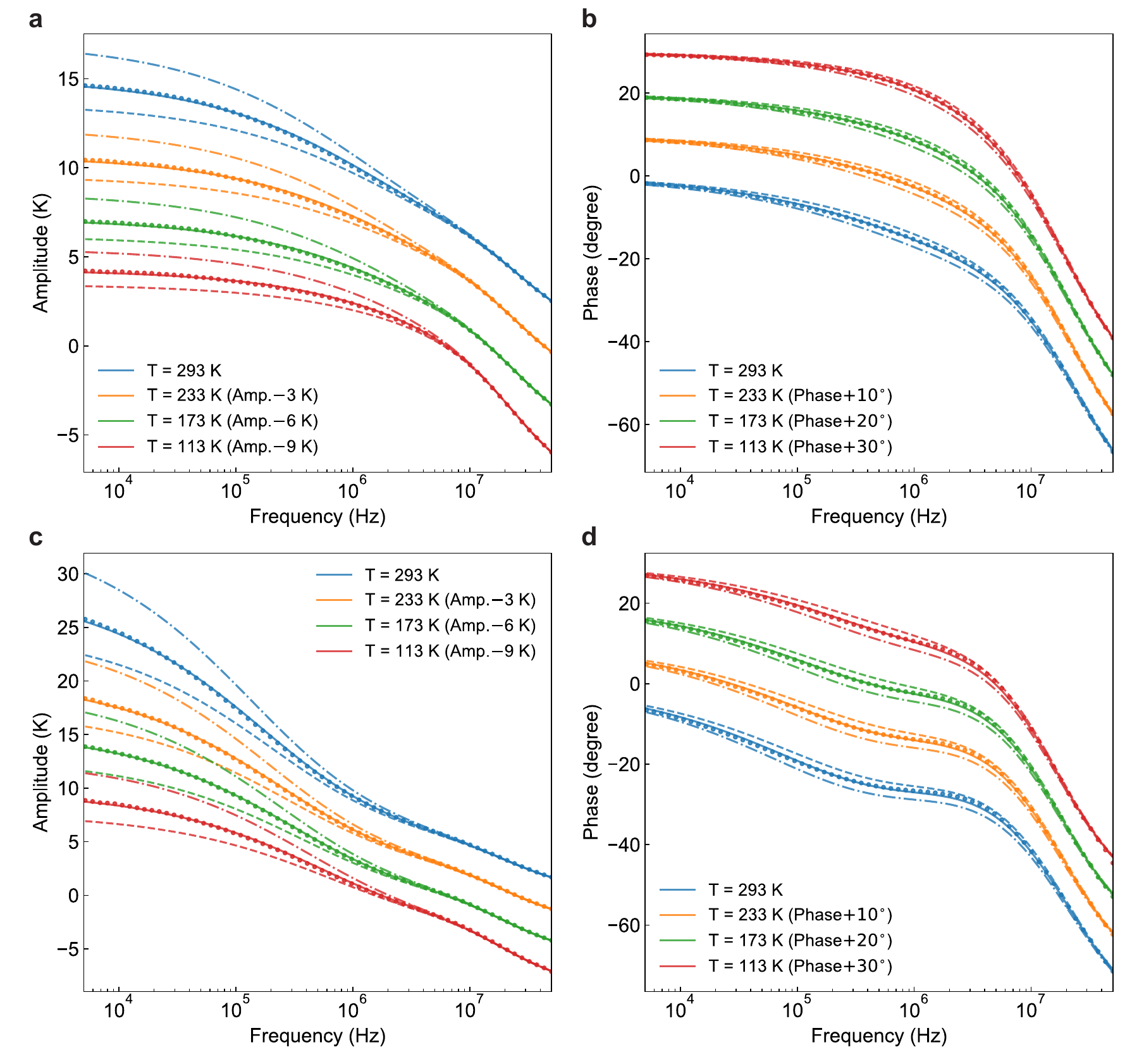}
\caption[Ge fits]{The experimental data and best fits for gold-coated Ge with $r_\mathrm{eff} = $ 2.91 $\mu$m (top row) and $r_\mathrm{eff} =$ 10.10 $\mu$m (bottom row). The dashed line and the dash-dotted line correspond to the fitted thermal conductivity deviated +20 \% and -20 \% from the best fit in solid line, correspondingly.}
\end{figure}

\begin{figure}[H]

\vspace{4 cm}
    \centering
\includegraphics[width=0.85\linewidth]{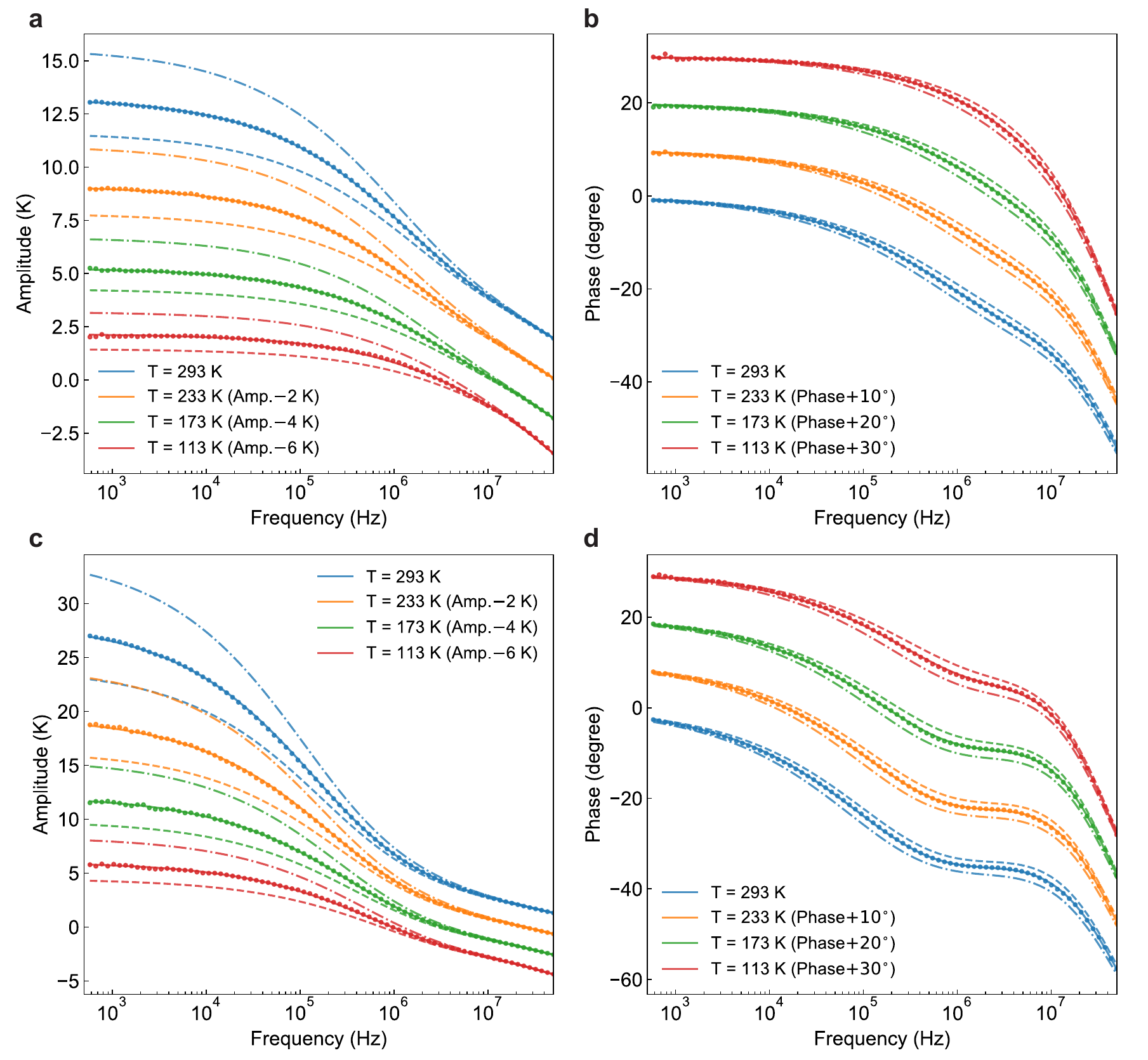}
\caption[GaAs fits]{The experimental data and best fits for gold-coated GaAs with $r_\mathrm{eff} = $ 2.91 $\mu$m (top row) and $r_\mathrm{eff} =$ 10.06 $\mu$m (bottom row). The dashed line and the dash-dotted line correspond to the fitted thermal conductivity deviated +20 \% and -20 \% from the best fit in solid line, correspondingly.}
\end{figure}

\begin{figure}[H]
    \centering
\includegraphics[width=\linewidth]{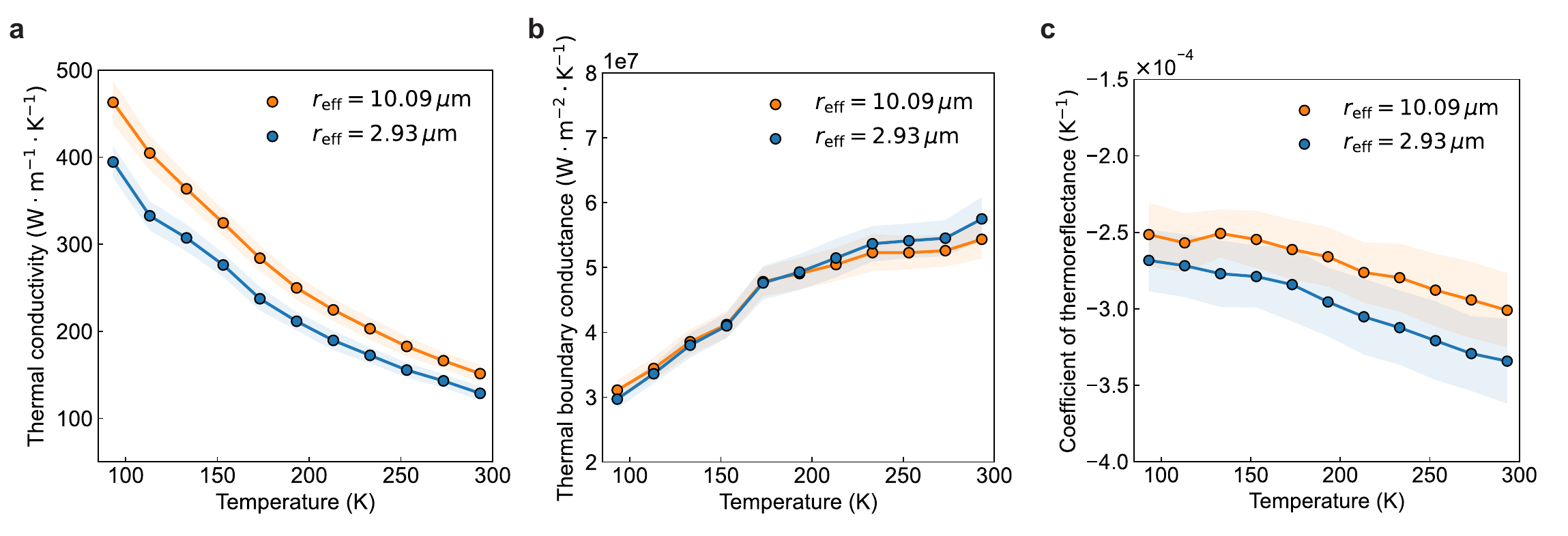}
\caption[Si transport]{The experimental and fits for Si with gold transducer.}
\end{figure}
\begin{figure}[H]
    \centering
\includegraphics[width=0.7\linewidth]{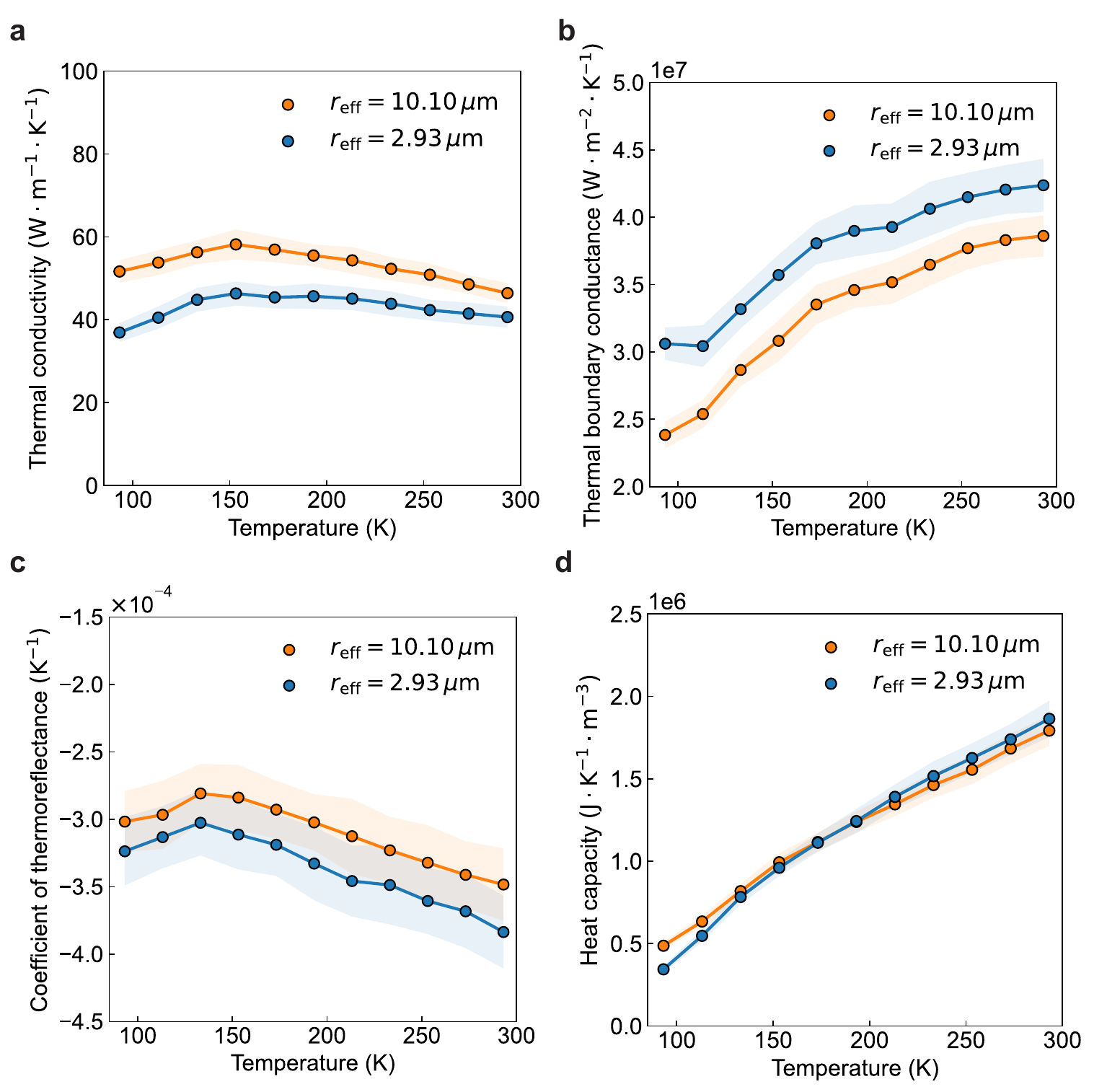}
\caption[SiGe]{The experimental and fits for Si$_{0.98}$Ge$_{0.02}$ with gold transducer.}
\end{figure}
\begin{figure}[H]
    \centering
\includegraphics[width=\linewidth]{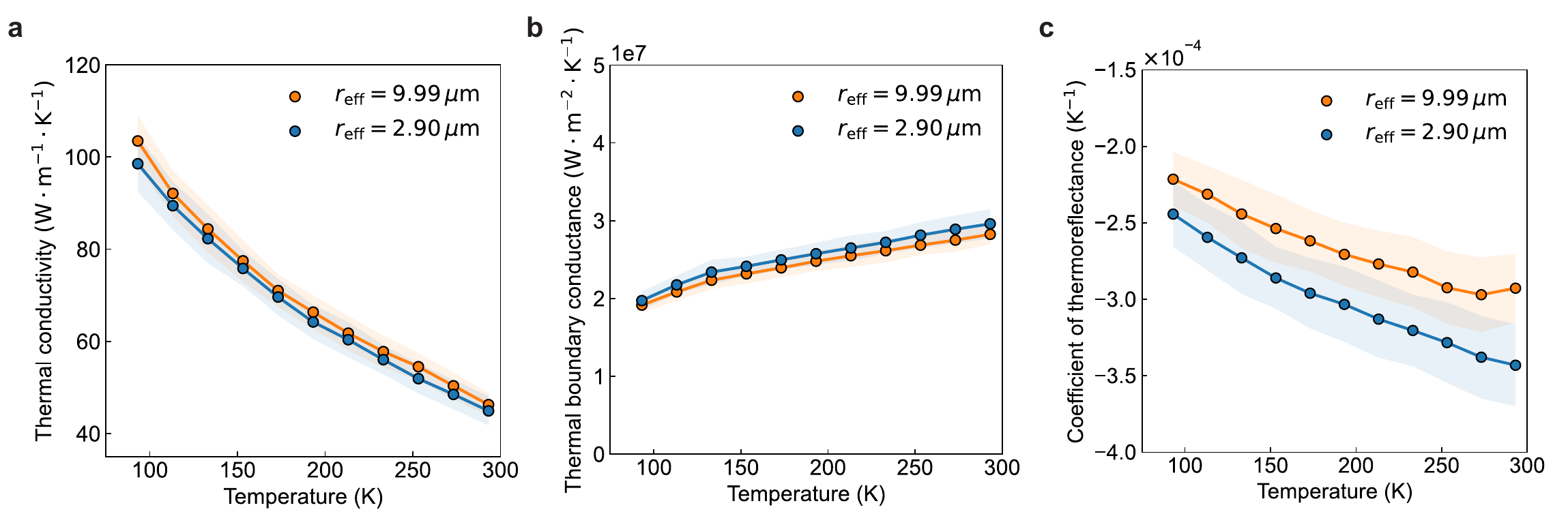}
\caption[Ge transport]{The experimental and fits for Ge with gold transducer.}
\end{figure}
\begin{figure}[H]
    \centering
\includegraphics[width=0.7\linewidth]{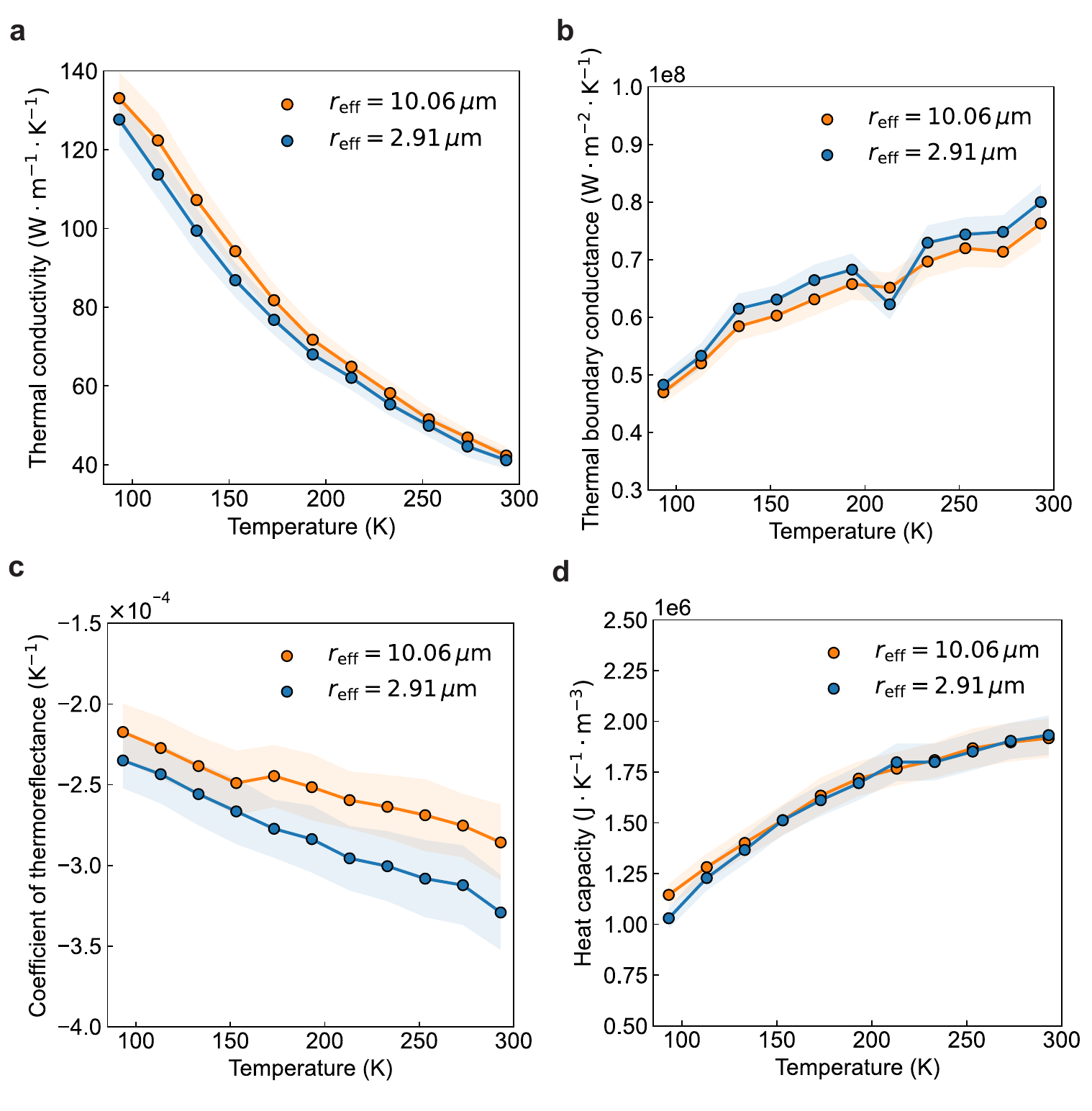}
\caption[GaAs transport]{The experimental data and fits for GaAs with Au transducer.}
\end{figure}

\clearpage

\section{Green's function approach to heat conduction in the small beam spot limit}

Our carrier and phonon model or the full model both explicitly consider the geometry of the pump laser beam. In a more general case, we want to know the Green's function so that we can easily evaluate the temperature response upon any type of heating geometry. In the section, we solve the temperature response under a modulated point heat source using the hyperbolic heat equation and Fourier heat equation. We aim to identify behaviors in the temperature distribution that are exclusively due to non-Fourier heat conduction.

We start with the hyperbolic heat equation in isotropic materials,
\begin{equation}
    \frac{\partial^2 T}{\partial t^2}+\frac{1}{\tau}\frac{\partial T}{\partial t}-c^2\nabla^2 T =0
    \label{hh3}
\end{equation}
where $c^2=\frac{\kappa}{C_p\tau}$.
The Green's function for the hyperbolic heat equation reads,
\begin{equation}
    \frac{\partial^2 G}{\partial t^2}+\frac{1}{\tau}\frac{\partial G}{\partial t}-c^2\nabla^2 G =\delta^3(\mathbf{r})\delta(t)
    \label{Ghhe}
\end{equation}
We further denote the Fourier transform of the Green's function,
\begin{equation}
    \tilde{G}(\mathbf{k},\omega) =\iint G(\mathbf{r},t) e^{-i(\mathbf{k}\cdot\mathbf{r}+\omega t)}d^3\mathbf{r}dt
\end{equation}
Subsequently, the Green's function can be then expressed by,
\begin{equation}
    G(\mathbf{r},t) = \frac{1}{(2\pi)^4}\iint \tilde{G}(\mathbf{k},\omega) e^{i\omega t}e^{i\mathbf{k}\cdot\mathbf{r}}d\mathbf{k}d\omega
\end{equation}

Under external perturbation $P(\mathbf{r},t)$, the hyperbolic heat equation in Eq.~(\ref{hh3}) becomes,
\begin{equation} 
\frac{\partial^2 T}{\partial t^2}+\frac{1}{\tau}\frac{\partial T}{\partial t}-c^2\nabla^2 T = P(\mathbf{r},t)
\label{hheext}
\end{equation}
The temperature response due to external excitation $P(\mathbf{r},t)$ can be expressed by,
\begin{equation}
    \begin{split}
    T(\mathbf{r},t)& = \int dt^\prime \int d\mathbf{r}^\prime G(\mathbf{r}-\mathbf{r}^\prime,t-t^\prime) P(\mathbf{r}^\prime,t^\prime)\\
    &=\frac{1}{(2\pi)^4}\iint \tilde{G}(\mathbf{k},\omega)\tilde{P}(\mathbf{k},\omega)e^{i(\mathbf{k}\cdot\mathbf{r}+\omega) t}d^3\mathbf{k}d\omega
    \end{split}
    \label{extF}
\end{equation}
where $\tilde{P}(\mathbf{k},\omega) = \iint P(\mathbf{r},t) e^{-i(\mathbf{k}\cdot\mathbf{r}+\omega t)}d^3\mathbf{r}dt$ is the Fourier of the external excitation.
Essentially, we can compute the temperature response of any external perturbation using the Green's function as a kernel function.
We can verify this by plug the first line of Eq.~(\ref{extF}) in Eq.~(\ref{hheext}),
\begin{equation}
    \begin{split}
    &\frac{\partial^2 T}{\partial t^2}+\frac{1}{\tau}\frac{\partial T}{\partial t}-c^2\nabla^2 T =\int dt^\prime \int d^3\mathbf{r}^\prime \left[\frac{\partial^2G}{\partial (t-t^\prime)^2}+ \frac{1}{\tau}\frac{\partial G}{\partial (t-t^\prime)}-c^2\nabla^2_{\mathbf{r}^\prime-\mathbf{r}} G\right]F(\mathbf{r}^\prime,t^\prime)=F(\mathbf{r},t)
    \end{split}
\end{equation}
which shows that the expression in Eq.~(\ref{extF}) is indeed the solution to the hyperbolic heat equation with external input Eq.~(\ref{hheext}).

According to Eq.~(\ref{Ghhe}), the Fourier transform of the Green's function should satisfy,
\begin{equation}
   \left(-\omega^2+\frac{i\omega}{\tau}+c^2|\mathbf{k}|^2\right)\tilde{G}(\mathbf{k},\omega)=1
\end{equation}
Hence, the Green's function can be written as,
\begin{equation}
    G(\mathbf{r},t) = \frac{1}{(2\pi)^4}\iint \frac{1}{-\omega^2+\frac{i\omega}{\tau}+c^2|\mathbf{k}|^2}e^{i(\mathbf{k}\cdot\mathbf{r}+\omega t)}d\mathbf{k}d\omega
\end{equation}
To seek the expression of Green's function,
we consider the one-dimensional case for the above integral,
\begin{equation}
    \begin{split}
    G(x,t) &= \frac{1}{(2\pi)^2}\iint \frac{1}{-\omega^2+\frac{i\omega}{\tau}+c^2k^2}e^{i\omega t}e^{ikx}dkd\omega\\
    &= \frac{1}{(2\pi)^2}\iint \frac{1}{c^2\left(k^2-z_0\right)}e^{i\omega t}e^{ikx}dkd\omega
    \end{split}
\end{equation}
where $z_0 = \frac{\omega^2}{c^2}-\frac{i\omega}{c^2\tau}=r_0e^{i\theta_0}$, $r_0 = \frac{\omega}{c^2}\sqrt{\omega^2+\tau^{-2}}$ and $-\frac{1}{2}\pi<\theta_0 = -\arctan{\frac{1}{\omega\tau}}<0$. Therefore, $\sqrt{z_0}=\sqrt{r_0}e^{i\theta_0/2}$.

We only need to solve the inner integral as the nature of frequency-domain photo-reflectance involves single-frequency heating,
\begin{equation}
    \begin{split}
    I& =  \int \frac{e^{ikx}}{(k-\sqrt{z_0})(k+\sqrt{z_0})}dk\\
    & = \frac{1}{2\sqrt{z_0}}\int \left(\frac{e^{ikx}}{k-\sqrt{z_0}}- \frac{e^{ikx}}{k+\sqrt{z_0}}\right)dk
    \end{split}
\end{equation}
Using Cauchy's residue theorem, when $x>0$, we have,
\begin{equation}
    I 
    =-\frac{\pi i}{\sqrt{z_0 }}e^{-i\sqrt{z_0}x}
\end{equation}
When $x<0$, we have,
\begin{equation}
    I 
    =-\frac{\pi i}{\sqrt{z_0 }}e^{i\sqrt{z_0}x}
\end{equation}
Hence, the Green's function in one-dimensional system at frequency $\omega$ writes,
\begin{equation}
    G(x,\omega)=-\frac {i}{2c\sqrt{\omega^2-\frac{i\omega}{\tau}}}e^{-i\sqrt{\omega^2-\frac{i\omega}{\tau}}\frac{|x|}{c}}
\end{equation}

In three-dimensional system with spherical symmetry, the hyperbolic heat equation is,
\begin{equation}
    \frac{\partial^2 T}{\partial t^2}+\frac{1}{\tau}\frac{\partial T}{\partial t}-\frac{c^2}{r}\frac{\partial^2}{\partial r^2}rT=\frac{P_0\delta(r)\delta(t)}{4\pi r^2}
\end{equation}
where $P_0$ is a constant that scales with the heating power.
In the frequency domain,
\begin{equation}
    -\omega^2T+\frac{i\omega}{\tau}T-\frac{c^2}{r}\frac{\partial^2rT}{\partial r^2}=\frac{P_0\delta(r)}{4\pi r^2}
    \label{3d}
\end{equation}
Let $\tilde{T}=r T$, we have,
\begin{equation}
    -\omega^2\tilde{T}+\frac{i\omega}{\tau}\tilde{T}-c^2\frac{\partial^2\tilde{T}}{\partial r^2}=\frac{P_0\delta(r)}{4\pi r}
\end{equation}
It is easy to show that solution has the following form while $r>0$,
\begin{equation}
    \tilde{T}(\mathbf{r},\omega) = Ae^{-i\sqrt{\omega^2-\frac{i\omega}{\tau}}\frac{r}{c}}
\end{equation}
Accordingly,
\begin{equation}
T(\mathbf{r},\omega) =\frac{A}{r}e^{-i\sqrt{\omega^2-\frac{i\omega}{\tau}}\frac{r}{c}}
\label{tr}
\end{equation}
We then need to validate if the solution still holds at $r = 0$.
Using Gauss's theorem over the domain of a sphere with radius of $a$ centered at the origin, we have,
\begin{equation}
    \int \nabla^2 T d^3\mathbf{r}= \int \nabla T \cdot \hat{n} dS = -4\pi Ae^{-i\sqrt{\omega^2-\frac{i\omega}{\tau}}\frac{a}{c}} - 4\pi Ai\sqrt{\omega^2-\frac{i\omega}{\tau}}\frac{a}{c} 
\end{equation}
When $a\to 0$, we have,
\begin{equation}
    \int \nabla^2 T d^3\mathbf{r}
    = -4\pi A
\end{equation}
As a result, we conclude that as $r \to 0$, the expression in Eq.~(\ref{tr}) satisfies,
\begin{equation}
   \nabla^2 T  = -4\pi A\delta^3(\mathbf{r})
\end{equation}
After substituting Eq.~(\ref{tr}) in the following hyperbolic heat equation in the frequency domain,
\begin{equation}
    \left(-\omega^2+\frac{i\omega}{\tau}-c^2\nabla^2\right)T(\mathbf{r},\omega)=P_0\delta^3(\mathbf{r})
\end{equation}
we obtain the expression of $A$ and the solution for temperature when the external heating is an infinitely fast pulse can be expressed by,
\begin{equation}
    \begin{split}
    T(\mathbf{r},\omega)& = \frac{P_0}{4\pi c^2}\frac{e^{-i\sqrt{\omega^2-\frac{i\omega}{\tau}}\frac{r}{c}}}{r}
    \end{split}
    \label{TG}
\end{equation}

We now consider the expression for the periodic heating source in the frequency-domain photo-reflectance experiment, $P(\mathbf{r},t) = P_0\left(\tau\frac{\partial}{\partial t}+1\right)e^{i\omega _0t}\frac{\delta(r)}{4\pi r^2}$, where we consider the cases when the beam radii for the pump and probe are much smaller that the heat diffusion length. Its Fourier transform in the frequency domain writes,
\begin{equation}
P(\mathbf{r},\omega)=2\pi P_0\delta(\omega-\omega_0)(1+i\omega\tau)\frac{\delta(r)}{4\pi r^2}
\end{equation}
Thereby, according to Eqs.~(\ref{TG}) and ~(\ref{extF}), the temperature under periodic heating of modulation frequency $\omega_0$ is as follows,
\begin{equation}
T(\mathbf{r},t) 
 = T(\mathbf{r})e^{i\omega_0 t}
\end{equation}
\begin{equation}
T(\mathbf{r})    
= \frac{P_0(1+i\omega_0\tau)}{4\pi c^2}\frac{e^{-i\sqrt{\omega_0^2-\frac{i\omega_0}{\tau}}\frac{r}{c}}}{r}
\end{equation}
Meanwhile, it is easy to write the temperature solution when $\tau\to\infty$, \textit{i.e.}, the solution to conventional heat equation under periodic heating, \begin{equation}
    T_0(\mathbf{r},t)=T_0(\mathbf{r})e^{i\omega_0 t} \propto \frac{e^{-\sqrt{i\frac{\omega_0 C_p}{\kappa}}r}}{r}e^{i\omega_0 t}
\end{equation} 

\begin{figure}[H]
    \vfill
    \centering
\includegraphics[width=0.85\linewidth]{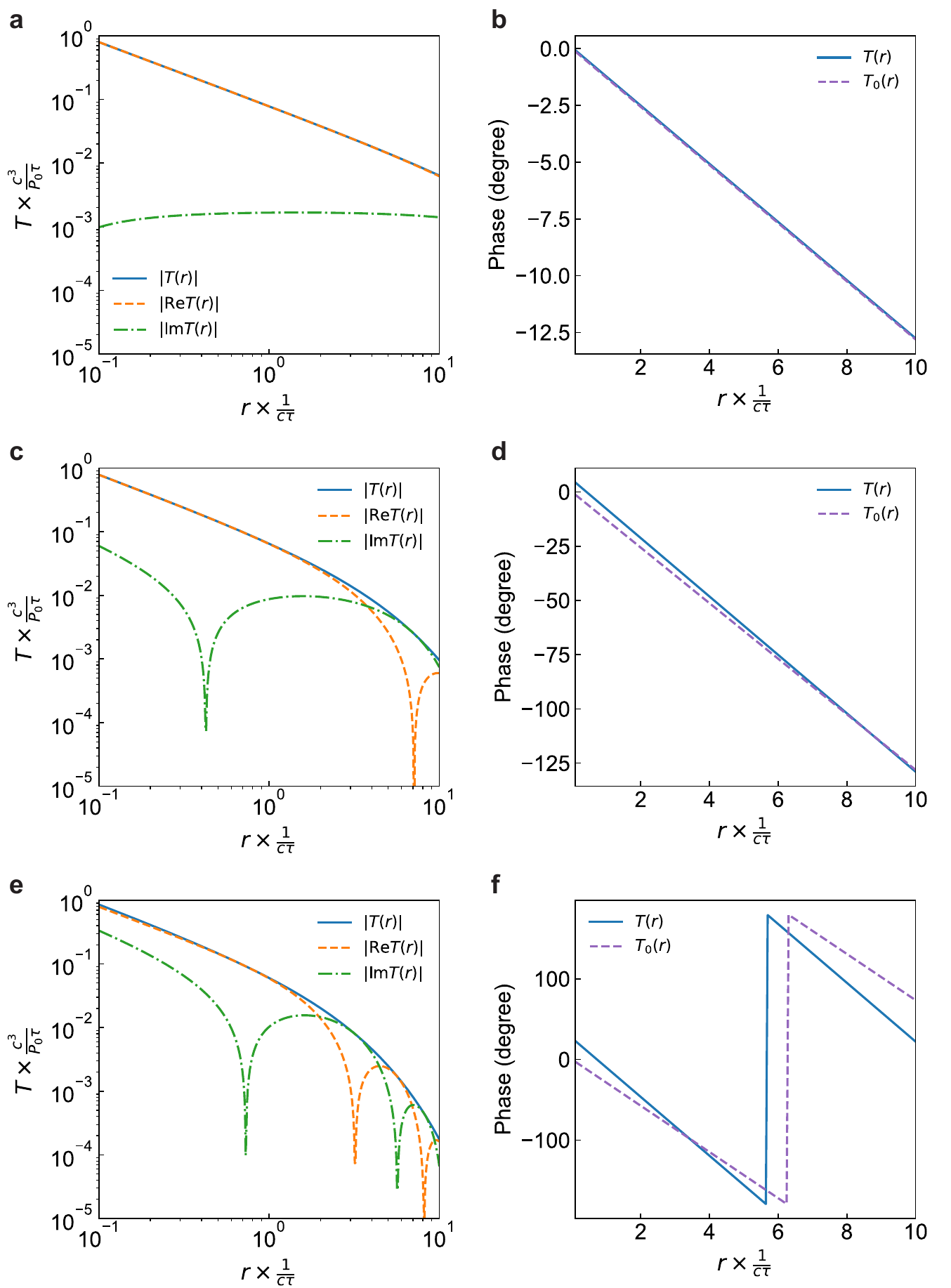}
\caption[The temperature profile in the small beam radius limit]{Left column: the absolute values of the complex temperature (blue line), the real part of the temperature (orange dashed line) and the imaginary part of the temperature (green dash-dotted line) as a function of distance $r$. Note that the dips in the real and imaginary parts of temperature correspond to a sign change and as the modulation frequency approaches $\tau^{-1}$, there are more oscillations across zero in the real and imaginary part of the temperature. Right column: the phase of temperature as a function of $r$ for the hyperbolic heat equation (blue line) and the conventional heat equation (purple line). In a-b, $\omega_0 = 0.001\times\tau^{-1}$. In c-d, $\omega_0 = 0.1\times\tau^{-1}$. In e-f, $\omega_0 = 0.5\times\tau^{-1}$. The parameters used in the model are $\tau = 1\,\mathrm{ns}$, $\kappa = 144\,\mathrm{W\cdot m^{-1}\cdot K^{-1}}$ and $C_p = 1.66\times 10^6\,\mathrm{J\cdot m^{-3}\cdot K^{-1}}$.}
\vfill
\label{green_appraoch}
\end{figure}

In Fig.~\ref{green_appraoch}, we calculate the temperature profile $T(r)$ at various modulation frequencies $\omega_0$. Generally, the amplitude of temperature $T(r)$ decreases with the distance $r$ and the phase almost linearly decreases with the distance $r$. We further find that when the modulation frequency is much smaller than $\tau^{-1}$, the phase of $T(r)$ recovers to the case of conventional heat equation. As the modulation frequency $\omega \sim \tau^{-1}$, the phase changes its sign at certain $r$, which can be considered as a \emph{unique} consequence of the hyperbolic heat conduction. It is important to note that the above analysis is rather qualitative as the small beam approximation usually does not hold unless the material has an extremely high thermal conductivity.

\clearpage
\clearpage

\section{Beyond Fourier's law of heat conduction: the phonon Boltzmann transport equation}
We follow the Green's function approach of modeling heat transfer using Boltzmann transport equation proposed in Ref.~\cite{PhysRevB.104.245424}, where a transient thermal grating geometry is discussed as an example, to derive the frequency-domain photo-reflectance signal with beam offset. The linearized phonon Boltzmann transport equation (BTE) takes the expression,
\begin{equation}
    \frac{\partial f_n}{\partial t}+\mathbf{v}_n\cdot\nabla f=Q_n\frac{N\nu}{\hbar\omega_n}+\sum_j W_{n,j} (f_j^0-f_j)
\end{equation}
where $\nu$ is the unit cell volume, $W_{n,j}$ is the scattering matrix, $Q_n$ is the volumetric heat generation (phonon excitation) associated with each phonon mode and $Q_n = p_n Q$ with $\sum_n p_n = 1$. A thermalized heating source satisfies, $p_n = c_n/C$, where $c_n = \frac{k_B}{N\nu}\left[\frac{\frac{\hbar\omega_n}{2k_B T_0}}{\sinh{\frac{\hbar \omega_n}{2k_B T_0}}}\right]^2$, $C = \sum_n c_n$ and $T_0$ is the background temperature. The effective temperature is defined to satisfy,
\begin{equation}
    \frac{1}{N\nu}\sum_n \hbar \omega_n f_{BE}\left(\frac{\hbar\omega_n}{k_B T}\right)=\frac{1}{N\nu}\sum_n \hbar\omega_n f_n
\end{equation}
Then, the equilibrium distribution can be linearized by and written in terms of a temperature rise,
\begin{equation}
    f_n^0 = f_{BE}\left(\frac{\hbar\omega_n}{k_B T_0}\right) + \frac{N\nu}{\hbar\omega_n}c_n\Delta T
\end{equation}

Define the phonon energy density deviation per mode, $g_n = \frac{\hbar\omega_n}{N\nu}\left[f_n - f_{BE}(\hbar\omega/k_B T_0)\right]$ and $g_n^0 = \frac{\hbar\omega_n}{N\nu}\left[f_n^0 - f_{BE}(\hbar\omega/k_B T_0)\right]=c_n\Delta T$. The linearized BTE in terms of the deviational phonon energy density  is,
\begin{equation}
    \frac{\partial g_n}{\partial t}+\mathbf{v}_n\cdot\nabla g_n = Q p_n + \underbrace{\sum_j \omega_n W_{n,j}\omega_j^{-1}\left(c_j\Delta T-g_j\right)}_\text{scattering matrix term for $n$}
    \label{z}
\end{equation}
The total deviational energy density can be used to compute the temperature change with respect to the background temperature change,
\begin{equation}
    C\Delta T = \sum_n g_n
\end{equation}
The summation of all scattering matrix terms should be zero, leading to the following sum rule,
\begin{equation}
    \sum_{i,j}\omega_i W_{i,j}\omega_j^{-1}\left(\frac{c_j}{C}-\delta_{jm}\right)=0
\end{equation}
Notably, under the relaxation time approximation, we have,
\begin{equation}
    \frac{1}{\tau_m} = \sum_i \frac{c_i}{C}\frac{1}{\tau_i}
\end{equation}

The Fourier transform of the deviational energy density gives rise to,
\begin{equation}
    i\omega \tilde{g}_n + i\mathbf{q}\cdot\mathbf{v}_n \tilde{g}_n  =  \tilde{Q}p_n+\sum_j\omega_nW_{n,j}\omega_j^{-1}\left(c_j\Delta \tilde{T}-\tilde{g}_j\right)
\end{equation}
Denoting,
\begin{equation}
    A_{n,j} = i(\omega+\mathbf{q}\cdot\mathbf{v}_n)\delta_{n,j}+\omega_nW_{n,j}\omega_j^{-1} 
\end{equation}
We then have,
\begin{equation}
    \begin{split}
    \tilde{g}_n &= \tilde{Q}\sum_jA^{-1}_{n,j}p_j+\Delta\tilde{T}\sum_j\left[\delta_{n,j}-A^{-1}_{n,j}i(\omega+\mathbf{q}\cdot\mathbf{v}_n)\right]c_j\\
    &=\tilde{Q}\sum_jA^{-1}_{n,j}p_j+\Delta\tilde{T}\left[c_n-i\sum_jA^{-1}_{n,j}(\omega+\mathbf{q}\cdot\mathbf{v}_n)c_j\right]
    \end{split}
\end{equation}

The temperature rise in the reciprocal space and time domain is given by,
\begin{equation}
    C\Delta \tilde{T} = \sum_n\tilde{g}_n= \tilde{Q}\sum_{n,j}A^{-1}_{n,j}p_j+\Delta\tilde{T}\sum_n\left[c_n-i\sum_jA^{-1}_{n,j}(\omega+\mathbf{q}\cdot\mathbf{v}_n)c_j\right]
\end{equation}
meaning that,
\begin{equation}
    \tilde{Q}\sum_{n,j}A^{-1}_{n,j}p_j-\Delta\tilde{T}\sum_n\left[i\sum_jA^{-1}_{n,j}(\omega+\mathbf{q}\cdot\mathbf{v}_n)c_j\right]=0
\end{equation}
Thereby, the temperature rise can be expressed as,
\begin{equation}
    \Delta \tilde{T}=\tilde{Q}\frac{\sum_{n,j}A_{n,j}^{-1}p_j}{i\sum_{n,j}A^{-1}_{n,j}\left(\omega+\mathbf{q}\cdot\mathbf{v}_n\right)c_j}
    \label{ft}
\end{equation}
where the second term can be regarded as a Green's function.
The deviational phonon energy density can be simplified as,
\begin{equation}
\tilde{g}_n =  \tilde{Q}\sum_jA^{-1}_{n,j}p_j+\tilde{Q}\frac{\sum_{n,j}A_{n,j}^{-1}p_j}{i\sum_{n,j}A^{-1}_{n,j}\left(\omega+\mathbf{q}\cdot\mathbf{v}_n\right)c_j}\left[c_n-i\sum_jA^{-1}_{n,j}(\omega+\mathbf{q}\cdot\mathbf{v}_n)c_j\right]   
\end{equation}

In the following, we want to discuss how to use the Green's function approach in the Boltzmann transport equation framework to evaluate the signal in FDTR geometry.
We first consider a simple pump laser heating geometry---a point source, $Q(\mathbf{r},t) = e^{i\omega t}\delta(\mathbf{r})$ and its Fourier transform is $\tilde{Q}(\omega) = 1$.
The temperature rise in the real space can be expressed as,
\begin{equation}
    \begin{split}
    \Delta T(\mathbf{r})& = \frac{1}{(2\pi)^3}\int \Delta \tilde{T}e^{i\mathbf{q}\cdot\mathbf{r}} d\mathbf{q} \\
    &= \frac{1}{(2\pi)^3}\int   \frac{\sum_{n,j}A_{n,j}^{-1}p_j}{i\sum_{n,j}A^{-1}_{n,j}\left(\omega+\mathbf{q}\cdot\mathbf{v}_n\right)c_j} e^{i\mathbf{q}\cdot\mathbf{r}} d\mathbf{q}
    \end{split}
    \label{deltaf}
\end{equation}
For material like Ge, the integrand in Eq.~(\ref{deltaf}) is isotropic. That is, one can simplify the integral into one-dimensional and only integrate along $(0,0,\xi)$.

Next, we consider the Gaussian beam. We assume the beam follows the expression $Q = e^{i\omega t} \times \frac{2}{\pi \sigma_x\sigma_y}e^{-\frac{2x^2}{\sigma^2_x}}e^{-\frac{2y^2}{\sigma^2_y}}\alpha e^{-\alpha |z|}$, where $\sigma_x$ and $\sigma_y$ are the $1/e^2$ radii along x and y directions. Its Fourier transform writes, $\tilde{Q}(\omega)=e^{-q_x^2\sigma_x^2/8}e^{-q_y^2\sigma_y^2/8}\frac{2\alpha^2}{\alpha^2+q_z^2}$. The temperature distribution writes,
\begin{equation}
    \Delta T = \frac{1}{(2\pi)^3}\int   \frac{\sum_{n,j}A_{n,j}^{-1}p_j}{i\sum_{n,j}A^{-1}_{n,j}\left(\omega+\mathbf{q}\cdot\mathbf{v}_n\right)c_j} e^{-q_x^2\sigma_x^2/8}e^{-q_y^2\sigma_y^2/8}\frac{2\alpha^2}{\alpha^2+q_z^2} e^{i\mathbf{q}\cdot\mathbf{r}}  d\mathbf{q}
\end{equation}
The probed signal writes,
\begin{equation}
    \begin{split}
    I & =\frac{1}{2} \int \int_{-\infty}^\infty \frac{2\beta}{\pi\sigma_x^\prime\sigma_y^\prime}T(\mathbf{r})\beta e^{-2(x-x_\mathrm{o})^2/\sigma_x^{\prime 2}} e^{-2(y-y_\mathrm{o})^2/\sigma_y^{\prime 2}} e^{-\beta |z|}  dz d\mathbf{r}_\parallel\\
      & = \frac{1}{2} \frac{1}{(2\pi)^3}\int \int \int \frac{\sum_{n,j}A_{n,j}^{-1}p_j}{i\sum_{n,j}A^{-1}_{n,j}\left(\omega+\mathbf{q}\cdot\mathbf{v}_n\right)c_j} e^{-q_x^2(\sigma_x^2+\sigma_x^{\prime 2})/8} e^{-q_y^2(\sigma_y^2+\sigma_y^{\prime 2})/8} e^{-iq_x x_\mathrm{o}}e^{-iq_y y_\mathrm{o}} d \mathbf{q}_\parallel \\
      &\times\frac{2\alpha^2}{\alpha^2+q_z^2} e^{iq_z z-\beta |z|}  dq_z  dz\\
      & = \frac{1}{2(2\pi)^3}  \int \frac{\sum_{n,j}A_{n,j}^{-1}p_j}{i\sum_{n,j}A^{-1}_{n,j}\left(\omega+\mathbf{q}\cdot\mathbf{v}_n\right)c_j} e^{-q_x^2(\sigma_x^2+\sigma_x^{\prime 2})/8} e^{-q_y^2(\sigma_y^2+\sigma_y^{\prime 2})/8} e^{-iq_x x_\mathrm{o}}e^{-iq_y y_\mathrm{o}}\frac{2\alpha^2}{\alpha^2+q_z^2}  \frac{2\beta^2}{\beta^2+q_z^2} d\mathbf{q}
    \end{split}
    \label{ilbte}
\end{equation}
where $\sigma_x^\prime$ and $\sigma_y^\prime$ are the probe radii.

To evaluate Eq.~(\ref{ilbte}), we need to compute the Green's function on a $\mathbf{q}$ point grid with $\mathbf{q}=\left(\mathbf{q}_\parallel,q_z\right)$. At each $\mathbf{q}$ point, a matrix inversion operation is involved and the matrix has the dimension of the number of phonon modes ($3N_xN_yN_zN_\mathrm{atom}$). This brings great computational challenges and the implementation goes beyond the scope of this work. However, we believe this is an intrinsically more accurate approach to model the temperature responses in a FDTR measurement. It is worth mentioning that in deriving Eq.~(\ref{ilbte}), we have implicitly assumed the Green's function does not depend on the boundary condition, which is generally not true. Including the boundary condition properly when solving linearized Boltzmann transport equation is not a trivial task and deserve a separate paper. Nevertheless, Eq.~(\ref{ilbte}) still has great advantages over Fourier's law, as it captures the essential physics of nonlocal thermal transport that Fourier's law fails to by construction.



\clearpage

\section{Measurement of the probe beam's radius}
\begin{figure}[H]
    \centering
\includegraphics[width=0.7\linewidth]{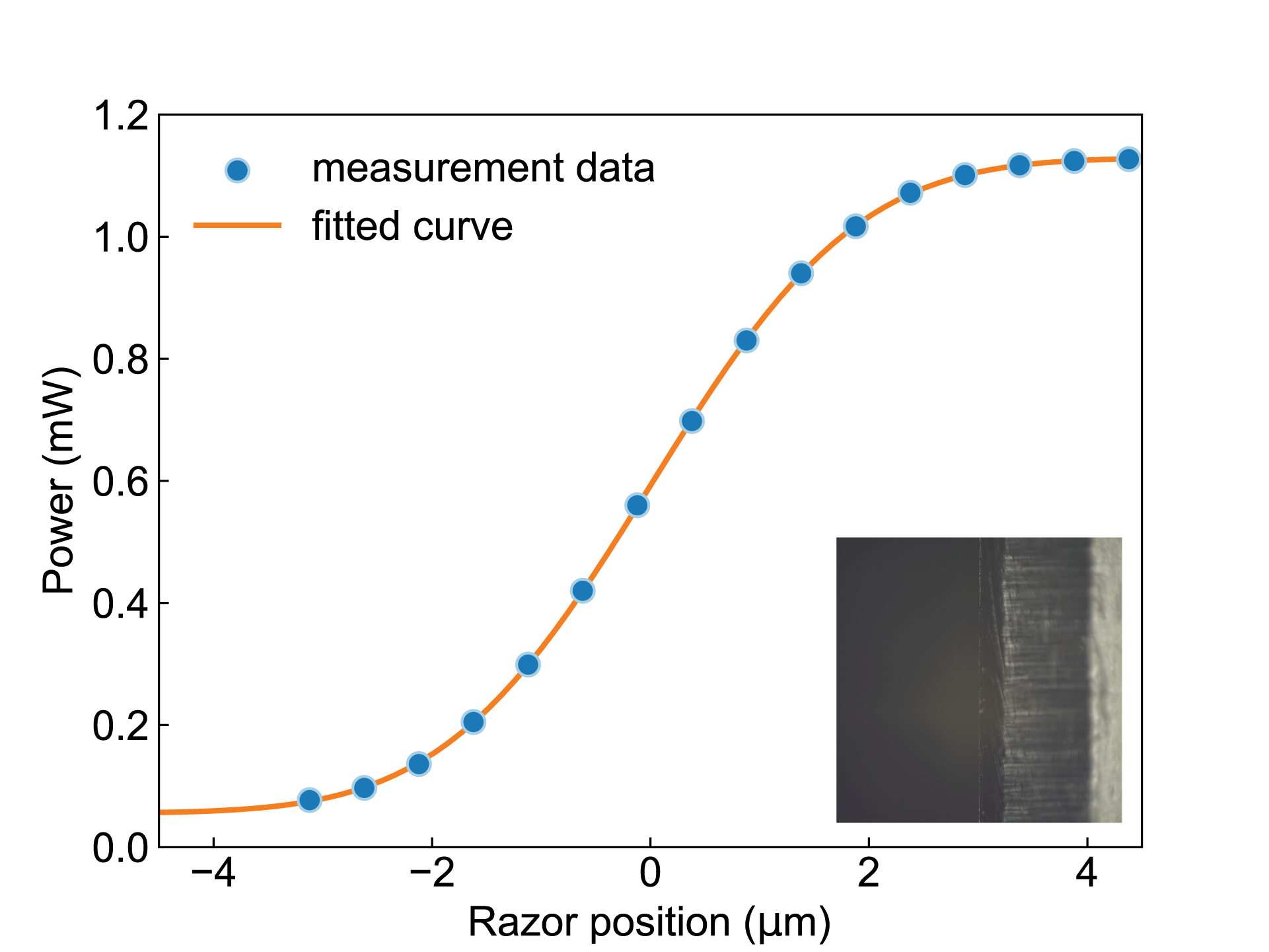}
\caption[Knife edge method to obtain the beam radius]{The knife edge method to obtain the probe radius. The razor, as shown in the inset image, is mounted onto the PInano XYZ piezo stage, such that when the stage moves, the razor moves as well. The laser power is measured using a Thorlabs PM100USB power meter. The function form used in curve fitting is $y(x)=A \mathrm{erf}\left[\frac{\sqrt{2}(x - x_0)}{r_0}\right] + C$, where $r_0$ is the $1/e^2$ beam radius. The best fit gives $A=0.54$ mW, $x_0=3.12$ $\mathrm{\mu}$m, $r_0=2.98$ $\mathrm{\mu}$m and $C=0.59$ mW. Note that the position of razor has been shifted by $-x_0$ so that the beam center is at the origin.}
\label{probe_radius}
\end{figure}

\bibliography{si}
\bibliographystyle{unsrt}